\documentclass[aps,twocolumn,superscriptaddress,floatfix,showpacs]{revtex4} 
\usepackage{amsmath,euscript,theorem,graphicx,rotating}
\usepackage[latin1]{inputenc}
\usepackage[T1]{fontenc}
\usepackage{txfonts}
\usepackage{stackengine}
\usepackage{url}
\usepackage{array,dcolumn}
\usepackage{xr} % for eXternal References
\usepackage{color}
\usepackage{dcolumn}% Align table columns on decimal point
\usepackage{bm}% bold math
\usepackage{longtable}% To allow splitting tables across multiple pages
\usepackage{xspace}% To preserve spaces after macros
\usepackage{booktabs}
\usepackage{version}
\bibstyle{achemso}

\usepackage{array}
\newcolumntype{L}{>{$}l<{$}}
\newcolumntype{R}{>{$}r<{$}}
\newcolumntype{C}{>{$}c<{$}}

\newcommand{\rr}{{\mathbf r}}

\newcommand{\etal}{\emph{et al.}\xspace}

\newcommand{\kJpermol}{\ensuremath{\rm{kJ\,mol}^{-1}}\xspace}

\newcommand{\rms}[0]{r.m.s.\xspace}

\newcommand{\EIND}[1]{\ensuremath{E^{(#1)}_{\rm IND}}\xspace}
\newcommand{\EpolMP}[1]{\ensuremath{V^{(#1)}_{\rm pol}[{\rm DM}]}\xspace}

\newcommand{\Eint}[1]{\ensuremath{E_{\rm int}^{(#1)}}\xspace}
\newcommand{\Esr}[1]{\ensuremath{E_{\rm sr}^{(#1)}}\xspace}

\newcommand{\deltaHF}[0]{\ensuremath{\delta^{\rm HF}_{\rm int}}\xspace}
\newcommand{\VSR}[1]{\ensuremath{V^{#1}_{\rm sr}}\xspace}
\newcommand{\Vsr}[2]{\ensuremath{V^{#1}_{\rm sr}[#2]}\xspace}
\newcommand{\Cn}[1]{\ensuremath{C_{#1}}\xspace}

\newcommand{\Cniso}[1]{\ensuremath{C_{#1}{\rm (iso)}}\xspace}

\newcommand{\ECT}[1]{\ensuremath{E_{\rm CT}^{(#1)}}\xspace}
\newcommand{\EPOL}[1]{\ensuremath{E_{\rm POL}^{(#1)}}\xspace}

\newcommand{\betapol}[0]{\ensuremath{\beta_{\rm pol}}\xspace}

\newcommand{\OpenMM}{{\sc OpenMM}\xspace}
\newcommand{\DLPOLY}{{\sc DL\_POLY}\xspace}
\newcommand{\CamCASP}{{\sc CamCASP}\xspace}
\newcommand{\ORIENT}{{\sc Orient}\xspace}
\newcommand{\DMACRYS}{{\sc DMACRYS}\xspace}
\newcommand{\Orient}{\ORIENT}

\newcommand{\vdw}{\mbox{\ensuremath{\cdot\!\cdot\!\cdot}}}

\mathchardef\lt="313C \mathchardef\gt="313E
\mathcode`<="4268 \mathcode`>="5269
\newcolumntype{d}[1]{D{.}{.}{#1}}
\newcolumntype{.}{D{.}{.}{-1}}
\newcolumntype{,}{D{,}{,}{2}}

\newcommand{\JCP}[0]{J. Chem. Phys.\ }
\newcommand{\JPCA}[0]{J. Phys. Chem. A\ }

\newcommand{\JCTC}[0]{J. Chem. Theory Comput.\ }

\newcommand{\CPL}[0]{Chem. Phys. Lett.\ }

\newcommand{\PCCP}[0]{Phys. Chem. Chem. Phys.\ }

\newcommand{\IRPC}[0]{Int. Revs. Phys. Chem.\ }

\newcommand{\CrystEngComm}[0]{CrystEngComm.\ }

\newcommand{\ChemPhysChem}[0]{ChemPhysChem\ }

\newcommand{\ChemComm}[0]{Chem. Commun.\ }

\newcommand{\paperA}{Part~I\xspace}

\begin{document}

\title[CamCASP]
{
  {\em Ab initio} atom--atom potentials using \CamCASP : 
  Many-body potentials for the pyridine dimer.
}

\author{Alston J. Misquitta}
\affiliation{School of Physics and Astronomy, Queen Mary, University of London,
London E1 4NS, U.K.}
\email{a.j.misquitta@qmul.ac.uk}
\author{Anthony J. Stone}
\affiliation{University Chemical Laboratory, Lensfield Road,
Cambridge, CB2 1EW, U.K.}

\date{\today}

\begin{abstract}
    In \paperA of this two-part investigation we described a methodology
    for the development of robust, analytic, many-body atom--atom potentials for
    small organic molecules from first principles and demonstrated how the
    \CamCASP program can be used to derive the damped, distributed multipole 
    models for pyridine.
    Here we demonstrate how the theoretical ideas for the short-range models
    described in \paperA, which are implemented in the \CamCASP
    suite of programs, can be used to develop a series of many-body
    potentials for the pyridine system. 
    Even the simplest of these potentials exhibit \rms errors of only
    about $0.6 \kJpermol$ for the low-energy pyridine dimers, significantly
    surpassing the best empirical potentials.
    Our best model is shown to support eight stable minima,  
    four of which have not been reported in the literature before.
    Further, the functional form can be made systematically more elaborate so as 
    to improve the accuracy without a significant increase in the human-time
    spent in their generation. 
    We investigate the effects of anisotropy, rank of multipoles, and
    choice of polarizability and dispersion models.
\end{abstract}

\maketitle

\section{Introduction}
\label{sec:introduction}

In \paperA we introduced our strategy for developing accurate, analytic, 
many-body potentials for molecular systems in manner that is both robust and
easy to implement. The key ideas of this strategy included the following:
(1) that all of the long-range potential parameters are derived from the
density and density response functions; 
(2) that the short-range parameters, including the atomic shape anisotropy,
are obtained via the distributed density-overlap model with a partitioning 
scheme based on the iterated stockholder atom (ISA) approach \cite{LillestolenW08,LillestolenW09}; 
(3) that we derive the potential in stages, with parameters derived or
fitted at each stage used as prior values for the next stage.
Uniquely in our scheme, the ISA density-partitioning method plays a central 
role in the potential development process; in particular, the ISA is key to
the robustness of the methodology used to determine the short-range parameters,
which pose a significant problem for standard fitting methods.
We have based our approach on the  basis-space ISA, or BS-ISA,
algorithm \cite{MisquittaSF14} that allows us to 
partition a molecular density uniquely into atomic domains and obtain analytic expansions
for the ISA atoms. The BS-ISA algorithm has many desirable
properties that make it ideal for our purpose \cite{MisquittaSF14}:
there is a well-defined basis-set limit to the ISA atoms; 
the distributed multipole moments resulting from this partitioning
have been shown to be amongst the most rapidly converging with rank;
and the ISA atoms are, in an information-theoretic sense, the most 
spherical atoms possible that simultaneously take into account atomic
electronegativity changes in the molecule.
It is because of these properties that we expect the short-range potential parameters
to be physical and well-defined.

In this paper we will apply the methodology presented in \paperA to develop
a set of many-body potentials for pyridine. 
In a study such as this is, it is important to use a system that 
simultaneously presents a challenge and also allows tests to be 
performed to validate the method sufficiently. We have chosen to use
the pyridine dimer
as our example as it is small enough to permit accurate interaction energy
calculations using SAPT(DFT) on as dense a grid as is needed, but 
large enough to exhibit a varied and complex potential energy surface (PES) 
with---as we shall see below---eight distinct minima. Additionally, the pyridine
molecule has a sizeable dipole moment and polarizability, so polarization effects are
expected to be important, and, as we shall see, the two-body
charge-transfer, or charge-delocalisation \cite{Misquitta13a}, energy
is also significant.
Finally, from the crystallographic studies by Price and co-workers \cite{AnghelDP02}
it is known that the crystal energy landscape of this molecule is complex and 
poses a significant challenge for seemingly accurate empirical potentials.
While we will not attempt to use the results of this study in a crystal structure
prediction, we intend to perform this test in later work.

This paper is organised as follows: we begin with a description of the numerical
details of the electronic structure calculations used in this work and 
a description of the data sets used in the potential development process. 
Next, in \S\ref{sec:short-range} we develop a set of short-range models
which are then combined with the long-range models derived in \paperA 
in \S\ref{sec:total}.
The resulting total energy potentials are analysed in \S\ref{sec:results}
where we compare pyridine dimer minima, vibrational frequencies and
the second virial coefficients on these surfaces. 
In \S\ref{sec:analysis} we critically analyse aspects of the methodology and
the potentials. In particular, we examine the stability of the potentials with 
respect to multipole rank. In the Conclusions we examine shortcomings of the method
and indicate possible directions for this work.

\section{Numerical details}
\label{sec:numerical}

The numerical details related to the distributed multipole
models and SAPT(DFT) calculations are described in \S\ref{A-sec:numerical}
in \paperA.
There we also describe the three data sets used in the development of
the pyridine potentials. 
Here we provide additional numerical details related to the weighting
schemes used in the fitting process.

The distributed density-overlap fits were performed using the 
\CamCASP program using the Gaussian/Log weighting scheme \cite{HodgesW00b}
in which $w_{\rm GL}(e) = \exp(-\alpha(\ln (e/E_0))^2)$, where $\alpha = 1/\ln 10$
and $E_0 = 100$ \kJpermol. 
Here the parameter $E_0$ sets the energy-scale for the fit, and it is usually
chosen to be some large multiple of the absolute global minimum dimer energy
so as to obtain a reliable fit to the repulsive wall.
The fits to individual site--site potentials \Vsr{}{ab} were performed
with the \Orient program using the same Gaussian/Log weighting scheme.

All relaxation steps were performed using the \Orient program
using the Boltzmann weighting function 
\begin{align}
  w_{\rm Bol}(e) = 
     \begin{cases}
         \exp{((e_{\rm low} - e)/E_0)}  & \text{for} \  e \gt e_{\rm low} \\
         1.0                            & \text{otherwise}.
     \end{cases}
\end{align}
Here $e_{\rm low}$ is typically set to the smallest energy in the data
set and the energy-scale for the fit is set by $E_0 = 40$ \kJpermol 
to increase the weight to lower energies.
We used $e_{\rm low} = 0$ \kJpermol for the relaxation of the repulsive energies,
and $-10$ \kJpermol in the final relaxation step involving the total 
interaction model.

\section{Short-range fit}
\label{sec:short-range}

\subsection{Fitting strategy and atomic shape}
\label{sec:sr-fitting-strategy}

We set out the fitting strategy for the short-range part of the
potential in some detail in \S\ref{A-sec:strategy} of \paperA. 
In this multi-stage approach we first fit to \Esr{1} calculated on 
the dense, pseudo-random set of 3515 dimers in Dataset(0).
This is done via the distributed density-overlap model
which allows us to partition \Esr{1} into contributions from pairs of
sites, and fit the terms in the potential for each atom-pair individually. 
However, if the atoms are close to spherical, as is the case for the
ISA atom densities, the atom-pair 
shape function $\rho_{ab}(\Omega_{ab})$ that appears in the
potential (see eq.~\eqref{A-eq:Vtot_Vsr} in \paperA) may be written to a
good approximation as
the sum of shape functions for the interacting atoms 
(see ch.\ 12\ in ref.~\citenum{Stone:book:13})
\begin{align}
    \rho_{ab}(\Omega_{ab}) \approx \rho_{a}(\Omega_{a}) + \rho_{b}(\Omega_{b}).
    \label{eq:shape-func-additive}
\end{align}
Here $\Omega_{a}$ is a generalised angular coordinate that describes the
direction of the vector from site $a$ to site $b$ in the local coordinate 
system of site $a$, and likewise for $\Omega_{b}$, and $\rho_a$ and $\rho_b$
are the atomic shape functions for atoms $a$ and $b$.
The atomic shape functions for all atoms of a given type should be the same. 

The shape-function additivity is observed in the first stage of the fitting
when the terms in \Vsr{(1)}{ab} are fitted individually via the density-overlap 
model, but it is not exact,
probably in part because of grid sampling variability around the sites.
It can however be exactly enforced in the next stage when the short-range
parameters are collectively relaxed in a constrained manner to the 
\Esr{1} energies in Dataset(1). 
We find it best to perform this relaxation iteratively, with only 
those parameters associated with a particular subset of sites relaxed at each step.
With this approach the constrained relaxation can be performed rapidly,
in a computationally efficient manner. At each step, shape-function additivity
is imposed by using pinning (prior) values for the parameters
from the averaged shape-function parameters from the previous step.

In a similar manner, we may relax the resulting potential parameters to include
effects from second and higher orders in the interaction operator.
However, there is no reason to expect the shape-function additivity to 
hold at this stage, as the higher-order short-range effect, 
which is predominantly the charge-transfer (or charge-delocalisation) energy,
depends on the pair of atoms involved in a non-additive manner.
In the absence of additivity, 
the number of independent parameters in the potential would depend quadratically
on the number of interacting atoms, but
fortunately, as we will demonstrate below, the higher-order correction
can be treated as isotropic. That is, the atom-pair shape function now becomes
\begin{align}
    \rho_{ab}(\Omega_{ab}) = \rho_{a}(\Omega_{a}) + \rho_{b}(\Omega_{b}) - \delta_{ab},
    \label{eq:shape-func-nonadditive}
\end{align}
where $\delta_{ab}$ is the isotropic higher-order correction.

We will now examine the effectiveness of this strategy in obtaining
a series of fits to the short-range potential for the pyridine dimer.

\subsection{Fitting using the distributed density overlap model}
\label{sec:dist-density-overlap}

In principle, it is straightforward to use the distributed
density-overlap model described above.  We have used this
approach\cite{MisquittaWSP08,TottonMK10a}, 
as have others \cite{NyelandT86,MitchellP00,HodgesW00b,PiquemalCRG06},
with a reasonable degree of success.
The problem lies in the choice of density partitioning method. 
There is no unique way of decomposing a density into atom-like domains,
yet the tacit assumption of the distributed density-overlap model is that
the partitioned density $\rho^{A}_{a}$ is well-behaved and may be used
to extract properties such as size and shape of the atom located on
site $a$. If this were not the case, then the potential parameters 
extracted from the model would be meaningless, and indeed, a fit to 
eq.~\eqref{A-eq:Vsr}  in \paperA could even be so poor as to be useless.
In the past we have used a density-fitting-based scheme to partition the
density \cite{MisquittaWSP08}. This works by expressing the electronic 
density as a single sum over an auxiliary basis set with functions located
on the atomic nuclei, which then naturally suggests a partitioning scheme:
\begin{align}
    \rho(\rr) &= \sum_k d_k \chi_k(\rr) \nonumber \\
              &= \sum_a \sum_{k \in a} d_k \chi_k(\rr) = \sum_a \rho_{a}(\rr).
    \label{eq:df-partitioning}
\end{align}
Here the $d_k$ are expansion coefficients and $\chi_k$ are Gaussian basis functions
from the auxiliary basis. We have previously argued \cite{MisquittaS06} that since the 
auxiliary basis sets are optimised on free atoms, or homo-diatoms, they may
be used in the above manner to partition the molecular density into atom-like 
parts. This does seem to work, but only if small enough auxiliary basis sets
are used, and even then, the resulting atomic domains may be meaningless.

\begin{figure}
    % Fig 6
    \begin{center}
        %\stackunder[-5pt]{\includegraphics[scale=0.23]{./figs/AIM/pyr_N_DFtzvpp_noQ_iso0p001.png}}{N}
        %\stackunder[-5pt]{\includegraphics[scale=0.23]{./figs/AIM/pyr_C1_DFtzvpp_noQ_iso0p001.png}}{C1}
        %\stackunder[-5pt]{\includegraphics[scale=0.23]{./figs/AIM/pyr_C2_DFtzvpp_noQ_iso0p001.png}}{C2}
        %\stackunder[-5pt]{\includegraphics[scale=0.23]{./figs/AIM/pyr_C3_DFtzvpp_noQ_iso0p001.png}}{C3}
        %\stackunder[-5pt]{\includegraphics[scale=0.23]{./figs/AIM/pyr_H1_DFtzvpp_noQ_iso0p001.png}}{H1}
        %\stackunder[-5pt]{\includegraphics[scale=0.23]{./figs/AIM/pyr_H2_DFtzvpp_noQ_iso0p001.png}}{H2}
        %\stackunder[-5pt]{\includegraphics[scale=0.23]{./figs/AIM/pyr_H3_DFtzvpp_noQ_iso0p001.png}}{H3}
        \stackunder[-5pt]{\includegraphics[scale=0.23]{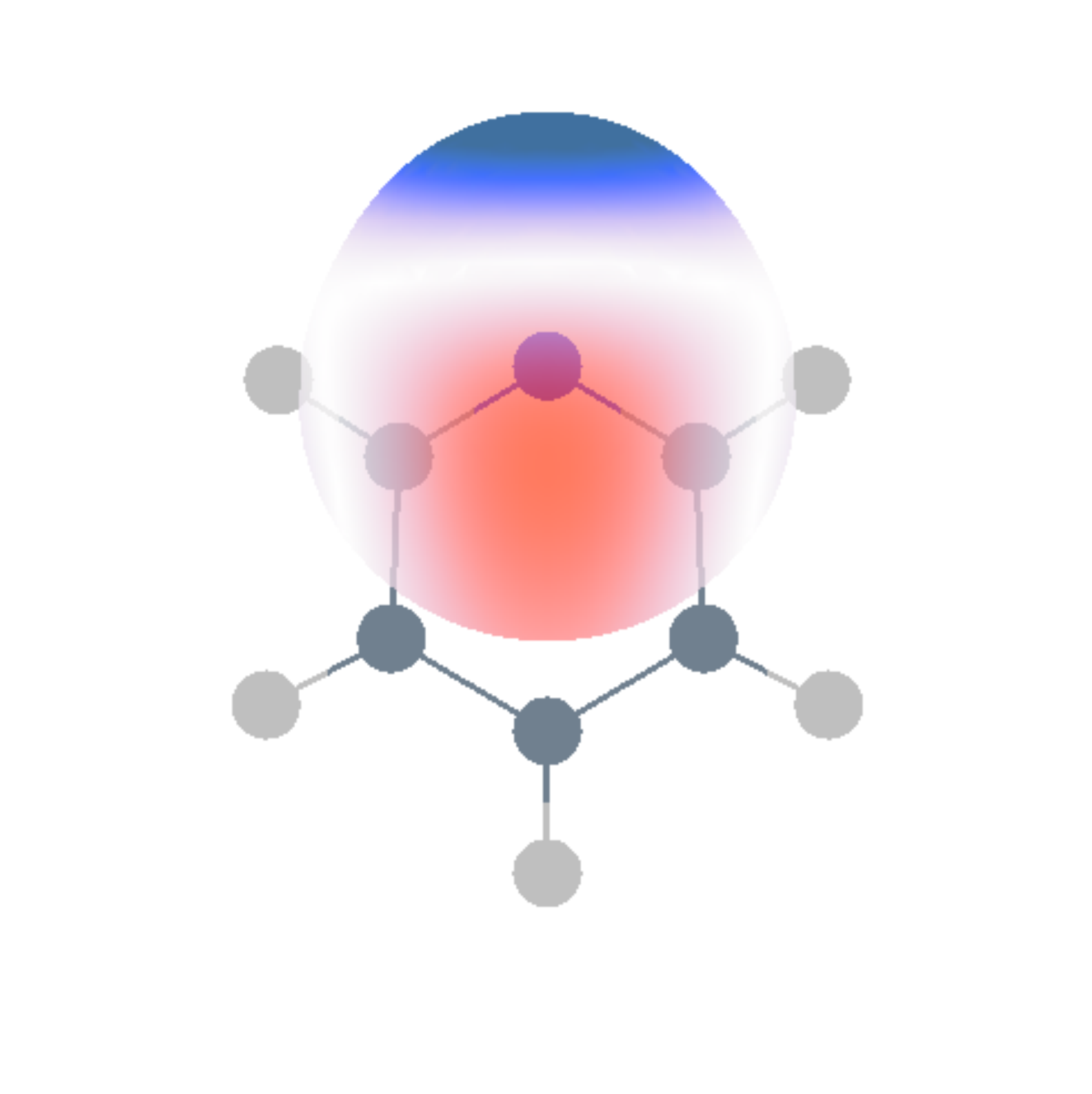}}{N}
        \stackunder[-5pt]{\includegraphics[scale=0.23]{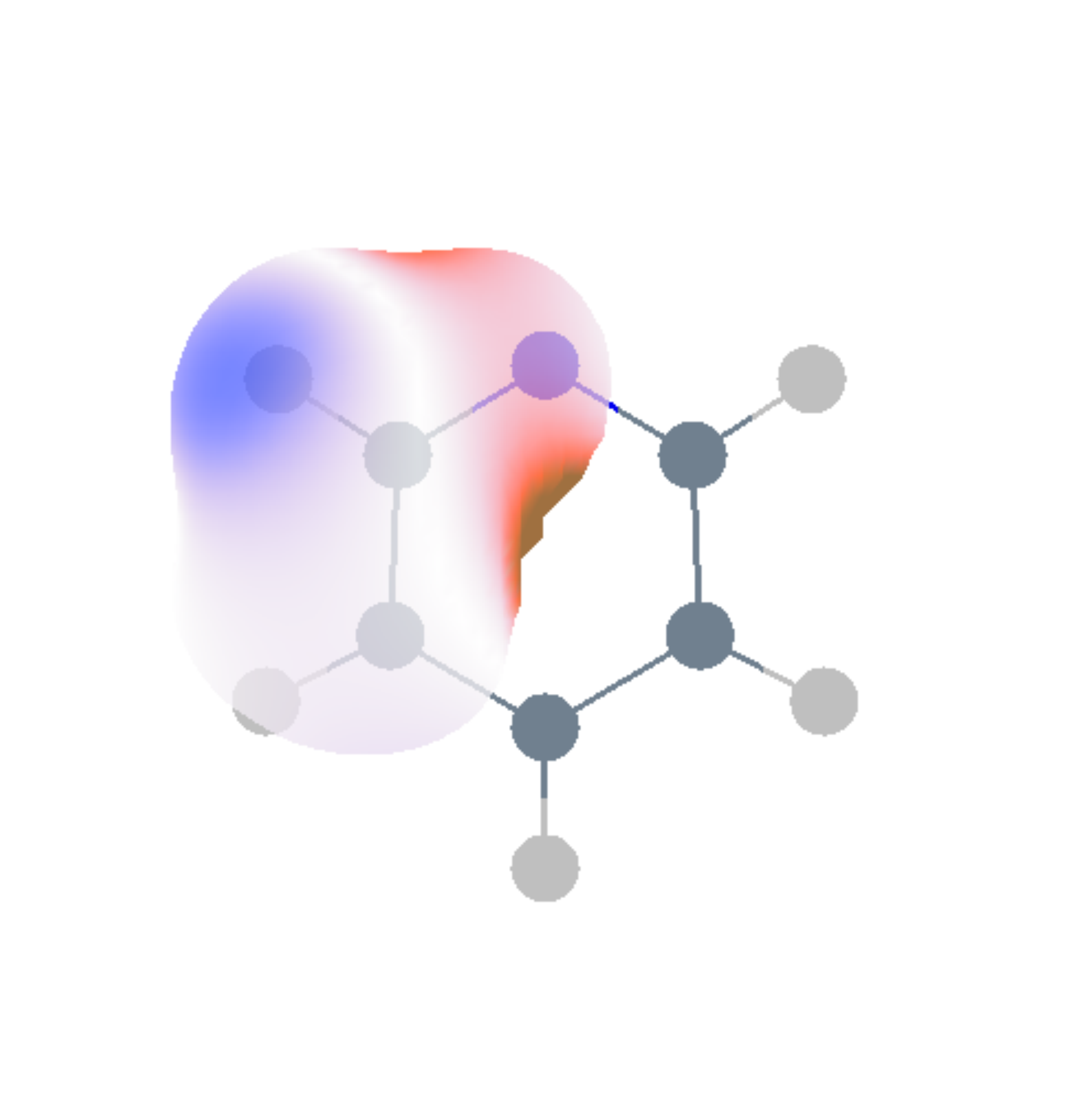}}{C1}
        \stackunder[-5pt]{\includegraphics[scale=0.23]{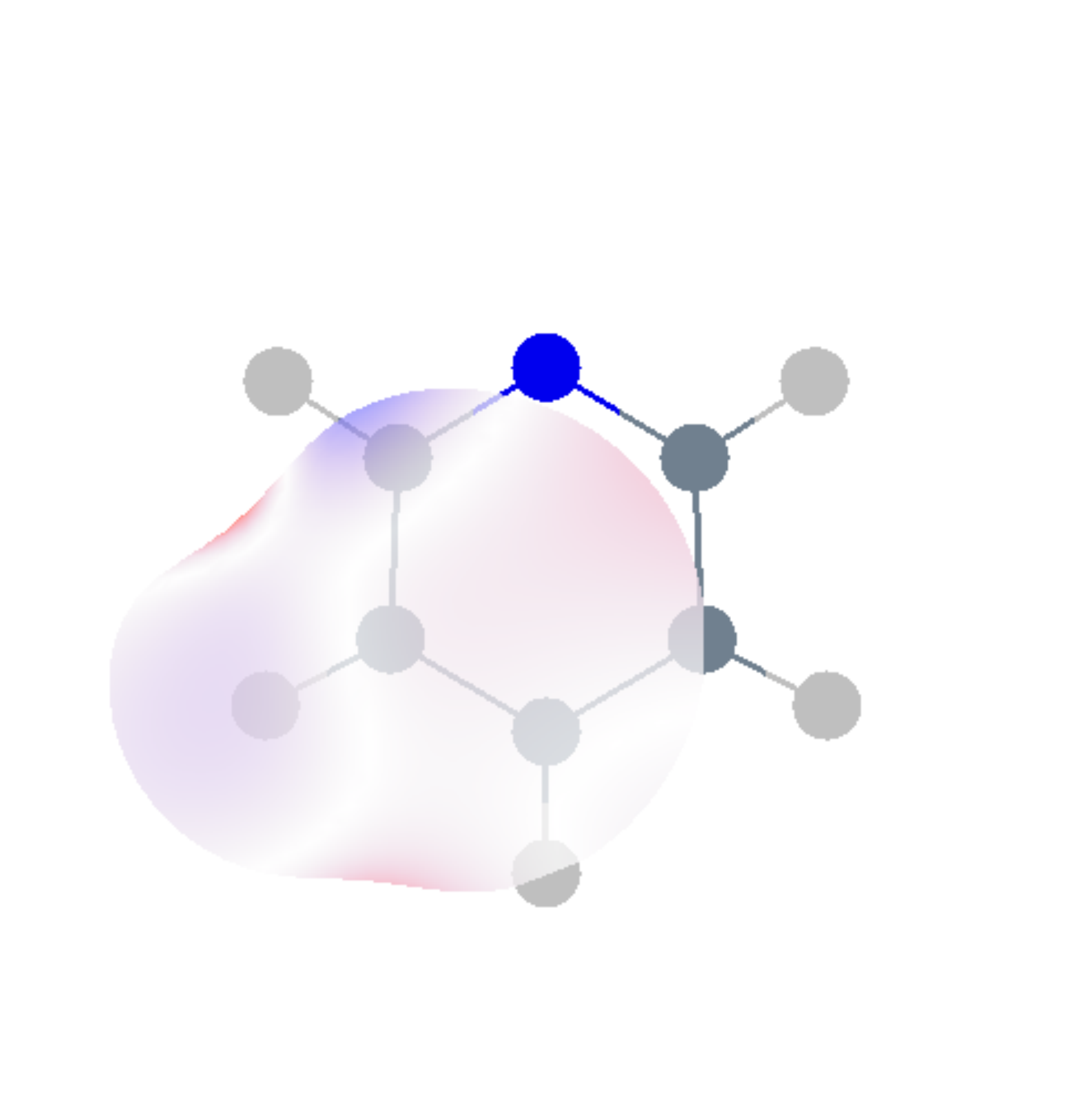}}{C2}
        \stackunder[-5pt]{\includegraphics[scale=0.23]{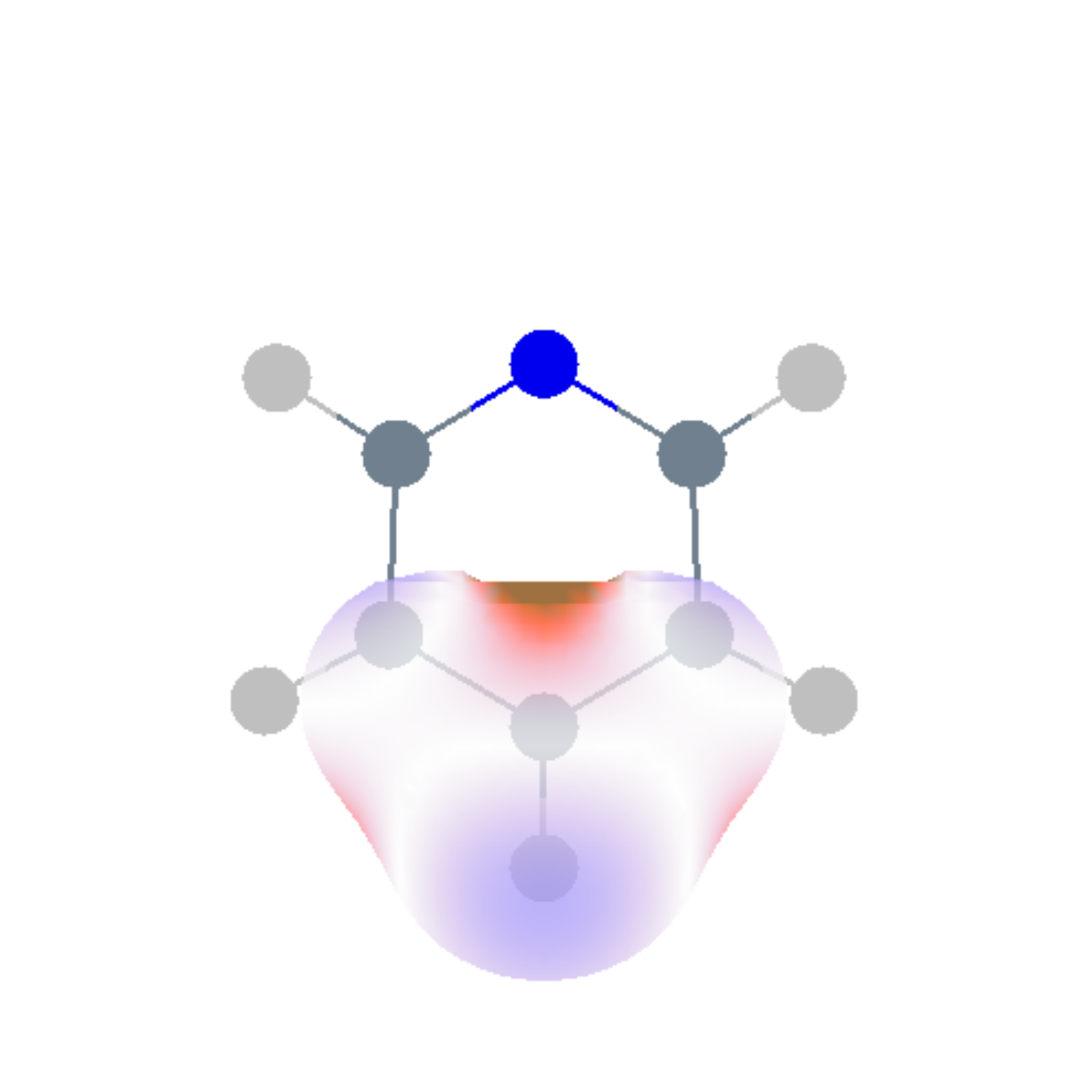}}{C3}
        \stackunder[-5pt]{\includegraphics[scale=0.23]{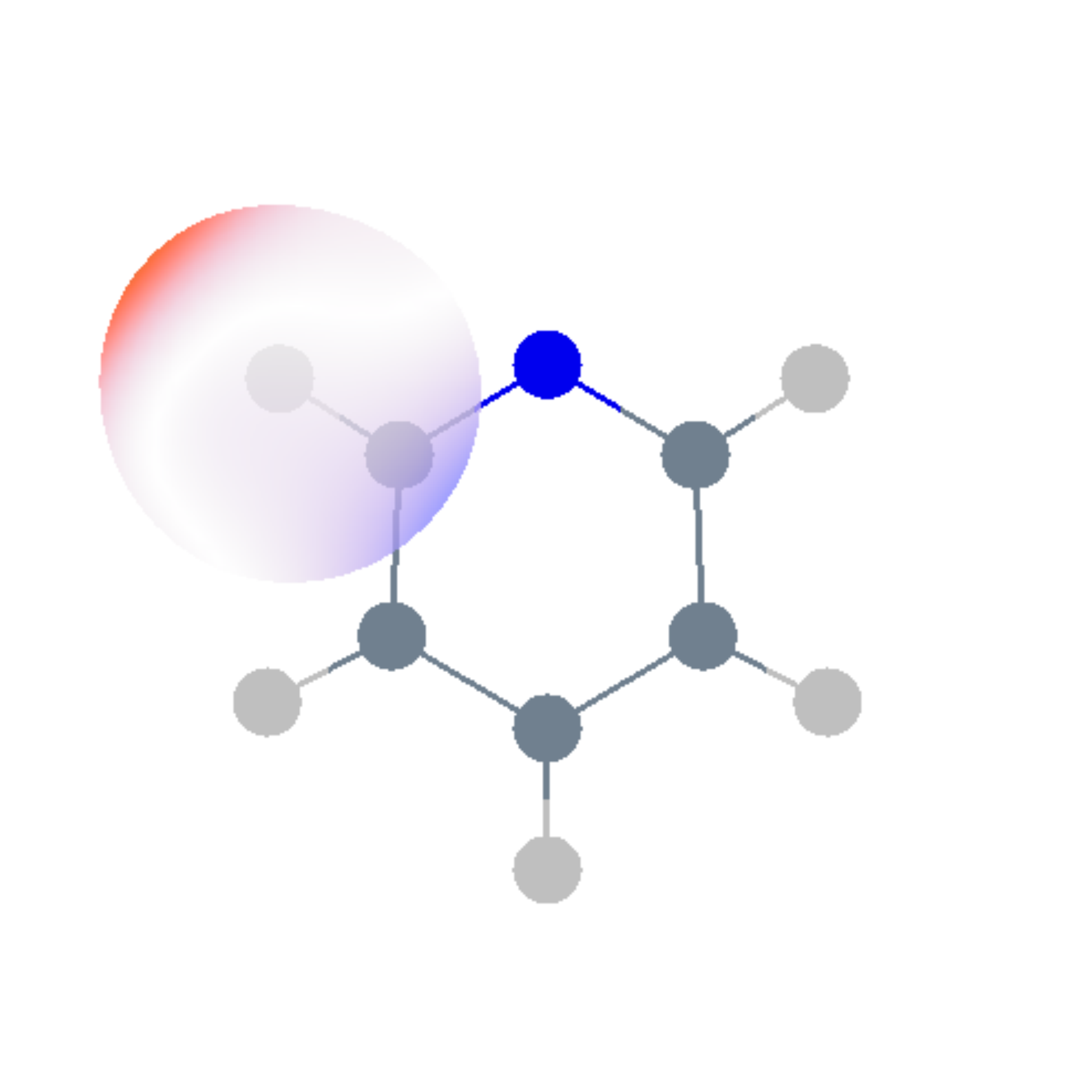}}{H1}
        \stackunder[-5pt]{\includegraphics[scale=0.23]{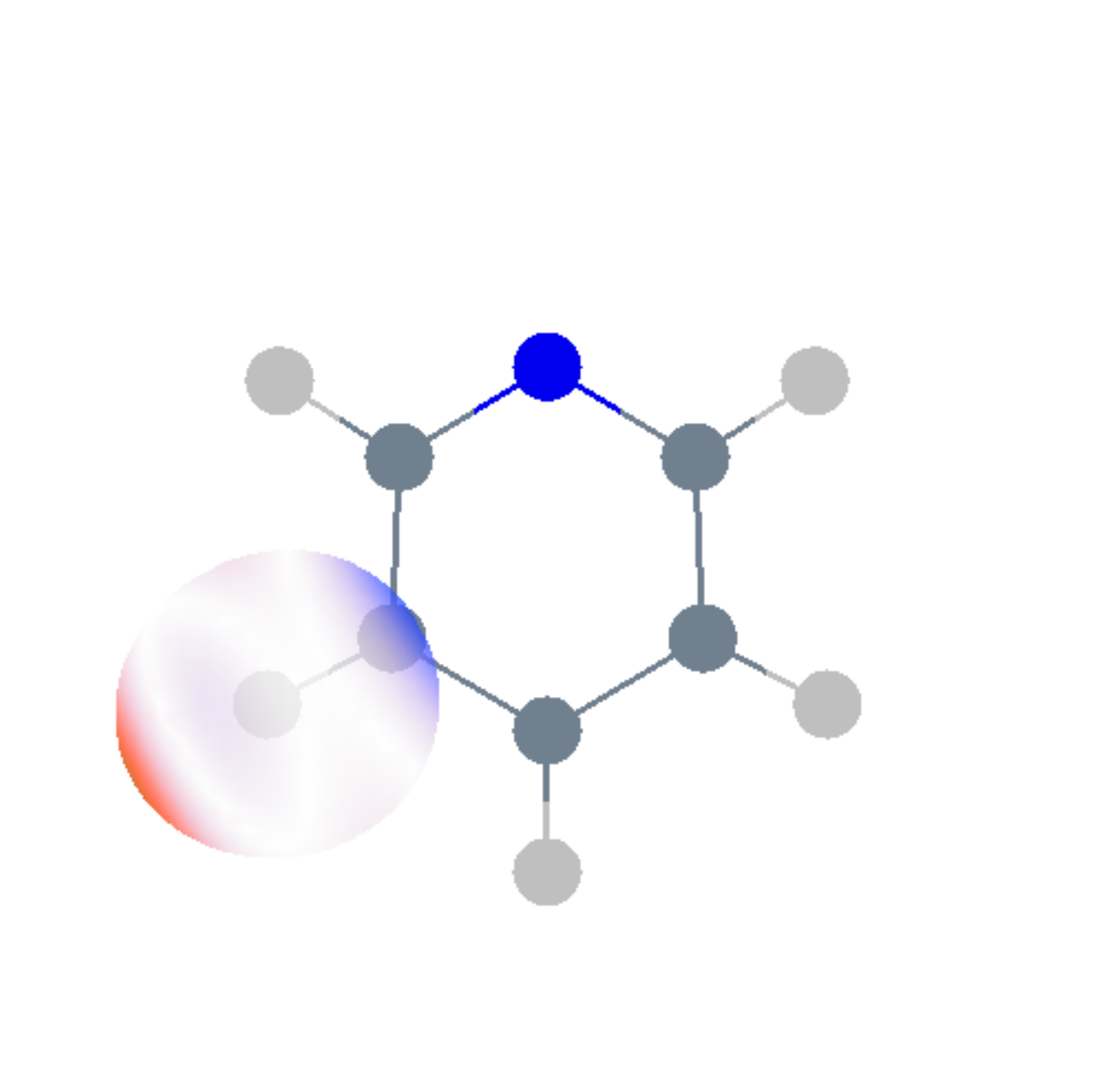}}{H2}
        \stackunder[-5pt]{\includegraphics[scale=0.23]{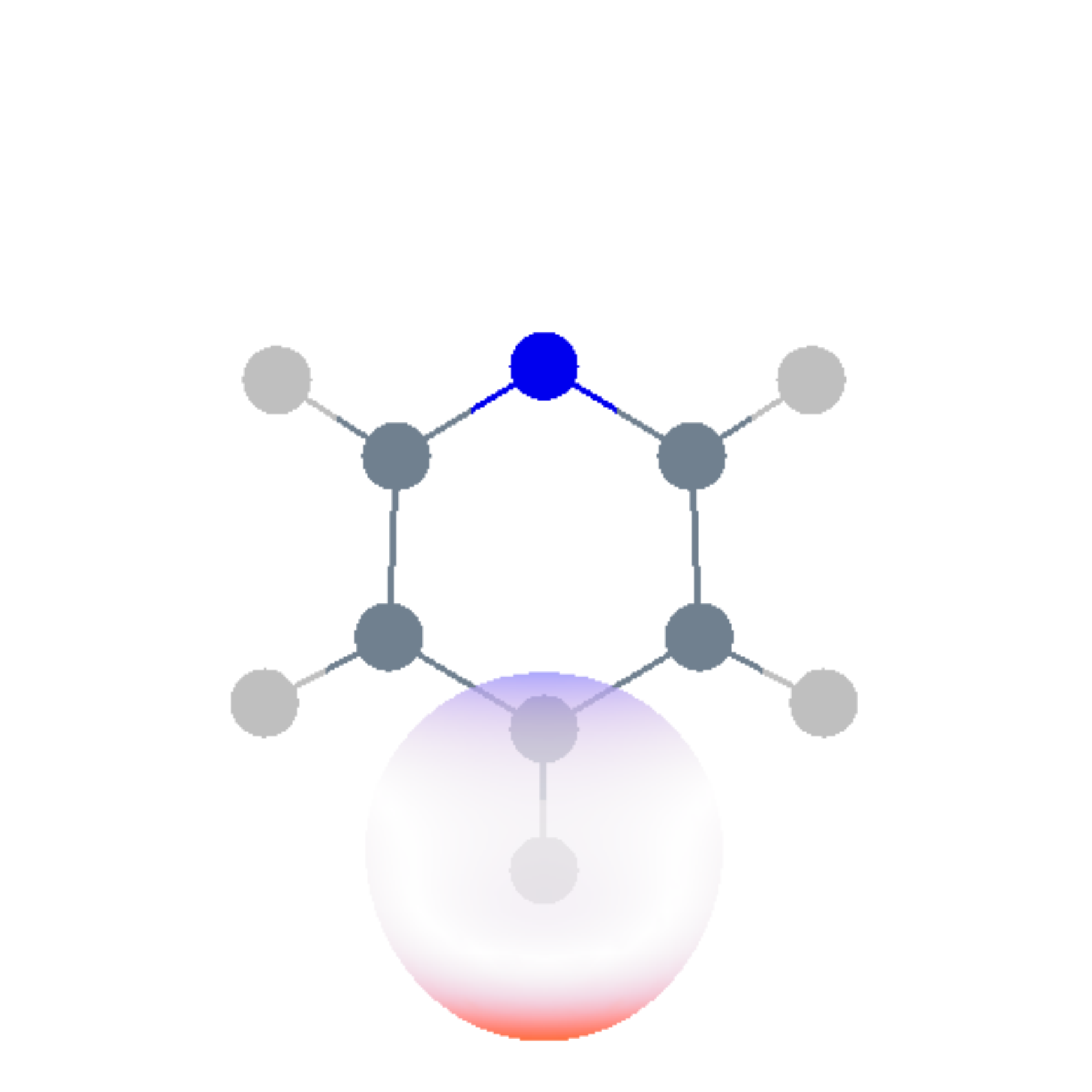}}{H3}
    \end{center}
\caption{
    The $10^{-3}$ a.u.\ iso-density surfaces of the density-fitting-based `atoms'
    in pyridine. The pyridine density was computed using a d-aug-cc-pVTZ basis, and
    the density-fitting was performed using the TZVPP auxiliary basis.
    The colour coding indicates the anisotropic component of the 
    electrostatic potential on the surface arising from ISA-based atomic
    multipoles located on the nuclei; that is, the atomic charge contributions
    are not included.
    The scale used varies from $-0.5$ V   (blue), through $0$ V (white),
    to $+0.5$ V (red).
}
\label{fig:pyr-DF-atoms}
\end{figure}

In Figure~\ref{fig:pyr-DF-atoms} we present the density-fitting-based
(DF-based) atomic isodensity
surfaces for the atoms in the pyridine molecule.
%\marginpar{Colour scale in figure?}
The total electronic density of pyridine was obtained with the d-aug-cc-pVTZ basis
using the PBE0/AC functional. We had to use the relatively less diffuse def-TZVPP 
basis for the density-fitting as results with any of the more diffuse 
RIMP2 auxiliary basis sets were so full of artifacts associated with the
basis set over-completeness as to lead to completely nonsensical results for the
density partitioning.
However, even with the relatively small def-TZVPP basis, the DF-based density 
partitioning results in carbon atoms with rather unusual shapes.
If this partitioning method is used to construct a short-range potential using the
density-overlap model as described above, we obtain potentials with spurious terms
in the atomic anisotropies and overall very poor fit qualities. 
% \marginpar{Obscure}
% These artifacts might be reduced or removed by relaxation to a sufficient
% amount of data, but it would be far better to remove them
% completely.

\begin{figure}
    % Fig 7
    \begin{center}
        %\stackunder[-5pt]{\includegraphics[scale=0.23]{./figs/AIM/pyr_N_ISA_noQ_iso0p001.png}}{N}
        %\stackunder[-5pt]{\includegraphics[scale=0.23]{./figs/AIM/pyr_C1_ISA_noQ_iso0p001.png}}{C1}
        %\stackunder[-5pt]{\includegraphics[scale=0.23]{./figs/AIM/pyr_C2_ISA_noQ_iso0p001.png}}{C2}
        %\stackunder[-5pt]{\includegraphics[scale=0.23]{./figs/AIM/pyr_C3_ISA_noQ_iso0p001.png}}{C3}
        %\stackunder[-5pt]{\includegraphics[scale=0.23]{./figs/AIM/pyr_H1_ISA_noQ_iso0p001.png}}{H1}
        %\stackunder[-5pt]{\includegraphics[scale=0.23]{./figs/AIM/pyr_H2_ISA_noQ_iso0p001.png}}{H2}
        %\stackunder[-5pt]{\includegraphics[scale=0.23]{./figs/AIM/pyr_H3_ISA_noQ_iso0p001.png}}{H3}
        \stackunder[-5pt]{\includegraphics[scale=0.23]{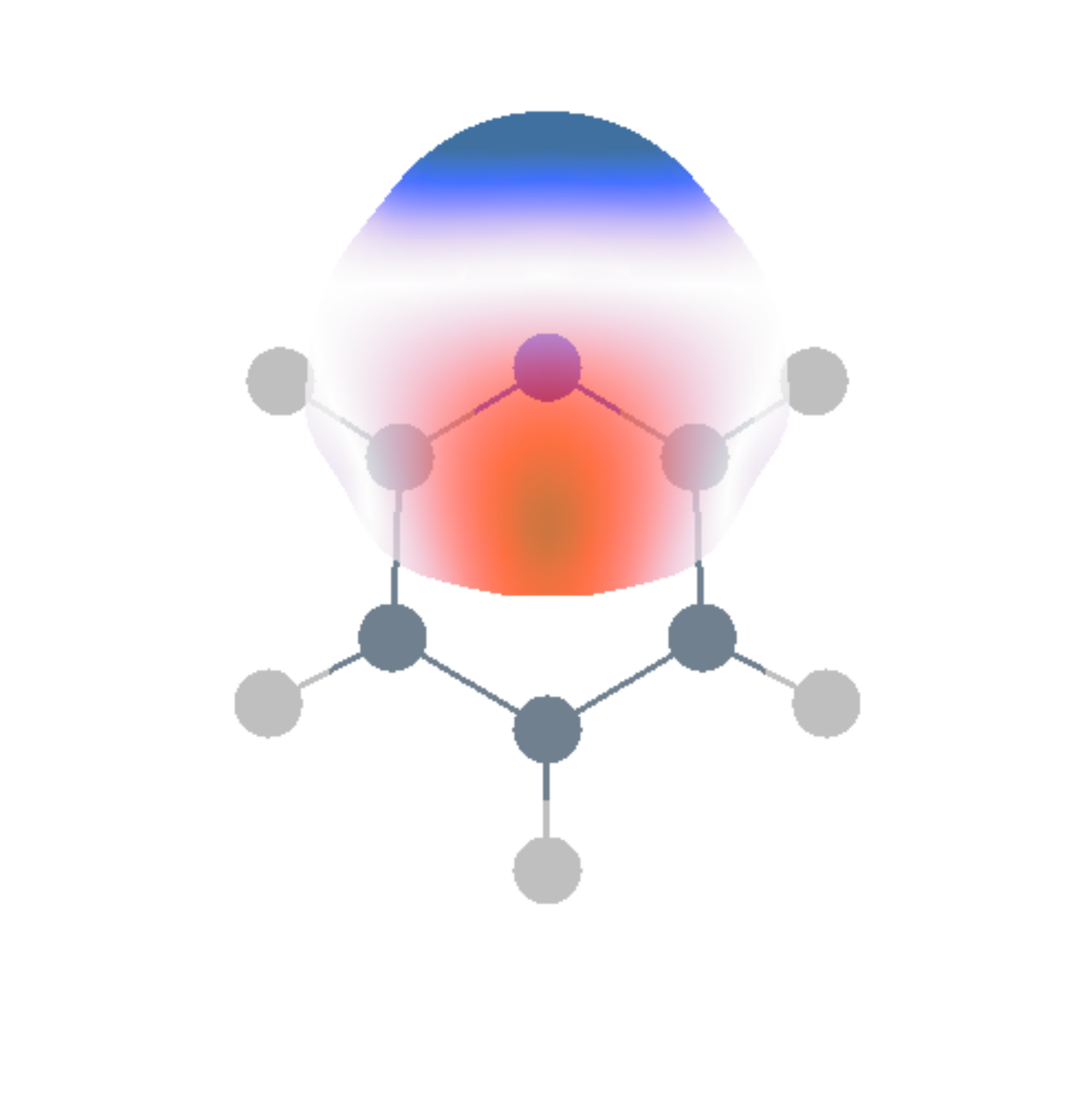}}{N}
        \stackunder[-5pt]{\includegraphics[scale=0.23]{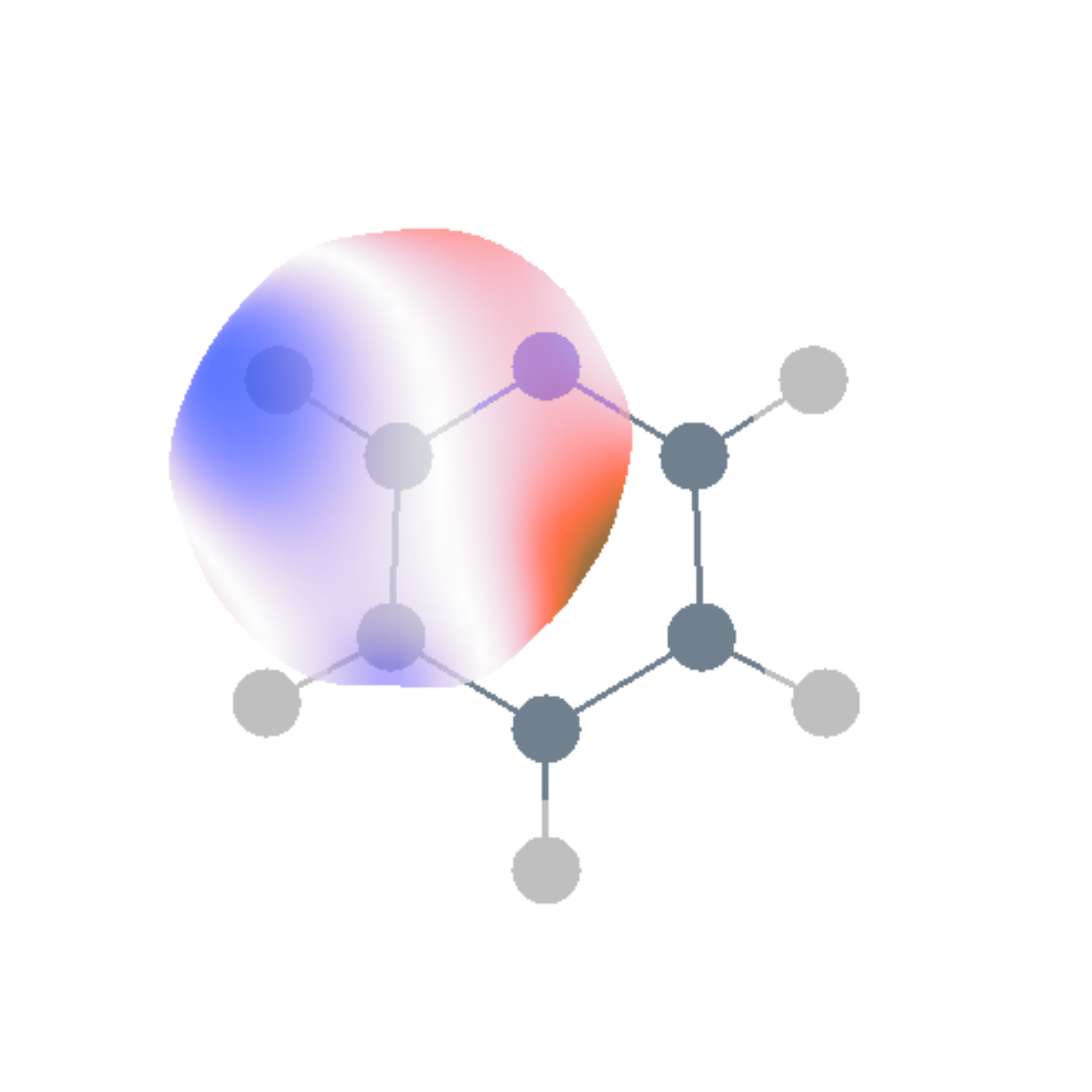}}{C1}
        \stackunder[-5pt]{\includegraphics[scale=0.23]{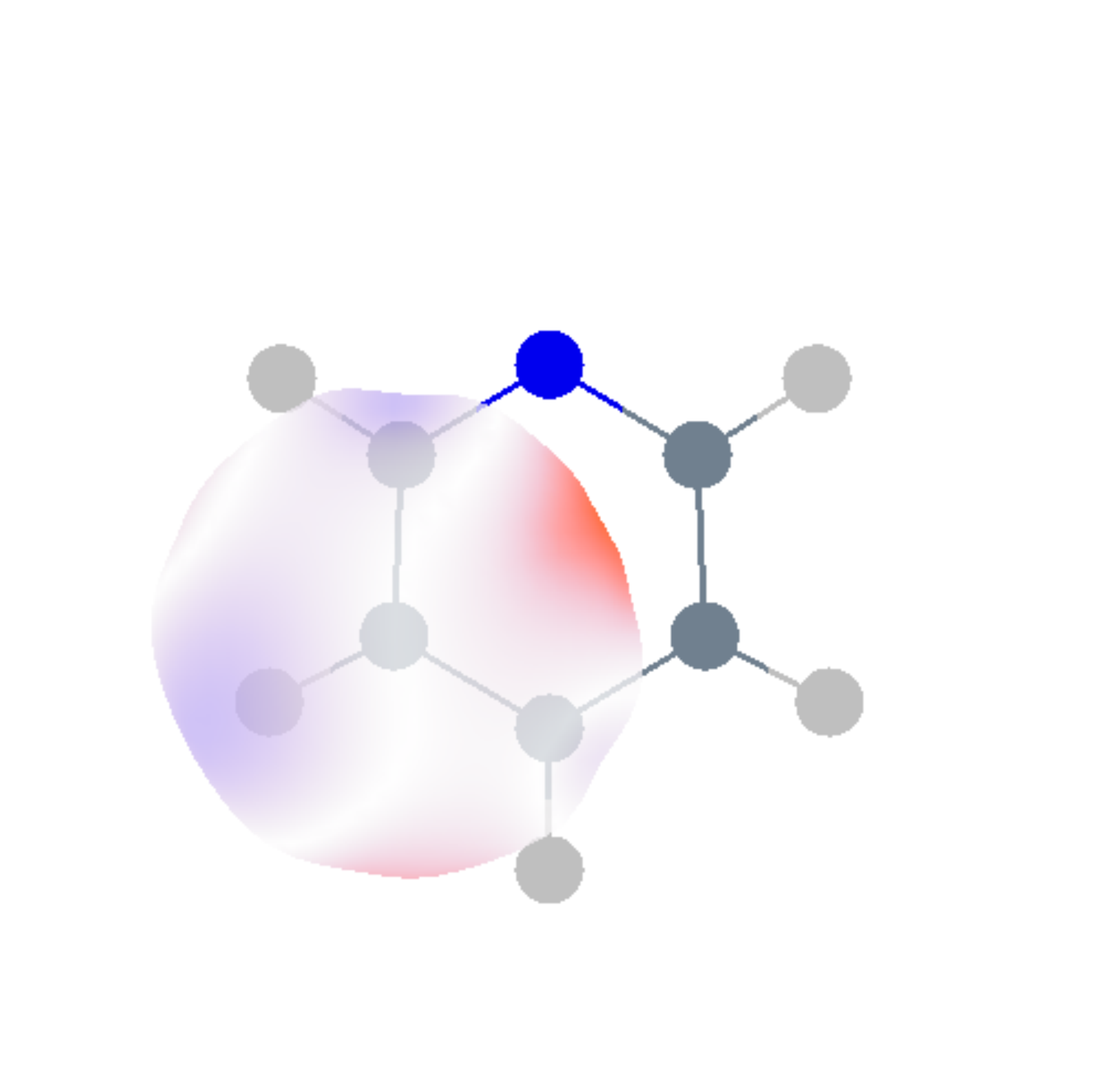}}{C2}
        \stackunder[-5pt]{\includegraphics[scale=0.23]{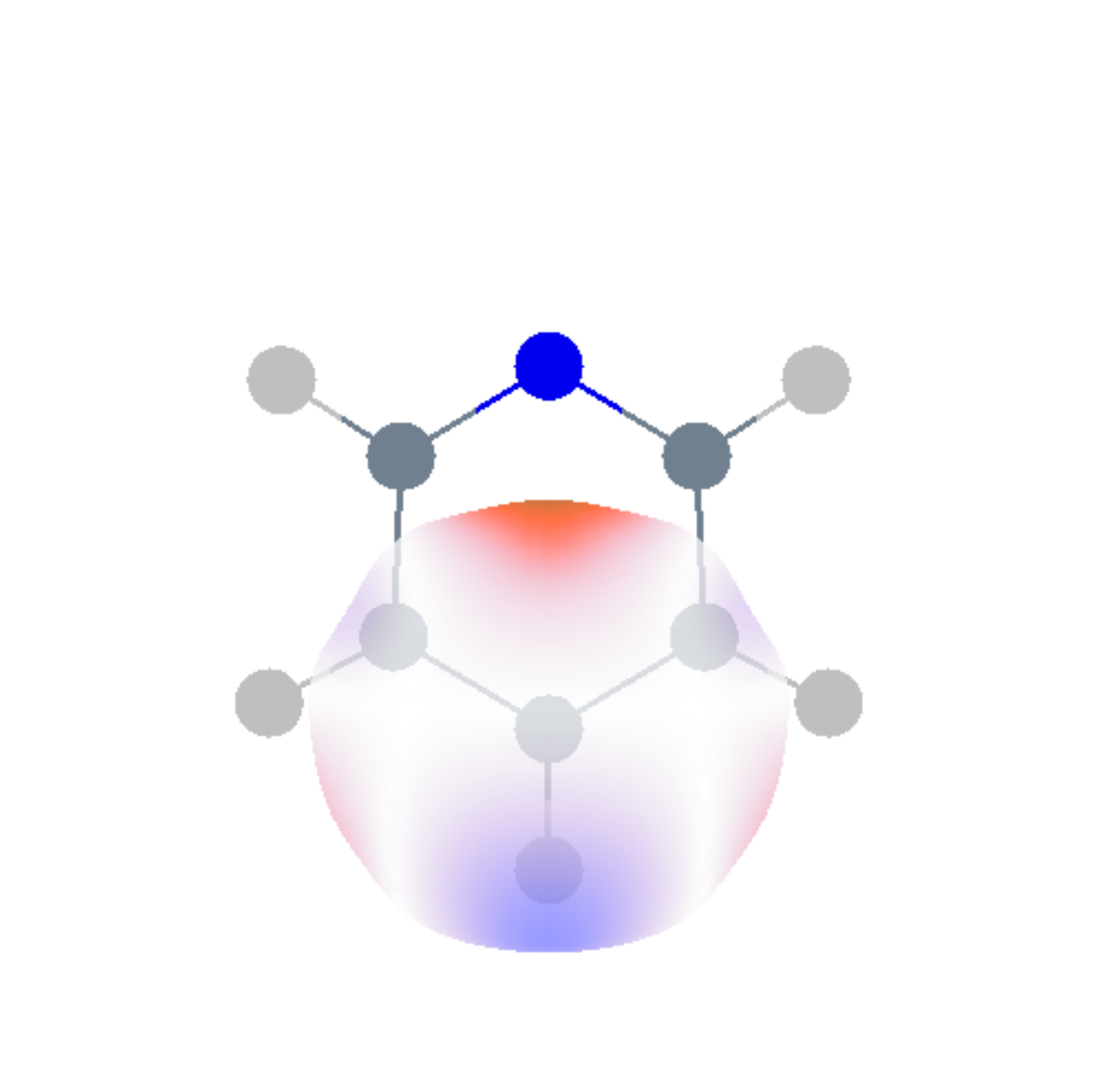}}{C3}
        \stackunder[-5pt]{\includegraphics[scale=0.23]{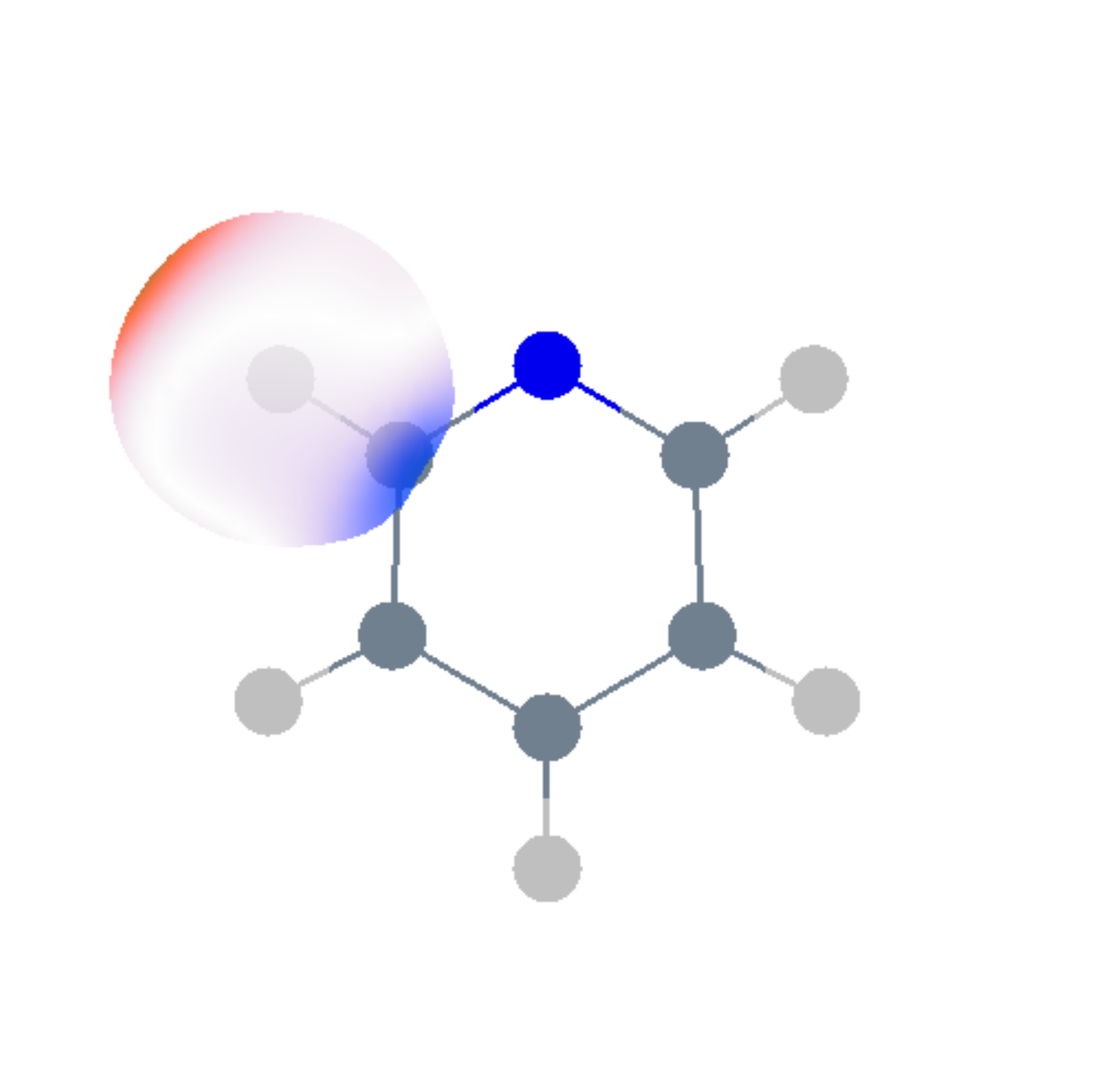}}{H1}
        \stackunder[-5pt]{\includegraphics[scale=0.23]{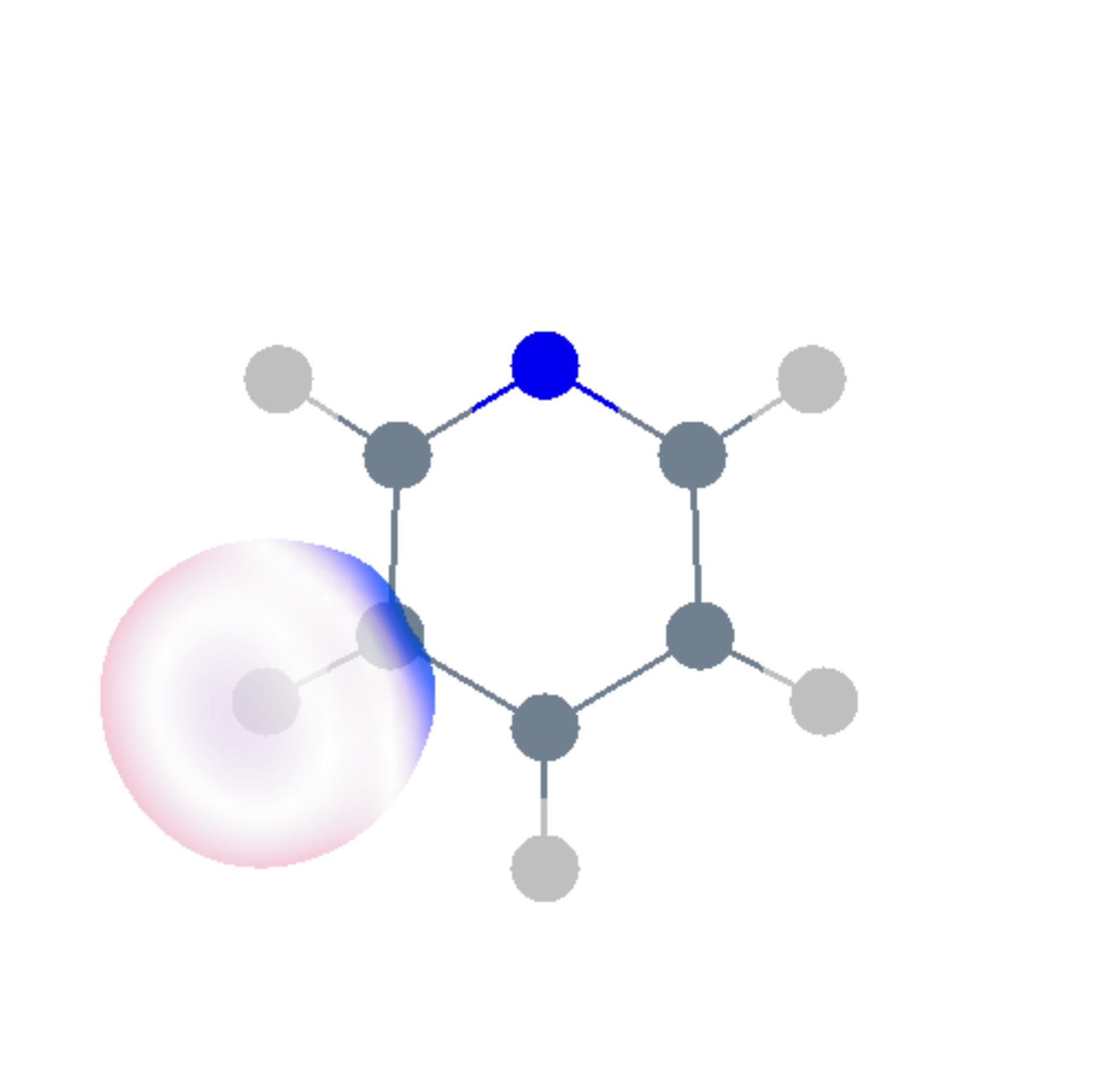}}{H2}
        \stackunder[-5pt]{\includegraphics[scale=0.23]{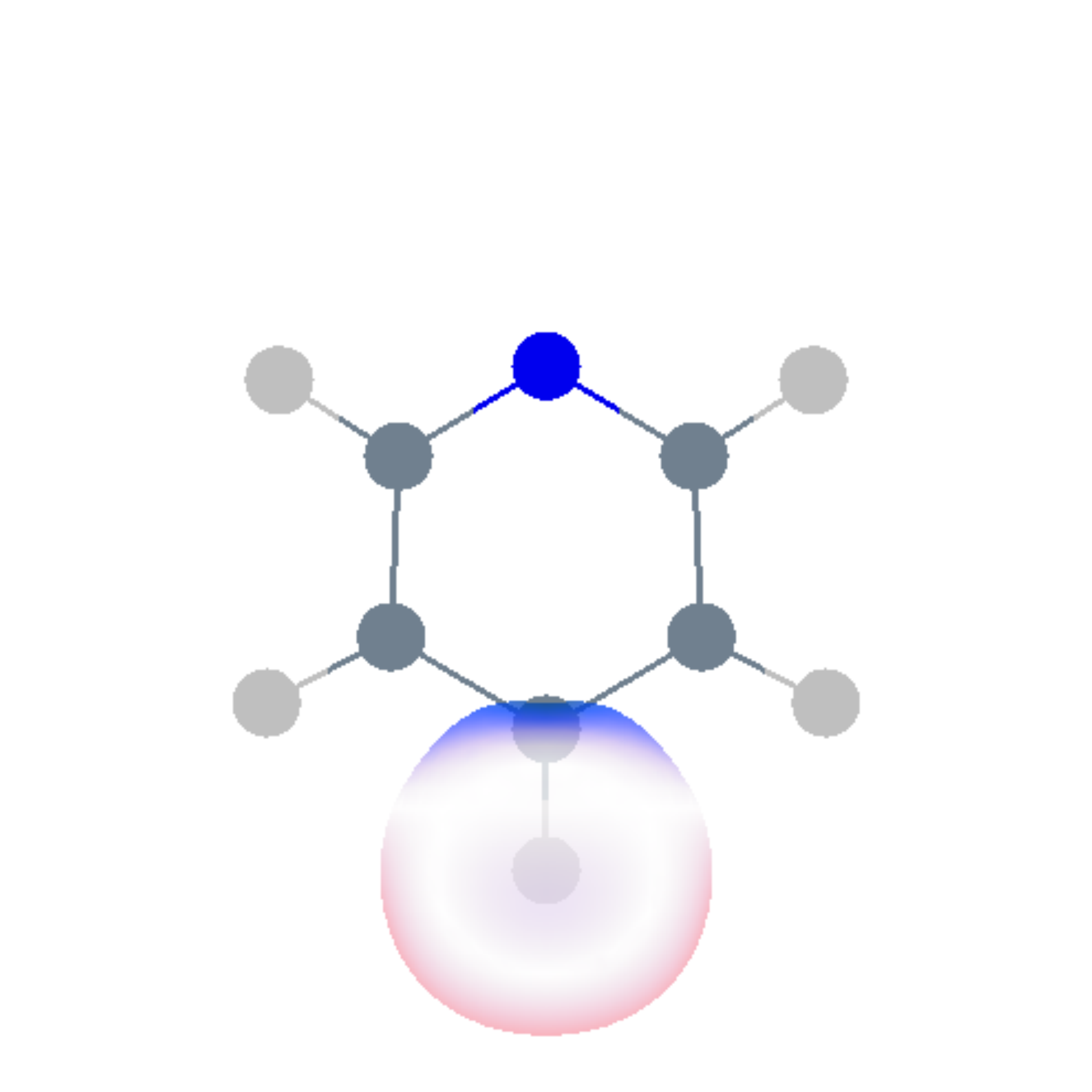}}{H3}
    \end{center}
\caption{
    The $10^{-3}$ a.u.\ iso-density surfaces of the ISA-based `atoms' in 
    pyridine. The pyridine density was computed using a d-aug-cc-pVTZ basis and
    the ISA calculations were performed using the aug-cc-pVQZ/ISA-set2 auxiliary basis set.
    Colour coding as described in Figure~\ref{fig:pyr-DF-atoms}.
}
\label{fig:pyr-ISA-atoms}
\end{figure}

In contrast, we can see in Figure~\ref{fig:pyr-ISA-atoms} that the ISA-based
atomic shapes obtained using the algorithm described in \S\ref{A-sec:ISA-density-overlap}
of \paperA are very well-behaved. These have been obtained with the significantly larger 
aug-cc-pVQZ/ISA-set2 fitting basis and show none of the artifacts seen with 
the DF-based scheme. Additionally, the ISA-based atoms do not show any significant
differences in shape when other basis sets are used, as long as these are large
and diffuse enough. This is a significant result: if we wish the atomic
shapes to be, in some sense, universal or transferable (properties we will not
explore in this paper), we must be able to calculate the atomic shapes 
with an algorithm that possesses a well-defined basis-set limit. 
The ISA approach is not the only such method, but for reasons discussed
in the Introduction of \paperA and in ref.~\citenum{MisquittaSF14}, it is one of the
few partitioning methods that has desirable numerical properties while
satisfying physical and chemical expectations.

\begin{figure}
    % Fig 8 
    \begin{center}
        %\stackunder[-5pt]{\includegraphics[scale=0.23]{./figs/AIM/pyr_N_ISA_iso0p001_contours.png}}{N: 11c,20,22c}
        %\stackunder[-5pt]{\includegraphics[scale=0.23]{./figs/AIM/pyr_C1_ISA_iso0p001_contours.png}}{C1: 11c,11s,20,22c}
        %\stackunder[-5pt]{\includegraphics[scale=0.23]{./figs/AIM/pyr_C2_ISA_iso0p001_contours.png}}{C2: 11c,11s,20,22c}
        %\stackunder[-5pt]{\includegraphics[scale=0.23]{./figs/AIM/pyr_C3_ISA_iso0p001_contours.png}}{C3: 11c,20,22c}
        %\stackunder[-5pt]{\includegraphics[scale=0.23]{./figs/AIM/pyr_H1_ISA_iso0p001_contours.png}}{H1: 11c,11s}
        %\stackunder[-5pt]{\includegraphics[scale=0.23]{./figs/AIM/pyr_H2_ISA_iso0p001_contours.png}}{H2: 11c,11s}
        %\stackunder[-5pt]{\includegraphics[scale=0.23]{./figs/AIM/pyr_H3_ISA_iso0p001_contours.png}}{H3: 11c}
        \stackunder[-5pt]{\includegraphics[scale=0.23]{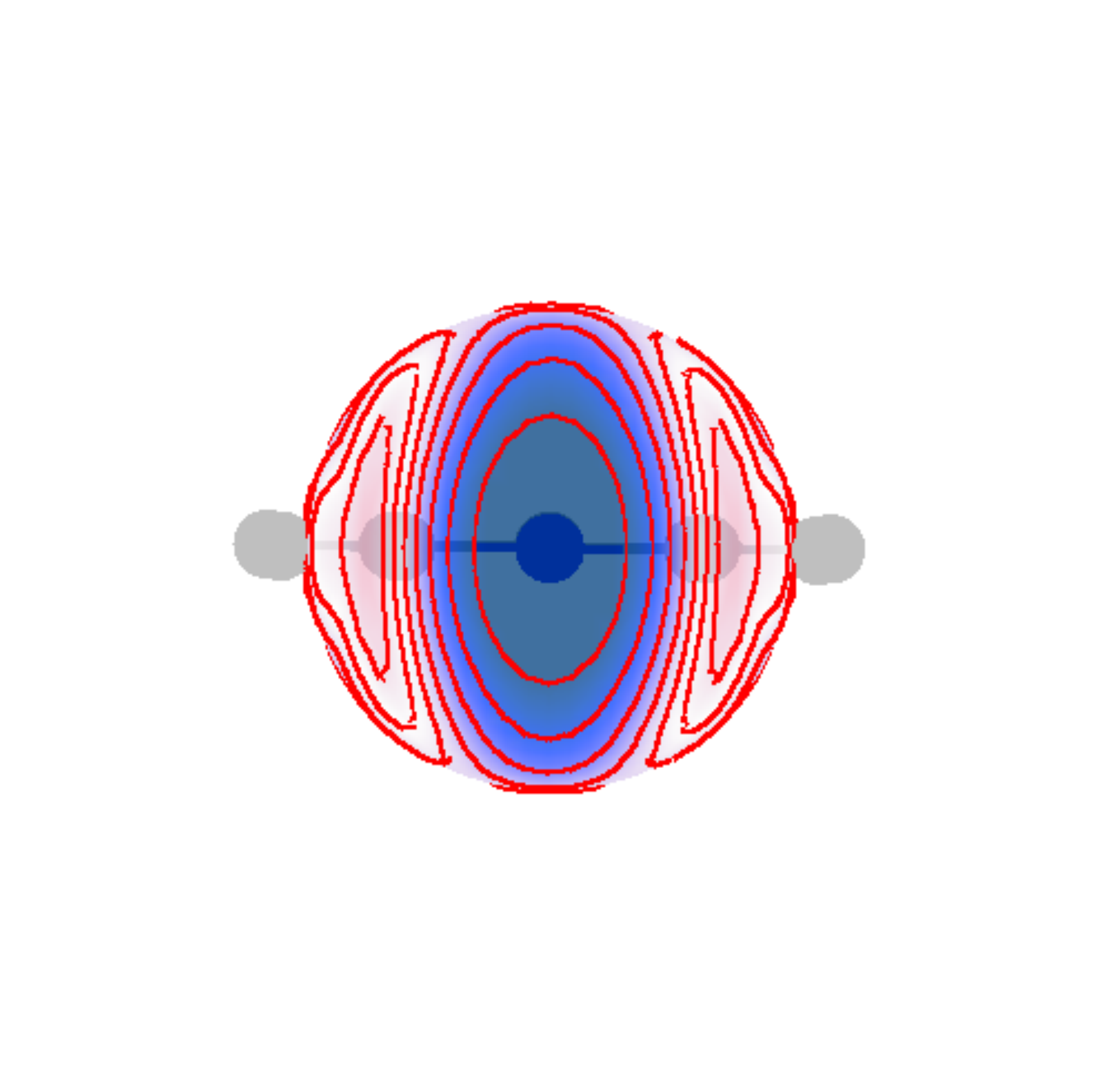}} {N: 11c,20,22c}
        \stackunder[-5pt]{\includegraphics[scale=0.23]{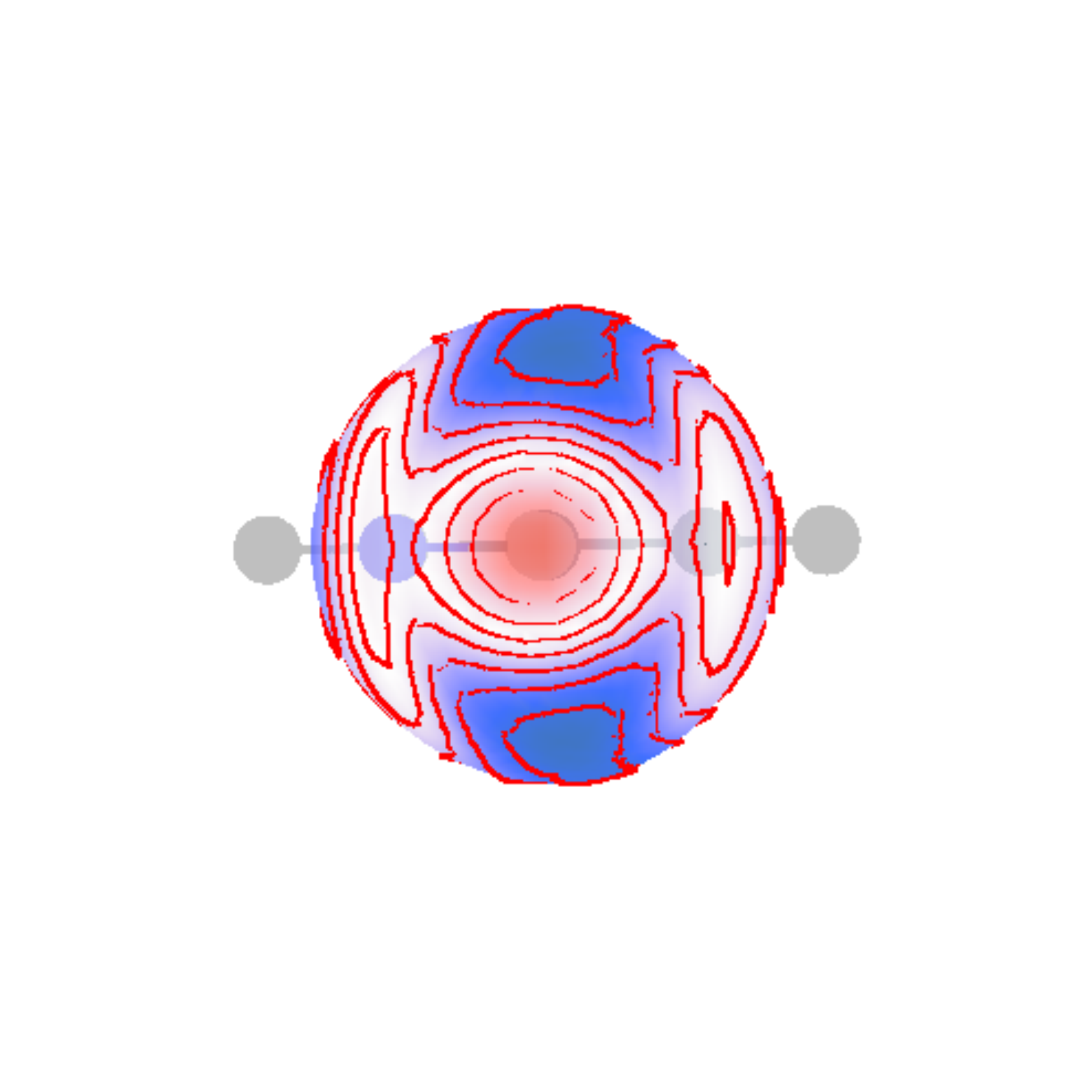}}{C1: 11c,11s,20,22c}
        \stackunder[-5pt]{\includegraphics[scale=0.23]{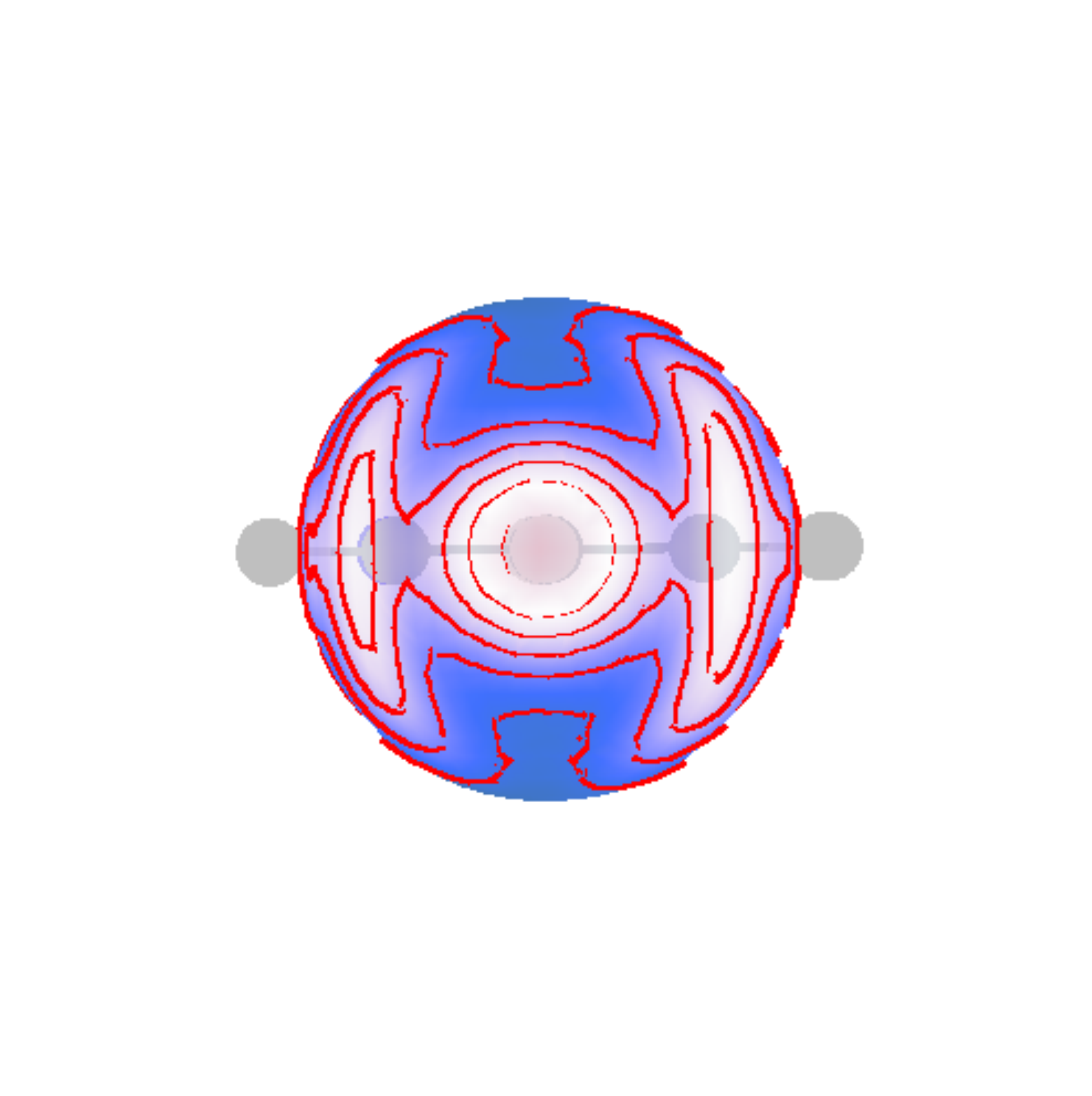}}{C2: 11c,11s,20,22c}
        \stackunder[-5pt]{\includegraphics[scale=0.23]{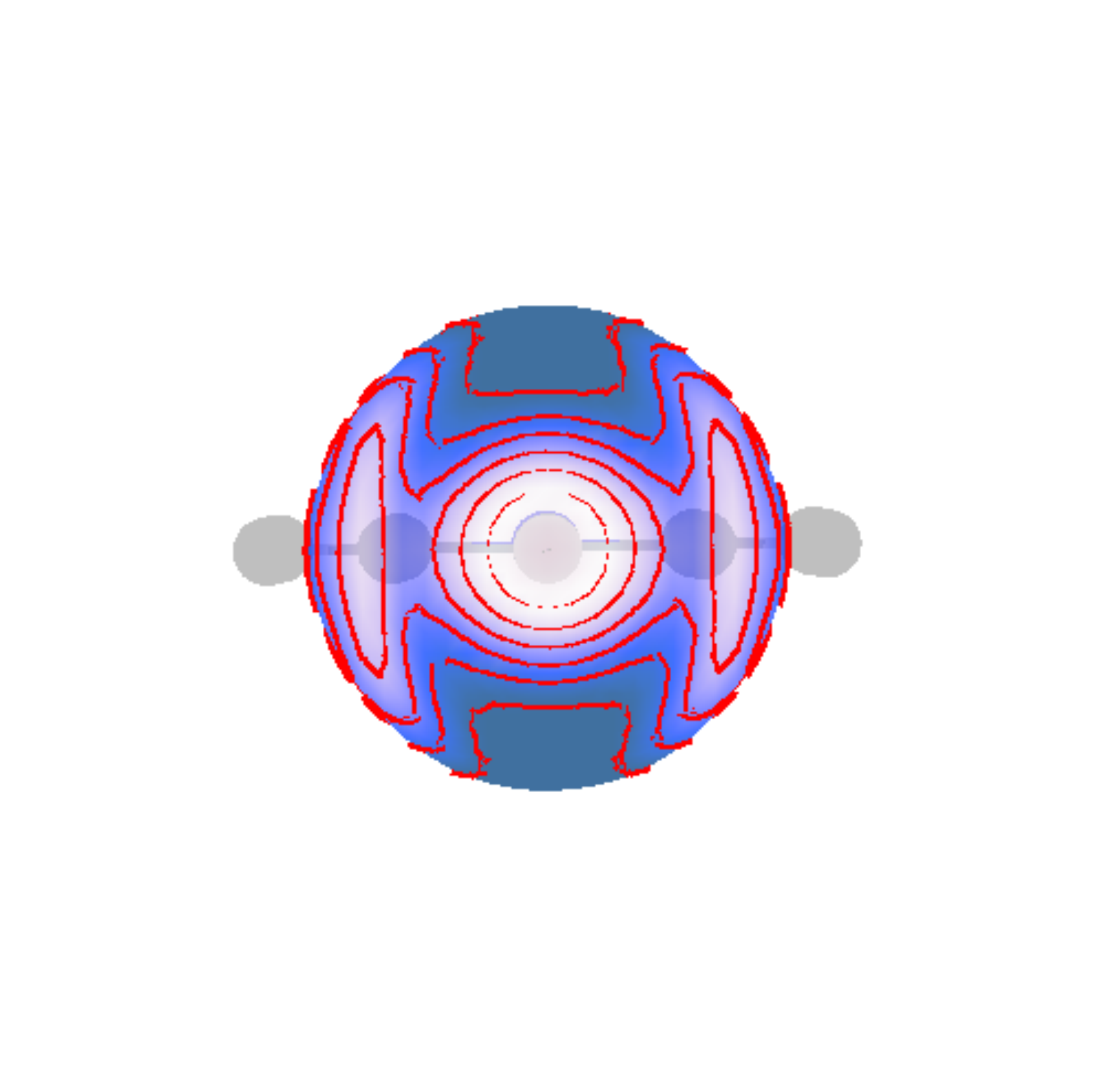}}{C3: 11c,20,22c}
        \stackunder[-5pt]{\includegraphics[scale=0.23]{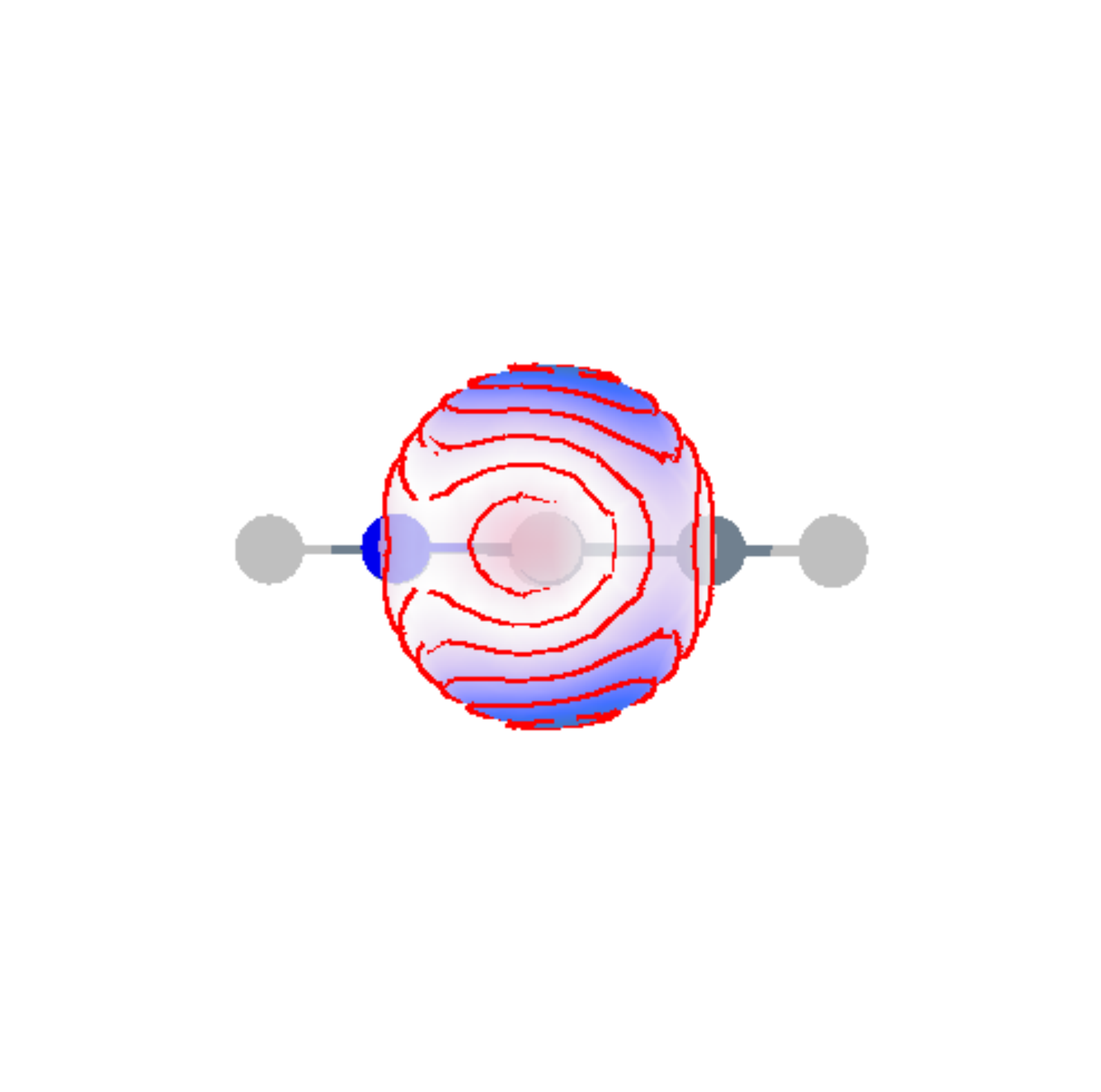}}{H1: 11c,11s}
        \stackunder[-5pt]{\includegraphics[scale=0.23]{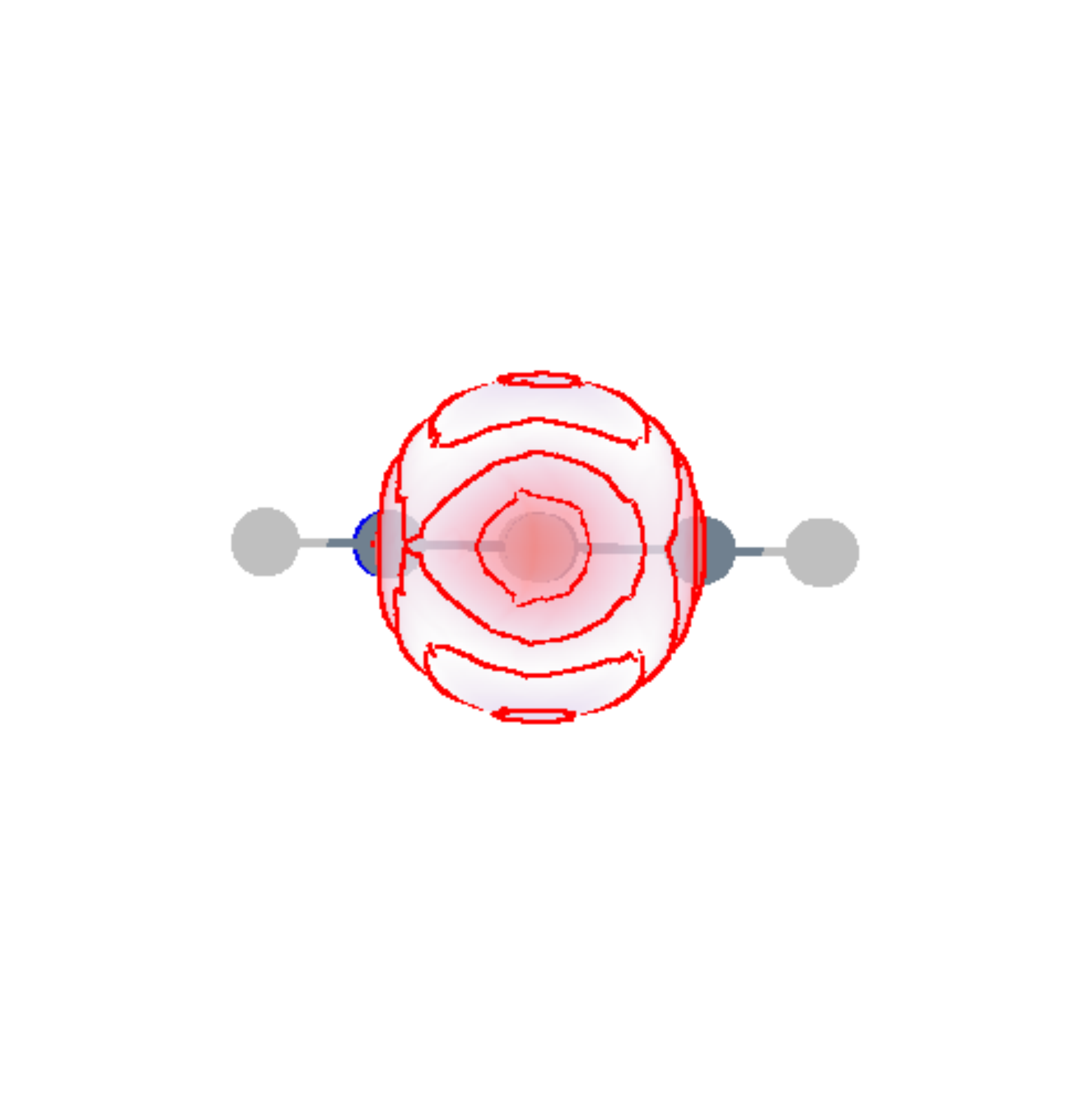}}{H2: 11c,11s}
        \stackunder[-5pt]{\includegraphics[scale=0.23]{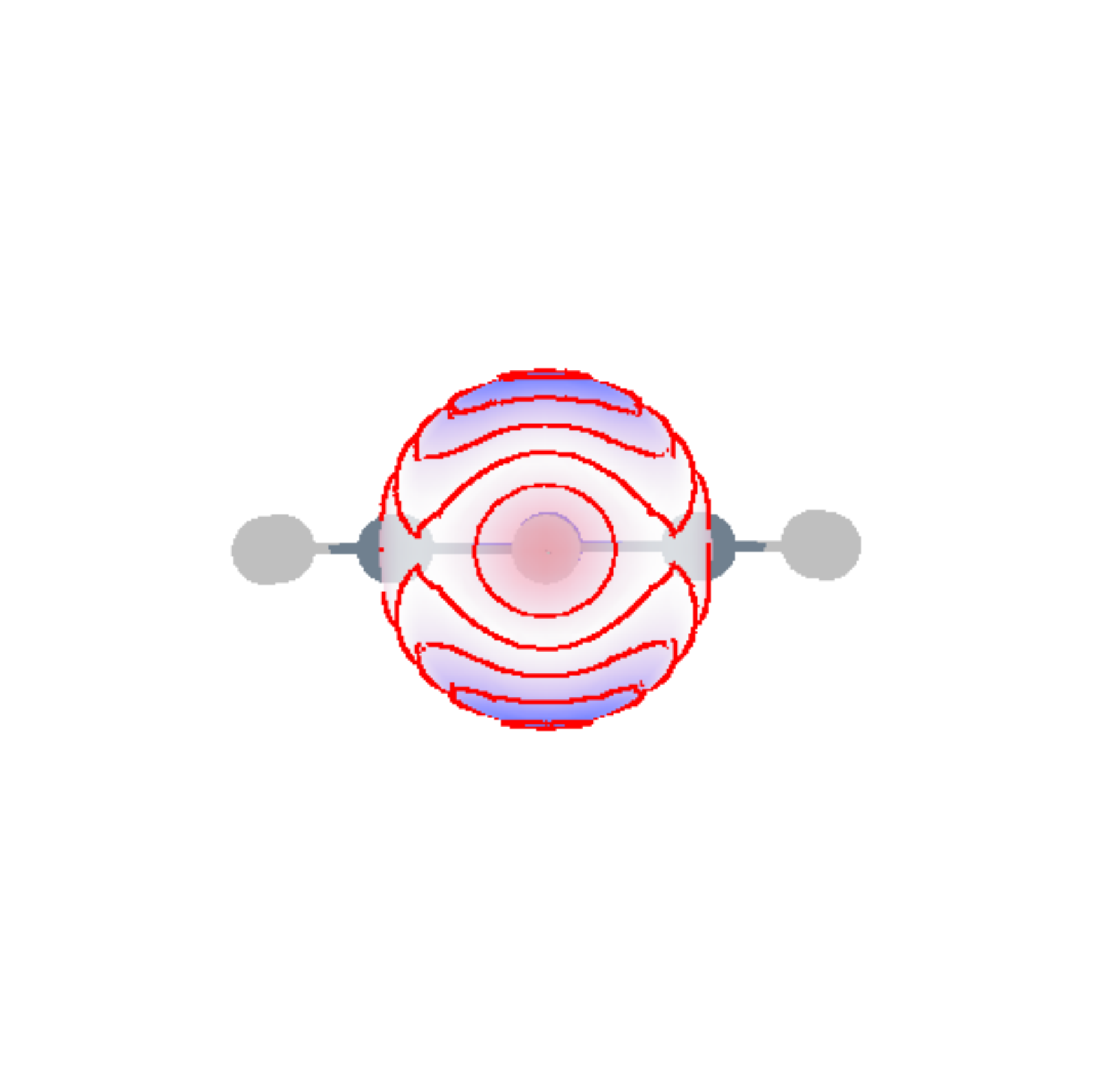}}{H3: 11c}
    \end{center}
\caption{
    Along-the-bond views of the ISA-based `atoms' in pyridine. Here we illustrate the
    anisotropy of the atom shapes by contours showing the intersections with
    spherical surfaces centred at the atomic nuclei. 
    The dominant anisotropy terms for each atom are listed for local axis frameworks
    with the $x$ axis pointing out of the bond (out of the page) and the $z$ axis
    normal to the plane of the molecule.
}
\label{fig:pyr-ISA-atom-contours}
\end{figure}

In Figure~\ref{fig:pyr-ISA-atom-contours} we present the ISA-atomic shapes viewed
in the molecular plane, along the bond axis, or, in the case of the nitrogen atom, 
along the N\vdw C3 axis. In order to highlight the atomic anisotropies
we have superimposed on the $10^{-3}$ a.u.\ isodensity surfaces some contours
showing the intersection with spheres centred on the 
atomic nuclei. These contours clearly illustrate the shape symmetries of each of the
atoms. Also included in the figure are the important shape anisotropies 
for these atoms. These have been calculated by fitting \Esr{1} via the
distributed overlap model using a set of local axis frames located on the
atomic centres with the $x$ axis pointing along and out of the bond,
and the $z$ axis perpendicular to and pointing out of the plane of the molecule.
During the relaxation step in this fit we eliminate all terms that are
less than a threshold, taken to be $0.01$ a.u.\
The picture that emerges is remarkably 
simple and convincing: 
\begin{itemize}
    \item {\em Nitrogen}: The largest anisotropy term for the nitrogen atom
        in pyridine is the 22c term that is associated with the lone pair. Additionally
        one may include the 11c and 20 terms on the nitrogen atoms, though these are smaller.
        All other terms are negligible. 
    \item {\em Carbon}: The 20 term associated with the $p_z$ orbitals is the 
        dominant source of anisotropy on all carbon atoms. Of the other symmetry-allowed
        terms, the 11c term associated with the C--H bond is relatively strong. 
        The 22c terms are present, but small. Finally, C1 and C2 contain 11s terms
        due to the proximity of the N atom. These terms describe the in-plane
        distortion of the C1/C2 densities due to N.
    \item {\em Hydrogen}: We have limited all hydrogen atoms to rank 1 terms
        only. All hydrogen atoms possess a 11c term to describe the
        electronic distortion along the C--H bond and, both H1 and H2 additionally
        have 11s terms.
\end{itemize}

\begin{figure}
    % Fig 9
    \begin{center}
        \includegraphics[scale=0.45]{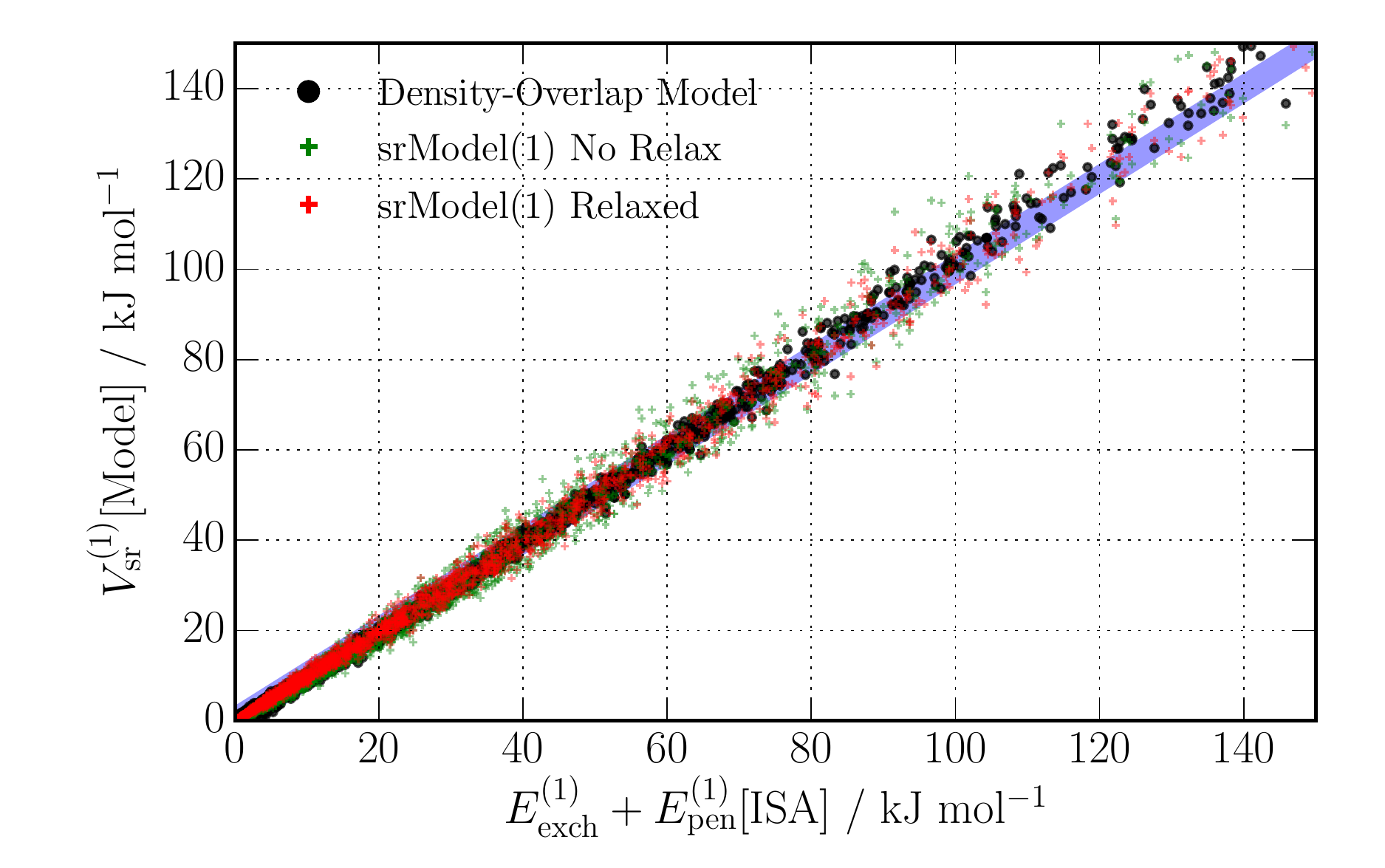}
        \includegraphics[scale=0.45]{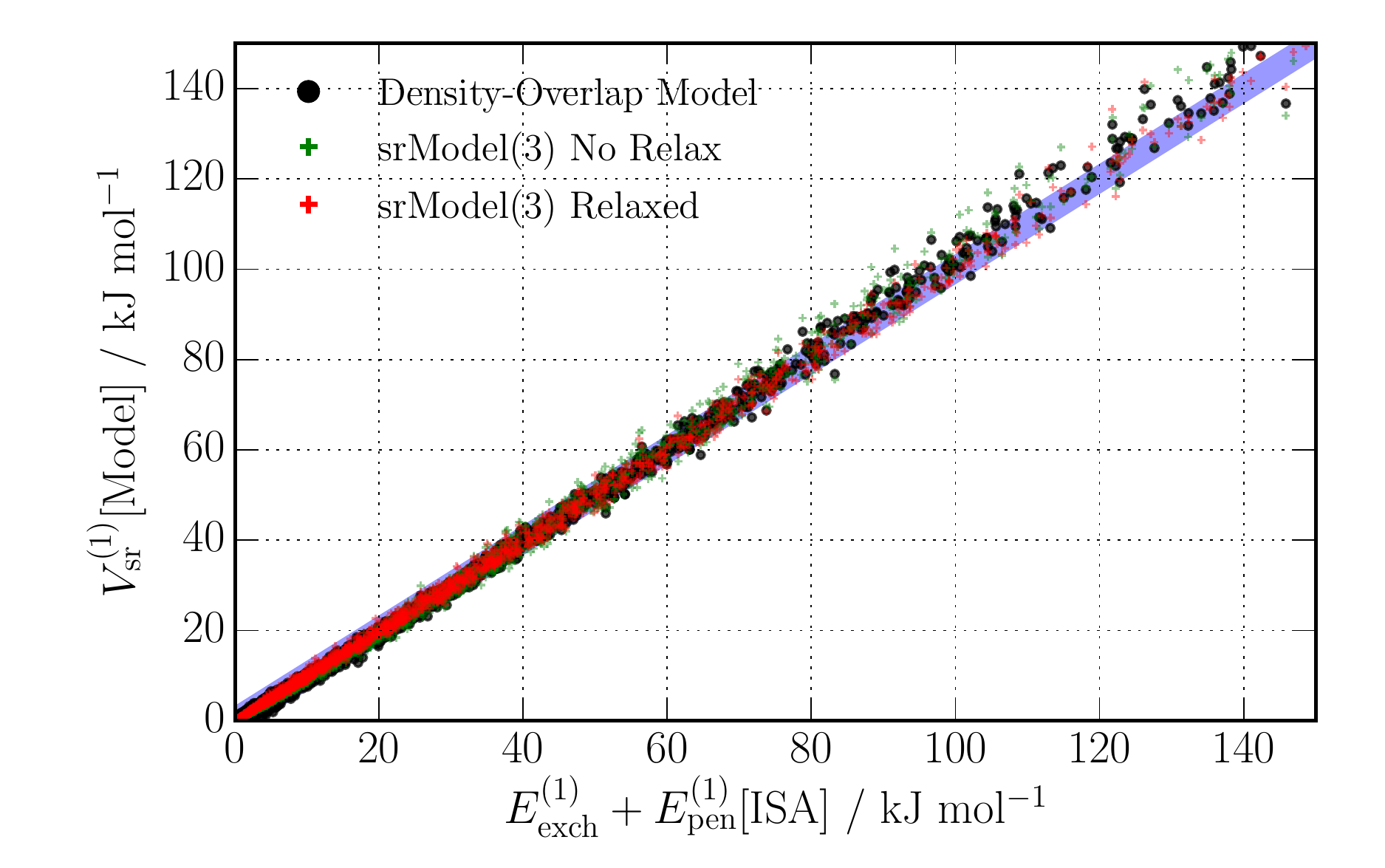}
    \end{center}
    \caption{
        Performance of two of the short-range models fitted to \Esr{1}. 
        srModel(1) is fully isotropic and srModel(3) contains the anisotropy terms
        described in the text and indicated in Figure~\ref{fig:pyr-ISA-atom-contours}. 
        srModel(2) results are only slightly better than those from srModel(1) and
        are not shown.
        The black circles are results directly from the distributed 
        density-overlap model; the green plus signs are data obtained
        from the model fitted to eq.~\eqref{A-eq:Vtot_Vsr}  in \paperA before relaxation,
        and the red plus signs are the same after relaxation to \Esr{1}.
        The blue bar represents the $\pm 1$ \kJpermol range.
    }
    \label{fig:pyr2-Esr1-scatter}
\end{figure}

We have developed three models for the short-range terms: srModel(1) contains
only isotropic terms, in srModel(2) we have included the 22c anisotropy term 
on the nitrogen atom, and  in srModel(3) we have used all the anisotropy terms
shown in Figure~\ref{fig:pyr-ISA-atom-contours}.
In all three models, the hardness parameters $\alpha_{ab}$ in 
eq.~\eqref{A-eq:Vtot_Vsr} in \paperA were kept isotropic.
The constrained relaxation was performed using eq.~\eqref{A-eq:constrained-opt}  in \paperA 
with constraint strength parameters $c_i$ chosen to be $0.1$ for the 
isotropic parameters and $1.0$ for the anisotropic terms in the 
$\rho_{ab}(\Omega)$ expansions. This choice was made empirically on
the basis that the appropriate parameters were those that when further reduced
did not result in any appreciable improvement in the fit quality.
The distributed density-overlap fits were performed using the 
\CamCASP program, and the fits to individual site--site potentials \Vsr{}{ab}
were performed using the \Orient program. 
The weighting schemes used in these fits are described in \S\ref{sec:numerical}.
The relaxation step was also performed using the \Orient program
but this time using the Boltzmann weighting function as described in \S\ref{sec:numerical}.
The scatter plots of these models at various stages in the fitting process
are shown in Figure~\ref{fig:pyr2-Esr1-scatter}. 
Weighted \rms errors at the final stage are $1.03$, $0.90$, and $0.61$ \kJpermol
for models 1, 2 and 3, respectively. 
These uncertainties are less than our target of 1 \kJpermol for all three models,
but the performance of srModel(3) is quite remarkable, with errors less than or close
to 1 \kJpermol for energies as large as 100 \kJpermol.

\subsection{Infinite-order charge transfer (delocalisation) energy}
\label{sec:CT}

The infinite-order charge-transfer energy is the dominant short-range
contribution at second and higher orders in the intermolecular
interaction operator.
While we can use regularised SAPT(DFT) \cite{PatkowskiJS01a,Misquitta13a}
to determine the second-order charge-transfer energy, the contributions
from higher orders cannot, at present, be computed within the SAPT framework.
Unfortunately, where charge-transfer is important, these higher-order
effects appear to be too large to be ignored, so we need to account for them, 
if only approximately.
As it turns out, the discussion of the infinite-order polarization in
\S\ref{A-sec:polarization} of \paperA readily suggests an approximation.
If we argue that the infinite-order induction energy is the sum of
just the infinite-order charge-transfer and polarization terms 
(i.e., assuming that there are no cross terms present), then if we 
know any two, we can compute the third.
Here we approximate the infinite-order induction energy as:
\begin{align}
    \EIND{2-\infty} \approx  \EIND{2} + \deltaHF
      \label{eq:Eind-infinite-order}
\end{align}
and define the two-body infinite-order charge-transfer energy to be
\begin{align}
    \ECT{2-\infty} &= \EIND{2-\infty} - \EPOL{2-\infty} \nonumber \\
                   &\approx \EIND{2} + \deltaHF - \EpolMP{2-\infty}.
    \label{eq:CT-infinite-order}
\end{align}
While this expression is readily implemented, it has a drawback in that the
definition depends on the type of polarization model used.

\begin{figure}
    % Fig 10
    \begin{center}
        \includegraphics[scale=0.45]{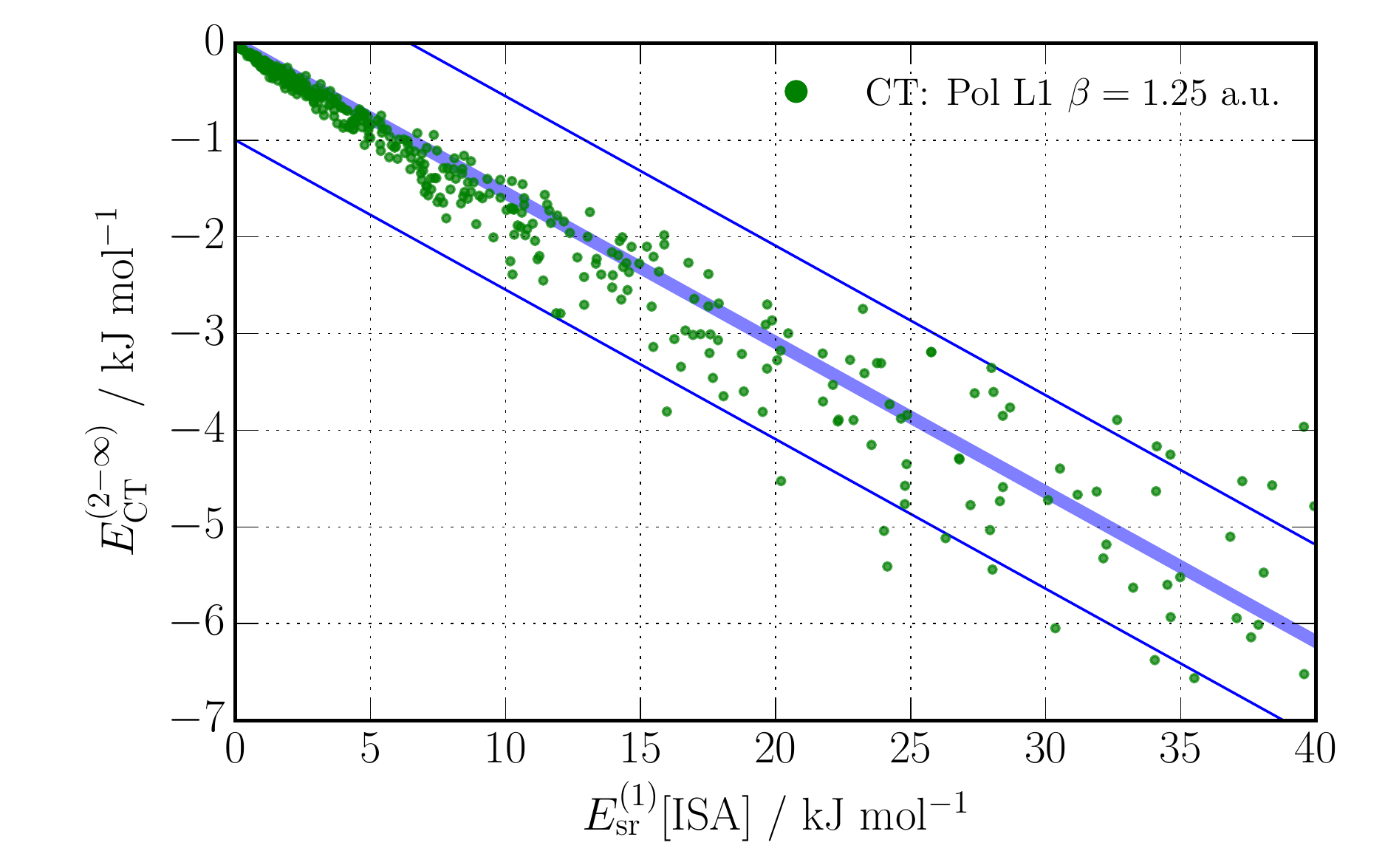}
    \end{center}
    \caption{
        The infinite-order charge delocalisation (charge-transfer) energy
        plotted against the first-order short-range energy \Esr{1}.
        The thin blue lines represent the $\pm 1$ \kJpermol limits.
    }
    \label{fig:CT-L1-scatter}
\end{figure}

In Figure~\ref{fig:CT-L1-scatter} we have plotted the infinite-order
charge-transfer energy calculated using eq.~\eqref{eq:CT-infinite-order}
against the first-order short-range energy \Esr{1}. 
First of all, at about 20\% of \Esr{1}, \ECT{2-\infty} is a significant
contribution to the short-range energy and it cannot be ignored.
Second, while these two energies are roughly proportional, there is a significant
scatter, particularly at the larger charge-transfer energies. 
Nevertheless, the scatter is rarely more than $\pm 1$ \kJpermol.
If we argue that the charge-transfer contribution to the intermolecular
interaction energy arises from a tunneling process \cite{Misquitta13a},
then it is natural to assume that the tunneling probability will be 
roughly proportional to the electron density overlap, but 
further work needs to be done to see whether this holds for other systems. 

We may include \ECT{2-\infty} into our models for the short-range energy
by constrained relaxation of the parameters in the models already obtained 
for \Esr{1}, or we may exploit the approximate proportionality of 
\Esr{1} and \ECT{2-\infty} and absorb the bulk of the charge-transfer
effects by scaling as follows.
If we assume a proportionality with constant $k \lt 0$:
\begin{align}
    \ECT{2-\infty} & \approx  k \Esr{1} \approx k \VSR{(1)} ,
\end{align}
then we can include \ECT{2-\infty} into the short-range energy model
by scaling it by $(1-k)$ yielding
\begin{align}
    \VSR{(1-\infty)} &\approx (1-k) \VSR{(1)} \nonumber \\
                   & \approx    (1-k) \sum_{a,b} 
                           G \exp{[-\alpha_{ab}(r_{ab} - \rho_{ab}(\Omega_{ab}))]}, 
\end{align}
then, re-writing $1-k = \exp{[-\alpha_{ab} \delta_{ab}]}$, where 
$\delta_{ab} = - \ln(1-k)/\alpha_{ab}$, we get
\begin{align}
    \VSR{(1-\infty)} = \sum_{a,b} G \exp{[-\alpha_{ab}(r_{ab} - (\rho_{ab}(\Omega_{ab})-\delta_{ab} ))]}.
\end{align}
That is, the isotropic atom-pair radii are reduced by $\delta_{ab}$
by the attractive effects of the charge delocalisation process. 
The atom-pair shape-function $\rho_{ab}(\Omega_{ab})$ 
remains additive in the sense of eq.~\eqref{eq:shape-func-additive},
but there is an isotropic non-additive correction $\delta_{ab}$,
as shown in eq.~\eqref{eq:shape-func-nonadditive}.

For the pyridine dimer we get $k \approx 0.16$ (it varies 
slightly with the type of polarization model used). 
Therefore the pair-radius reduction is of the order $0.05$ Bohr,
which is small but not negligible as it leads to an overall 
reduction in the intermolecular separation of a few tenths of a
Bohr in some dimer orientations.
These effects may be expected to be larger in more 
strongly hydrogen-bonded systems where the charge-delocalisation
is stronger.

The above scaling absorbs the bulk of the charge-transfer
energy into our short-range energy models. The remainder
may be included in a subsequent relaxation step, but we find that this 
is not necessary as it is usually small, and in any case, 
this and all other errors against the SAPT(DFT) reference energies will
be accounted for in the final relaxation stage that we 
describe next.

\section{Total energy fits: combining the terms}
\label{sec:total}

The analytic fits to the various components of the total 
interaction energy model may be combined as appropriate, 
and optionally relaxed, using constraints, to the total
SAPT(DFT) interaction energies calculated for a suitable
set of dimer geometries.
These models have been obtained with a significant amount
of data derived directly from the density and
transition densities using various partitioning methods.
The limited amount of fitting has been largely restricted to the
short-range energy model, and even here, our approach ensures that
the parameters are well-defined and physically meaningful, with little
of the uncertainty usually associated with fits to sums of exponentials.
Further, the target residual error for each of the models 
has been $0.5$ to $1$ \kJpermol, and we have largely succeeded
in achieving this target. Consequently, as we shall see,
these models may be combined without further relaxation
to produce reasonably accurate models for the total interaction 
energy.

In this paper, we have reported the following models:
\begin{itemize}
    \item \emph{Short-range}: Three models have been obtained.
        srModel(1) is fully isotropic; srModel(2) contains a 22c anisotropy
        term on the nitrogen atoms; and srModel(3) contains all the 
        dominant anisotropy terms needed.
        These short-range energy models include the first-order 
        exchange, the electrostatic penetration, and
        infinite-order charge-transfer energies.
    \item \emph{Electrostatic}: A rank 4 ISA-based distributed
        multipole model.
    \item \emph{Polarization}: Three distributed polarization models obtained
        from the WSM procedure. The L1(iso) and L1 models include rank 1 
        polarizabilities, with the former being isotropic, and the
        L2 model includes terms to rank 2 on the heavy atoms. 
        All these models are damped. The many-body contributions are
        obtained through the polarization models.
        We will consider only the L1 model in this paper.
    \item \emph{Dispersion}: Two damped isotropic dispersion models have been obtained.
        The \Cniso{6} model contains only (scaled) isotropic \Cn{6}
        coefficients for all pairs of atoms.
        And the \Cniso{12} model consists of isotropic terms to \Cn{12} between pairs of
        heavy atoms, isotropic terms to \Cn{10} between any hydrogen atom and a heavy 
        atom, and only isotropic \Cn{6} terms between pairs of hydrogen atoms.
        As the \Cn{12} terms in the \Cniso{12} are found to have a minimal effect
        on the quality of the model, we will instead use the equivalent \Cniso{10}
        in the remainder of this work.
        All models are damped. At present we do not include any three-body dispersion 
        non-additivity.
\end{itemize}
This gives us $18$ possible ways of combining these
models into total interaction energy potentials. Of these, we explore
three combinations in this paper:
\begin{itemize}
    \item \emph{Model(1)}: Isotropic short-range model, with rank 4 ISA-DMA,
        L1 polarizability model, and \Cniso{10} dispersion model.
    \item \emph{Model(2)}: Short-range model containing isotropic terms on all 
        atoms and an additional 22c term on the nitrogen atoms, 
        with rank 4 ISA-DMA, L1 polarizability model, and \Cniso{10} dispersion model.
    \item \emph{Model(3)}: Anisotropic short-range model, combined with
        rank 4 ISA-DMA, L1 polarizability model,
        and \Cniso{10} dispersion model.
\end{itemize}
These models differ only in their description of the short-range repulsion.

In Table~\ref{tab:models-rms} we report \rms errors made by these models
before relaxation against the SAPT(DFT) interaction energies.
The \rms errors are remarkably small at this stage, with models
(1) and (3) exhibiting errors less than $1$ \kJpermol for the most
energetically important dimers. Surprisingly, Model(2) fares slightly
worse than the simpler Model(1) with \rms errors of $1.5$ \kJpermol in
this energy range.
All models fare reasonably well for the positive energy dimers,
with \rms errors between $1.8$ to $2.9$ \kJpermol.

\begin{figure}
    % Fig 11
    \begin{center}
        \includegraphics[scale=0.45]{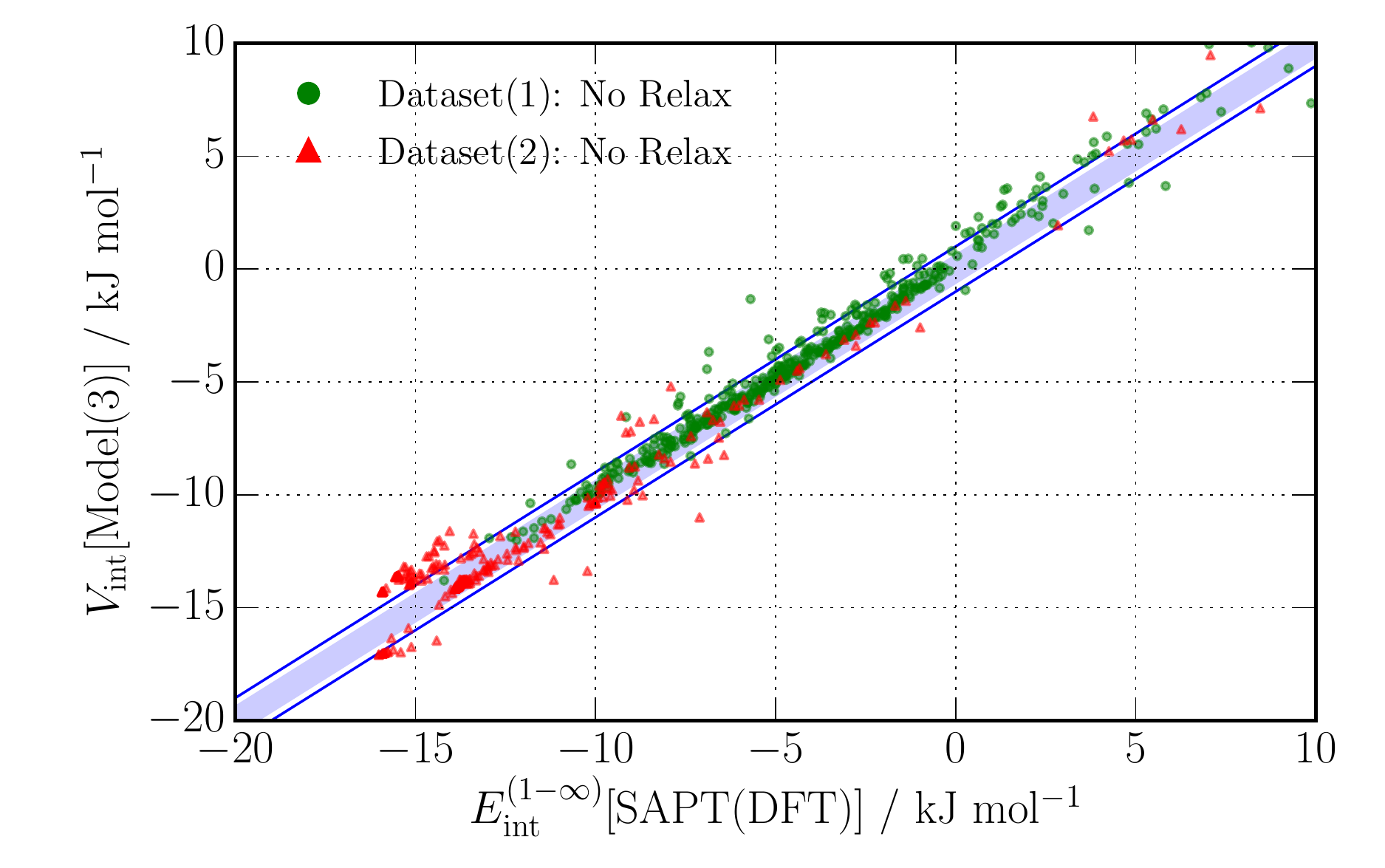}
        \includegraphics[scale=0.45]{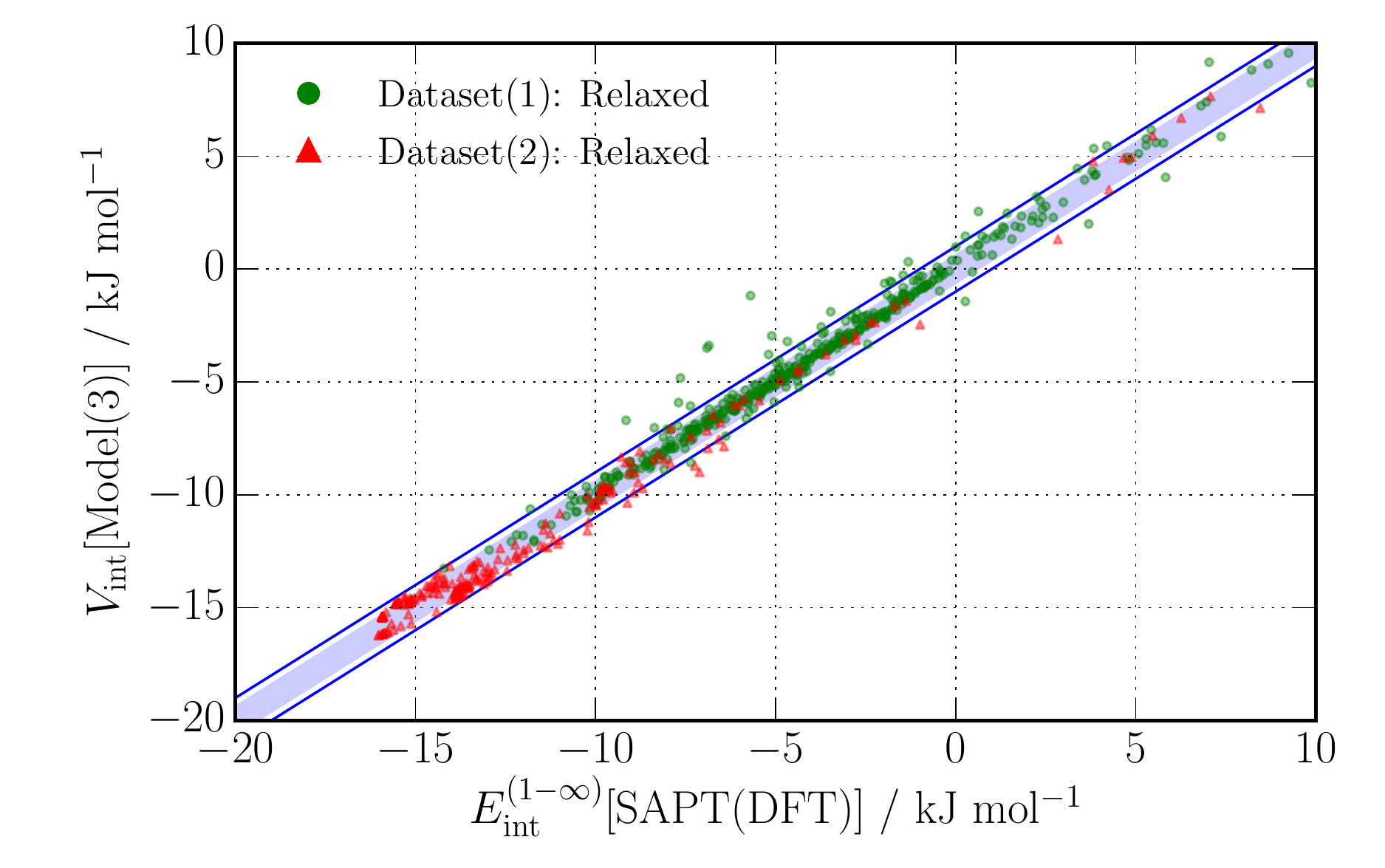}
    \end{center}
    \caption{
        The total interaction energy models for Model(3). The upper panel
        shows energies from Model(3) before relaxation to the dimers in 
        Dataset(1) and Dataset(2), and the lower panel shows
        model energies after relaxation. In both cases these energies are
        plotted against the total SAPT(DFT) interaction energy \Eint{1-\infty}.
        The blue bar represents the $\pm 1$ \kJpermol deviation from SAPT(DFT).
    }
    \label{fig:Eint-Model3-scatter}
\end{figure}

The models may be improved by constrained relaxation to SAPT(DFT) total
interaction energies.
We initially relaxed the models against energies from the random dimers in 
Dataset(1), but this led to a reduction in the quality of the fits for the test
set of low-energy dimers.
It appears that while the random dimers are suitable for an unbiased 
parametrization of the individual components of the model, 
they are not suitable for relaxing the sum of these components.
The principal reason for this seems to be that the random dimer set
does not contain low-energy dimers, as can be seen in 
Figure~\ref{fig:Eint-Model3-scatter}. Consequently, relaxing to this set
causes the models to represent these relatively high-energy 
dimers better at the cost of the more physically important low-energy 
configurations. 
Because of this, we have performed the relaxation of the models using
both Dataset(1) and Dataset(2).

The constrained relaxation was performed using the \Orient program
with the weighting scheme described in \S\ref{sec:numerical}. 
Constraints were imposed using eq.~\eqref{A-eq:constrained-opt}  in \paperA 
with tight constraint strength parameters $c_i$ chosen to be $1.0$ for the 
isotropic parameters and the \Cn{8} and \Cn{10} parameters. 
The \Cn{6} terms were kept unaltered so as to preserve the long-range dispersion
interaction.
The anisotropy parameters were not allowed to vary. 
Rather than relax all parameters simultaneously, the relaxation was 
performed in stages, with parameters associated with particular sites
allowed to vary in each stage. 
This procedure, though computationally efficient, needed to be iterated
to ensure that the relaxation was adequate.

In Table~\ref{tab:models-rms} we also report \rms errors made by the relaxed
models. After relaxation, all three models show \rms errors of only
$0.5$ to $0.6$ \kJpermol for the most strongly bound dimers, and
somewhat larger errors for the higher energy dimers.
Perhaps unsurprisingly, Model(3) fares best, with \rms errors less than $1$ \kJpermol 
for all dimers with energies less than or equal to $20$ \kJpermol. 

In Figure~\ref{fig:Eint-Model3-scatter} we display scatter plots of the
interaction energies calculated with Model(3) against SAPT(DFT) energies
both before and after relaxation. 
The excellent performance of the unrelaxed Model(3) is evident. At no stage 
in the development of Model(3) were the total interaction energies from 
Dataset(1) included; rather we only used the charge-transfer energies in 
the development of this model. Additionally, none of the low-energy dimers
in Dataset(2) were used in any way in the construction of Model(3), yet these
energies are accurately predicted by the unrelaxed Model(3), with very few
outliers.
This model may be improved by relaxing it to the dimer energies in both data
sets. This relaxation was performed with the anisotropic terms in the potential
frozen and only the isotropic parameters, including the low-ranking dispersion 
coefficient, allowed to vary with tight anchors imposed (see the SI for additional
information).
As seen in Figure~\ref{fig:Eint-Model3-scatter}, this relaxed model exhibits an
excellent correlation with the SAPT(DFT) reference energies, and has fewer low 
energy outliers compared with the unrelaxed model. In the remainder of this
paper by `Model(3)' we will refer to this relaxed model.

Similar figures for Model(1) and Model(2) can be found in the SI.
As may be expected from the \rms errors reported in Table~\ref{tab:models-rms},
the performance of the unrelaxed Model(1) is excellent given the simplicity of
the model, but the unrelaxed Model(2) shows somewhat larger errors for the 
most strongly bound dimers. However, both of these models improve considerably on 
relaxation.

In Table~\ref{tab:models-rms} we also report \rms errors for a model
functionally identical to Model(3), but created using the DF-AIM approach
and with DMA4 multipoles. 
Apart from these two differences, this model, termed Model(3)-DF-DMA4,
has been created in an identical manner to the others reported in this 
paper. 
This is the kind of model that might have been 
created using the approach we have described in an earlier paper
on atom--atom potentials\cite{StoneM07}. 
We see that across the $-20:20$ \kJpermol energy range
the \rms errors made by this model are twice as large as those
from Model(3). This is mainly a consequence of the unphysical 
AIM atoms that result from the DF-AIM approach that are shown in 
Figure~\ref{fig:pyr-DF-atoms}. This approach results in the wrong
atomic anisotropies that the fit cannot correct with the limited
amount of SAPT(DFT) data in Datasets (1) and (2). 
This is an inevitable consequence of the Bayes-like approach we
have adopted: the role of the first step in the fitting process 
--- the first-order fits through the distributed density overlap model ---
is to determine prior values for the fitting parameters 
(see \S\ref{A-sec:strategy} in \paperA). The subsequent relaxation steps 
merely refine these prior values. However, if the prior values are 
very poor, as they are with the DF-AIM approach, then we require a 
considerable amount of data to move them to the correct values.
This is not needed with the ISA-based AIM approach, and demonstrates
the superiority of this method.

\begin{table*}%[Ht]
    \setlength{\tabcolsep}{7pt}
    \caption[Pyr2-models-rms]{
        R.m.s.\ errors (\kJpermol) for the total interaction energy models
        for the pyridine dimer. 
        Errors are calculated against SAPT(DFT) total interaction energies,
        and are reported both for the models relaxed to the set of SAPT(DFT)
        energies in Dataset(1) and Dataset(2), and for the models
        obtained by combining the different terms in the potential as described
        in the text. 
        The errors for these unrelaxed models are reported in parantheses. 
        %The columns headed `No Relax' denote errors obtained from unrelaxed 
        %models, that is those obtained by combining the various terms as described
        %in the text, while the columns headed `Relaxed' denote the same for 
        %the models after relaxation to the set of SAPT(DFT) interaction energies.
        Model(3)-DF-DMA denotes a model functionally similar to Model(3) but 
        created using the DF-AIM approach with multipoles from the DMA4 model.
%       Energies and \rms errors are reported in \kJpermol.
        % AJM May16
    }
        \label{tab:models-rms} 
    \begin{center}
        \begin{tabular}{lcccc}
            \toprule
            Energy range                            &\multicolumn{1}{c}{Model(1)}&\multicolumn{1}{c}{Model(2)}&\multicolumn{1}{c}{Model(3)}
            & \multicolumn{1}{c}{Model(3)-DF-DMA4}\\
            \midrule
            $\phantom{-10 \lt \,\,} E \le -10$      &  0.59 (1.26)          &  0.59 (1.22)         &  0.53 (1.08)         &  0.97 (1.85)          \\                                                             
            $-10 \lt E \le \phantom{-1}0$           &  0.80 (0.99)          &  0.72 (0.95)         &  0.56 (0.70)         &  1.21 (1.39)          \\                                                             
            $\phantom{-1}0 \lt E \le \phantom{-}20$ &  1.69 (2.71)          &  1.19 (2.58)         &  0.95 (1.53)         &  2.17 (3.16)          \\

%       \begin{tabular}{lcccccccc}
%           Energy range                            & \multicolumn{2}{c}{Model(1)}  & \multicolumn{2}{c}{Model(2)}  & \multicolumn{2}{c}{Model(3)}  
%           & \multicolumn{2}{c}{Model(3)-DF-DMA4}\\
%                                                   & No Relax     &  Relaxed       &  No Relax     &  Relaxed      &  No Relax     &  Relaxed      
%           & No Relax     &  Relaxed             \\ 
%           \midrule
%           $\phantom{-10 \lt \,\,} E \le -10$      & 1.26         & 0.59           & 1.22          & 0.59          & 1.08          & 0.53          
%           & 1.85         & 0.97                 \\
%           $-10 \lt E \le \phantom{-1}0$           & 0.99         & 0.80           & 0.95          & 0.72          & 0.70          & 0.56          
%           & 1.39         & 1.21                 \\
%           $\phantom{-1}0 \lt E \le \phantom{-}20$ & 2.71         & 1.69           & 2.58          & 1.19          & 1.53          & 0.95          
%           & 3.16         & 2.17                 \\
            \bottomrule
        \end{tabular}
    \end{center}
\end{table*}

\section{Results}
\label{sec:results}

\subsection{Minima}
\label{sec:minima}

We have used the basin-hopping algorithm (see Ref.~\citenum{Wales:book:03} for a review)
as implemented in the \Orient program to search for stable dimers on
the potential energy surfaces.
In contrast to the rather simple PES of the benzene dimer
\cite{PodeszwaBS06b,TottonMK10a} which supports only three minima, 
we have found eight minima for the pyridine dimer.
The minimum-energy structures, which are illustrated in Figure~\ref{fig:minima-structures}, 
may be classified according to their bonding:
\begin{itemize}
    \item Hydrogen-bonded: These include Hb1, Hb2 and Hb3. Of these, Hb1 is doubly
        hydrogen-bonded and has been found in a DFT-D (BLYP+Grimme D1 correction) 
        search \cite{PiacenzaG05} and has also been investigated at the CCSD(T)/CBS 
        level of theory \cite{HohensteinS09} to be around $-15.5$ \kJpermol 
        (estimated from Figure 5 in Ref.~\citenum{HohensteinS09}). This compares
        well with our SAPT(DFT) energy of $-16.6$ \kJpermol.
        The Hb2 and Hb3 structures do not appear to have been reported in prior literature. 
    \item Stacked: The S1 and S2 minima are the stacked dimers which are largely
        dispersion-bound. Both these structures have been found in the DFT-D 
        search, however we see no evidence of the two other stacked structures
        reported in that study.
    \item T-shaped: None of these minimum energy dimers are exactly T-shaped,
        but the T1 and T2 minima are nearly so, and the bT minimum is a 
        very bent-T-shaped structure. 
        The bT structure is similar to one of the T-shaped structures found in the DFT+D search.
        We do not find the `T-shaped 1' structure in the DFT+D search by Piacenza and
        Grimme \cite{PiacenzaG05}. 
\end{itemize}
The minimum configurations are displayed in
Figure~\ref{fig:minima-structures} and their energies are reported in 
Table~\ref{tab:Eint-minima} and displayed visually in Figure~\ref{fig:Eint-minima-corr}.
%\marginpar{Evidence that these are true minima? Do the references provided above suffice?}
For comparison, we have calculated
SAPT(DFT) interaction energies for the dimer configurations 
obtained from the relaxed Model(3) PES. 
Not all of the models support all the minima. Model(2) does not support
the Hb3 minimum, which instead relaxes to the T1 structure on this model PES.
The relaxed Model(3)-DF-DMA4 supports only five of the eight minima, and
two of those (S2 and T1) differ in structure from the corresponding structures 
on the ISA-based surfaces: in the S2 structure on this surface the molecules are 
not parallel, and the T1 is bent.
The three missing structures relax to either the Hb2 or the T1 structures.
The Hb2 minimum is the global energy minimum on this PES.

For Model(3) we have reported energies for the minima on both
the unrelaxed and relaxed models.
These largest energy differences in the minima on these two PESs differ 
by just over $1.1$ \kJpermol (just over 7\% of the interaction energy).
This is a remarkable result as it indicates that the unrelaxed models
can be predictive without the need for fitting to the SAPT(DFT) 
total interaction energies, in particular, no information about total 
interaction energies of the stable, low-energy dimers was used in creating
the three unrelaxed models.  Further, the similarity of the relaxed 
and unrelaxed models suggests that the procedure used here appears 
to be free of artifacts usually introduced by fitting procedures,
and is robust to the inclusion of additional data. However this data needs
to be biased to low energy dimers, as has been noted above.
We will explore this issue in a forthcoming paper \cite{VanVleetMSS15}.

%\marginpar{Similar in geometry? Yes.}

\begin{figure}
    % Fig 12
    \begin{center}
        \includegraphics[scale=0.45]{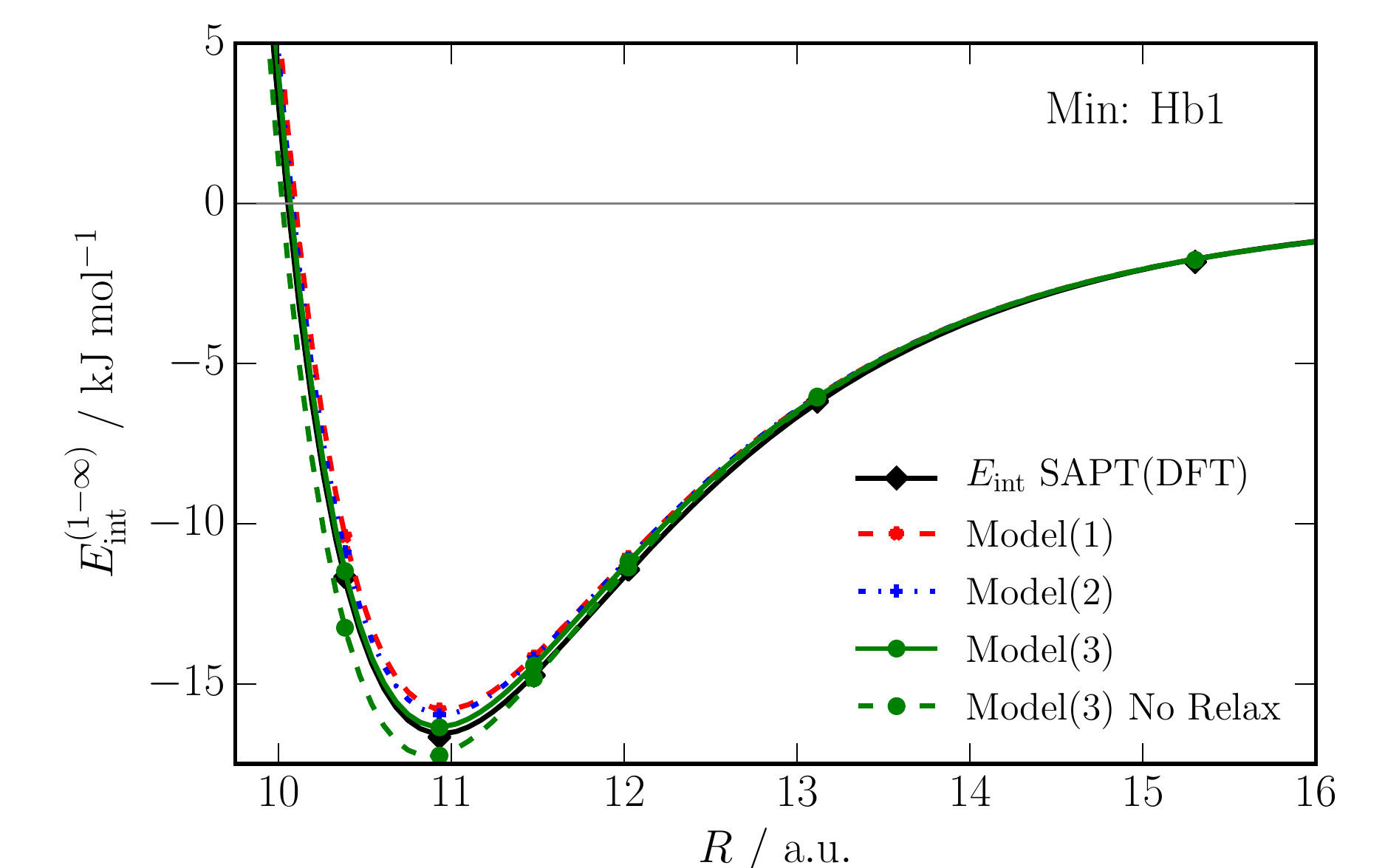}
        \includegraphics[scale=0.45]{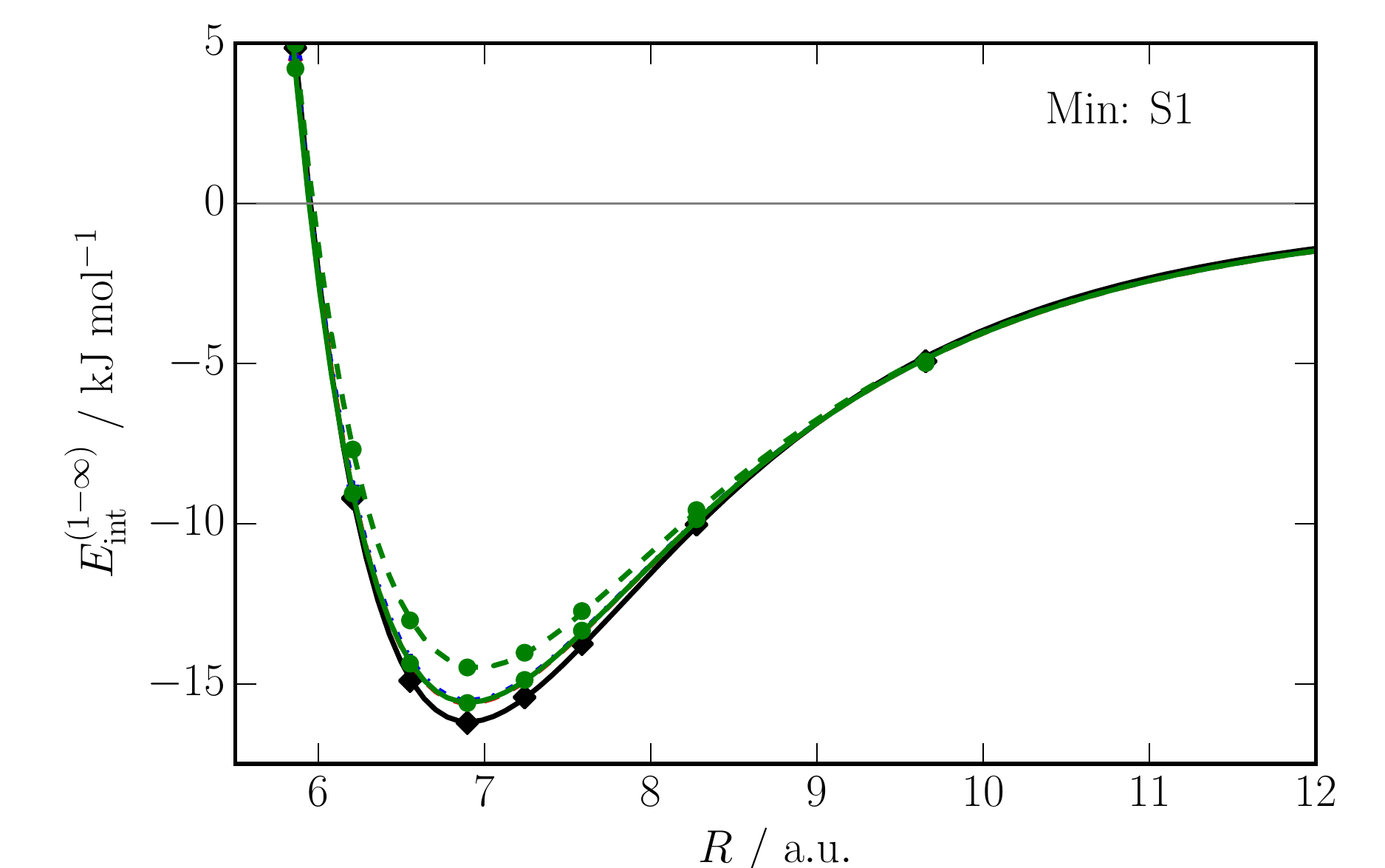}
        \includegraphics[scale=0.45]{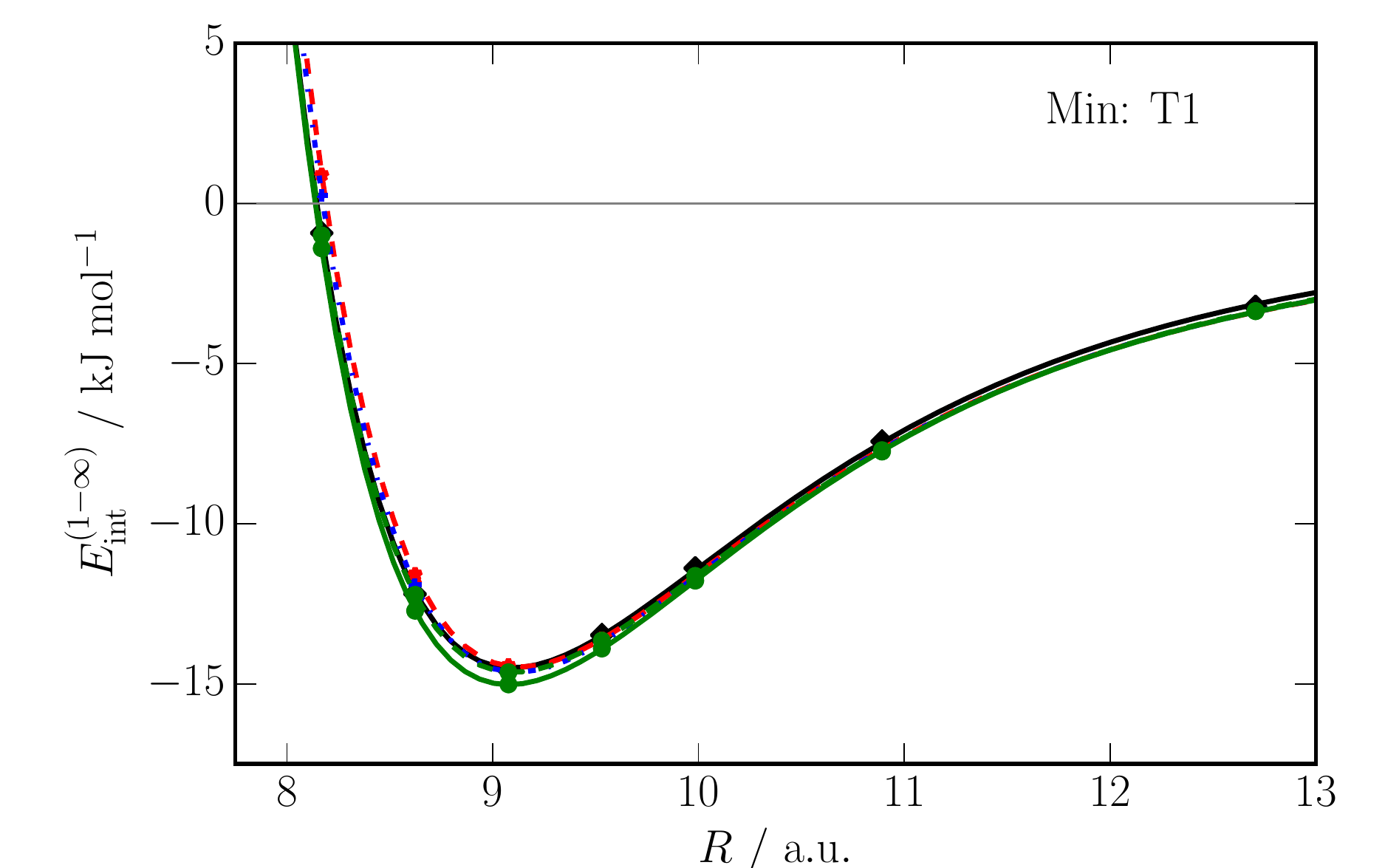}
    \end{center}
    \caption{
        PES sections at the Hb1, S1 and T1 dimer orientations.
        Sections at the other minima are provided in the supplementary information.
    }
    \label{fig:Eint-sections}
\end{figure}

\begin{figure}
    % Fig 13
    \begin{center}
        %\stackunder[-1pt]{\includegraphics[scale=0.13]{./figs/minima/Min-1-Hb1.png}}{(1) Hb1}
        %\stackunder[-1pt]{\includegraphics[scale=0.13]{./figs/minima/Min-2-S1.png}}{(2) S1}
        %\stackunder[-1pt]{\includegraphics[scale=0.13]{./figs/minima/Min-3-S2.png}}{(3) S2}
        %\stackunder[-1pt]{\includegraphics[scale=0.13]{./figs/minima/Min-4-T1.png}}{(4) T1}
        %\stackunder[-1pt]{\includegraphics[scale=0.13]{./figs/minima/Min-5-T2.png}}{(5) T2}
        %\stackunder[-1pt]{\includegraphics[scale=0.13]{./figs/minima/Min-6-Hb2.png}}{(6) Hb2}
        %\stackunder[-1pt]{\includegraphics[scale=0.13]{./figs/minima/Min-8-bT.png}}{(7) bT}
        %\stackunder[-1pt]{\includegraphics[scale=0.13]{./figs/minima/Min-7-Hb3.png}}{(8) Hb3}
        \stackunder[-1pt]{\includegraphics[scale=0.13]{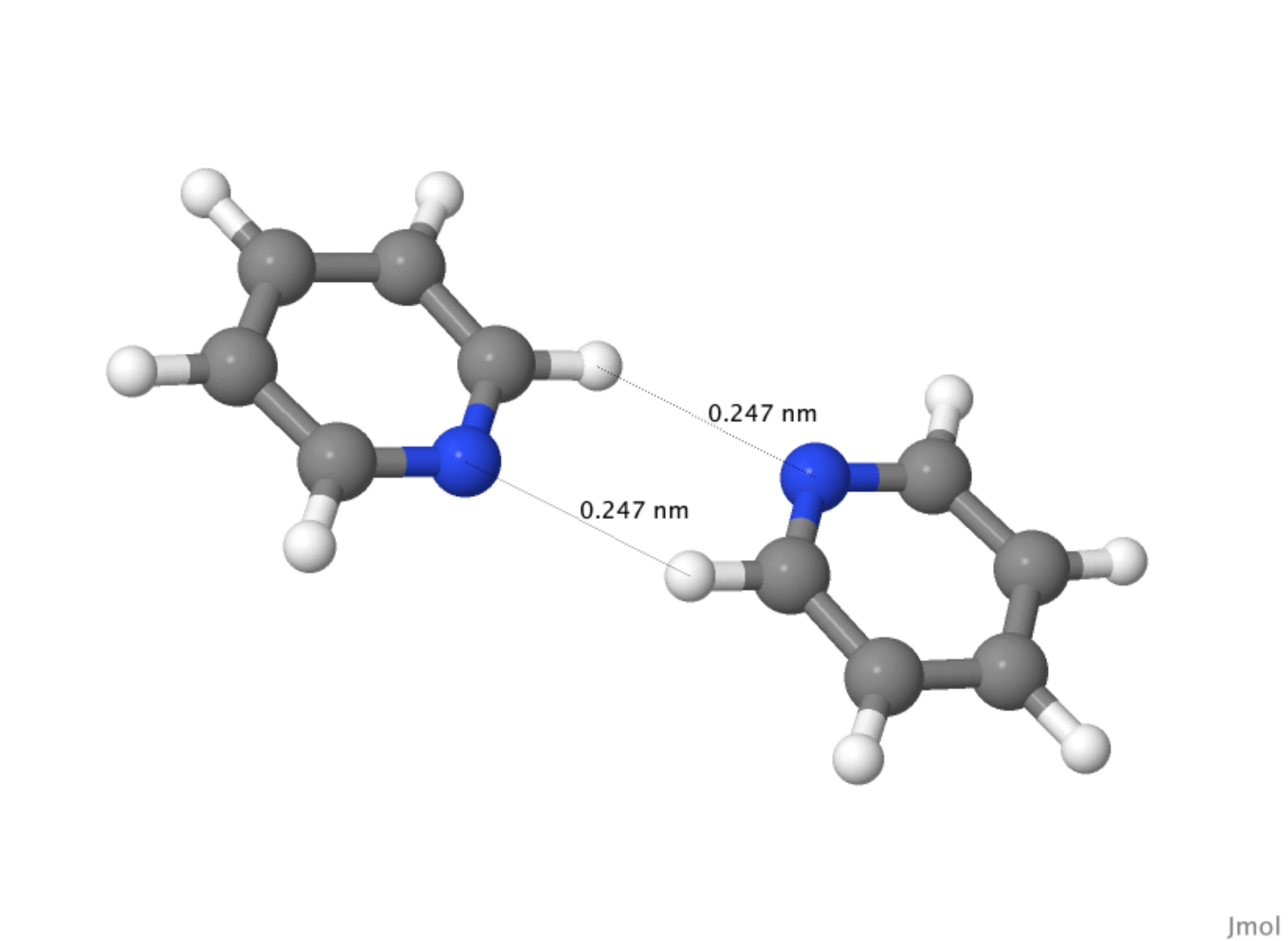}}{(1) Hb1}
        \stackunder[-1pt]{\includegraphics[scale=0.13]{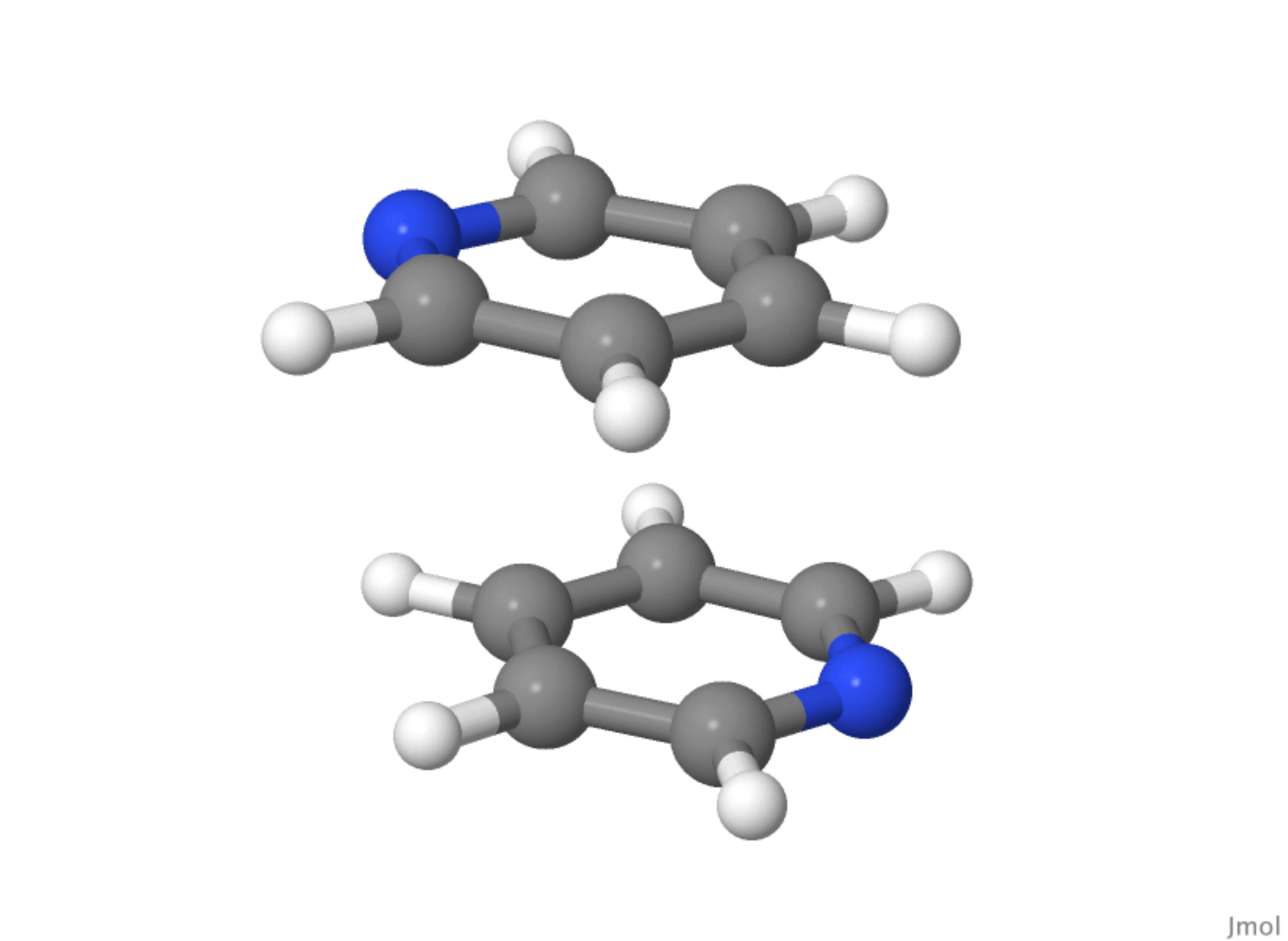}}{(2) S1}
        \stackunder[-1pt]{\includegraphics[scale=0.13]{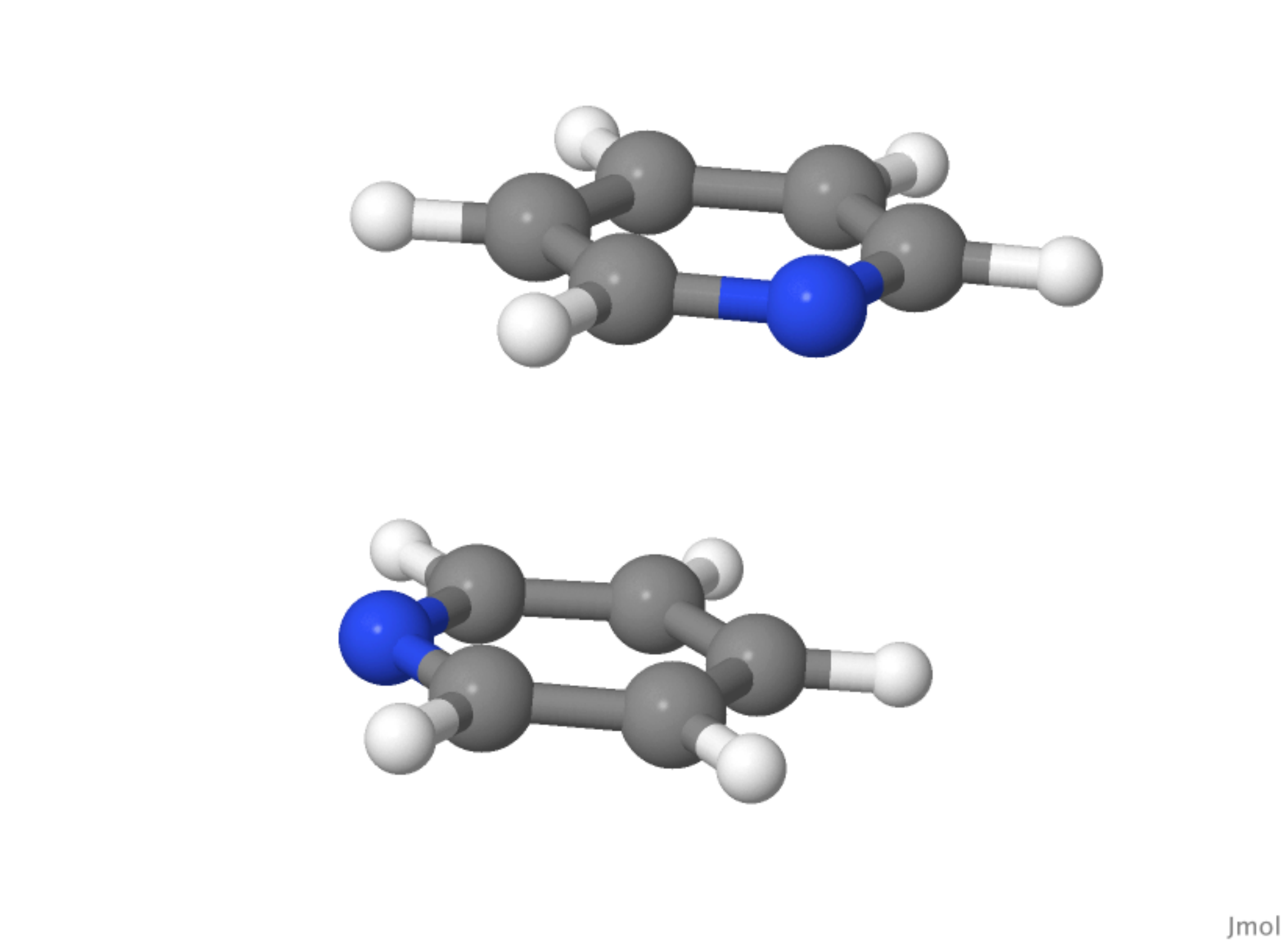}}{(3) S2}
        \stackunder[-1pt]{\includegraphics[scale=0.13]{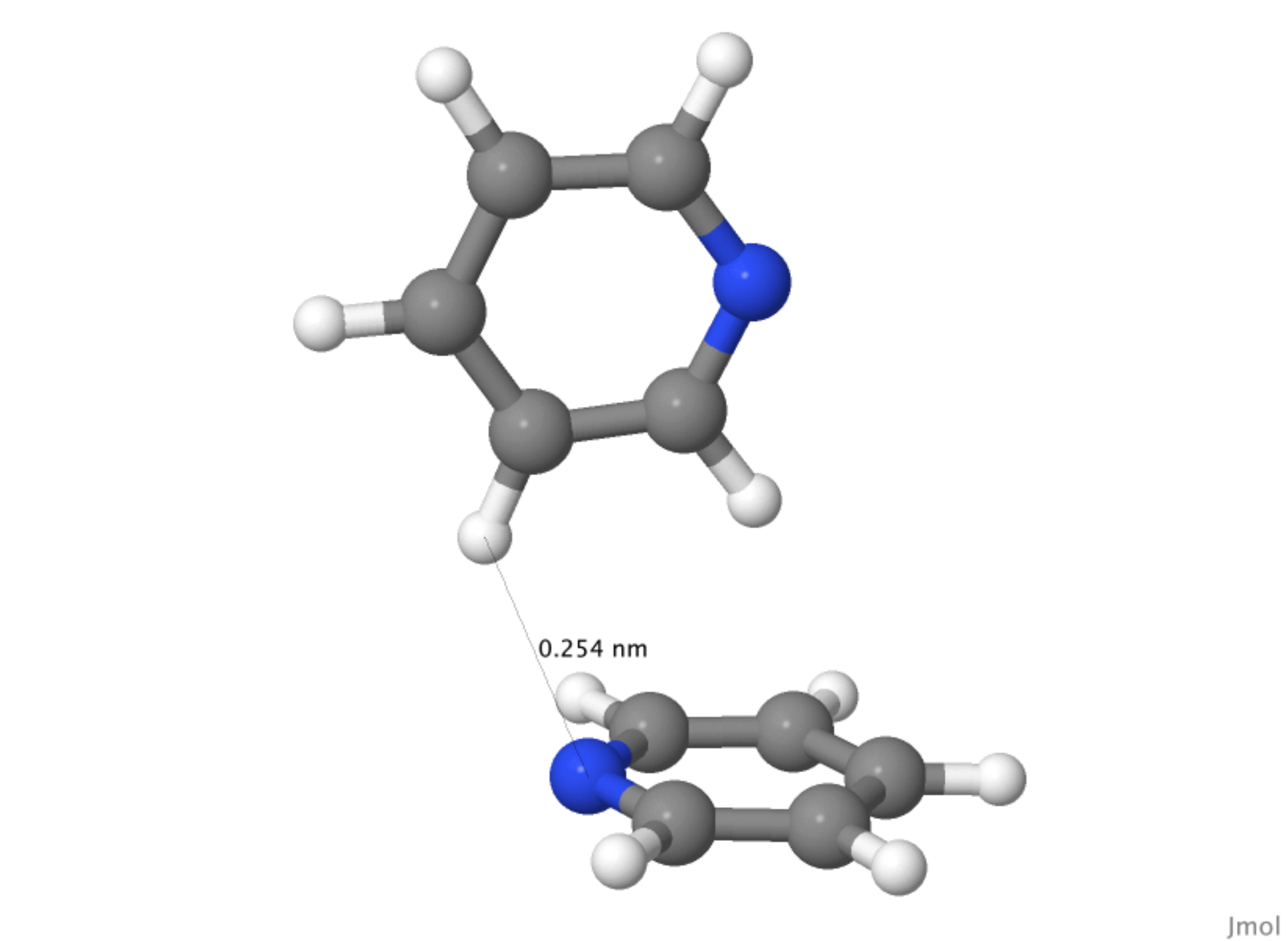}}{(4) T1}
        \stackunder[-1pt]{\includegraphics[scale=0.13]{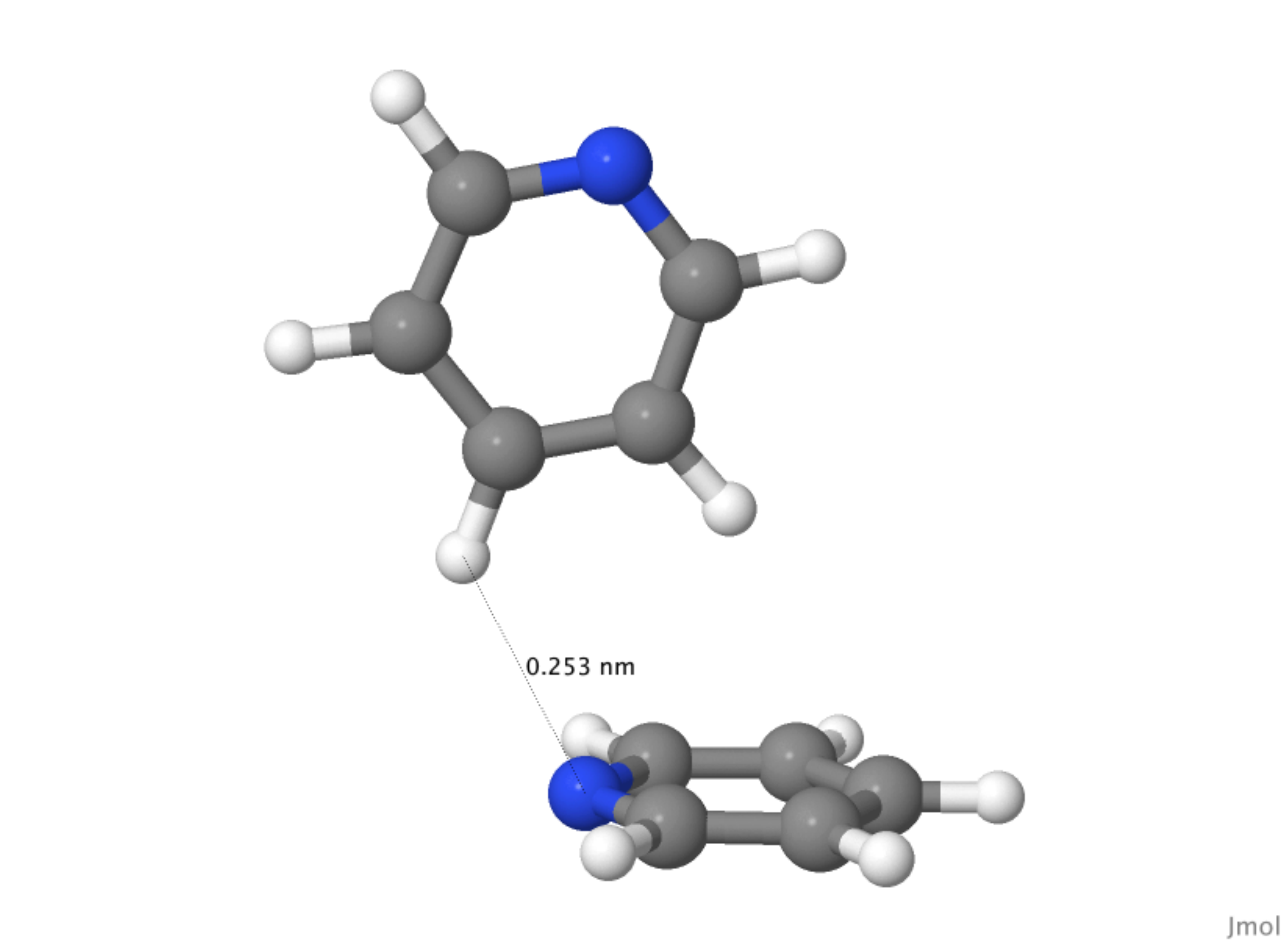}}{(5) T2}
        \stackunder[-1pt]{\includegraphics[scale=0.13]{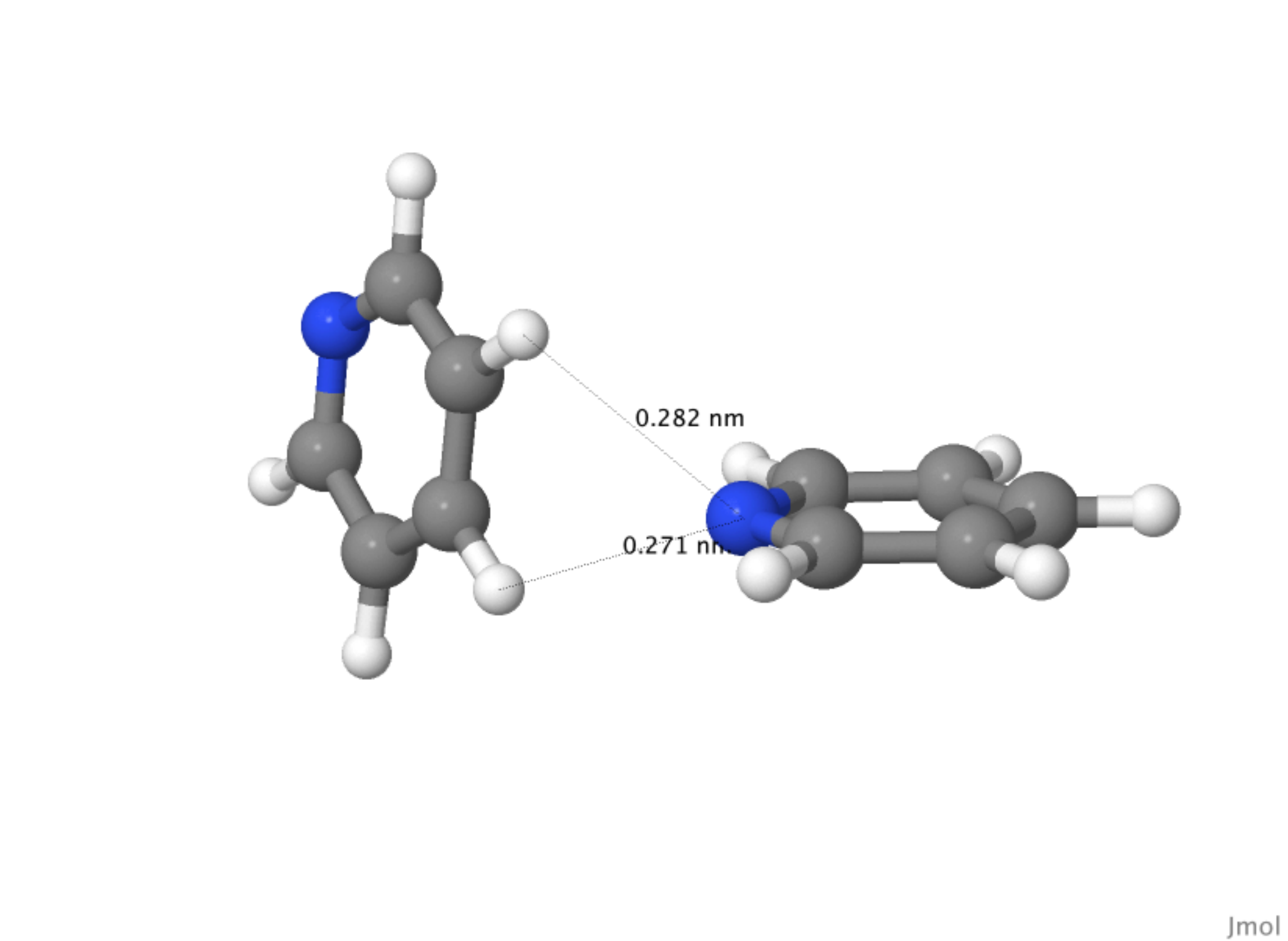}}{(6) Hb2}
        \stackunder[-1pt]{\includegraphics[scale=0.13]{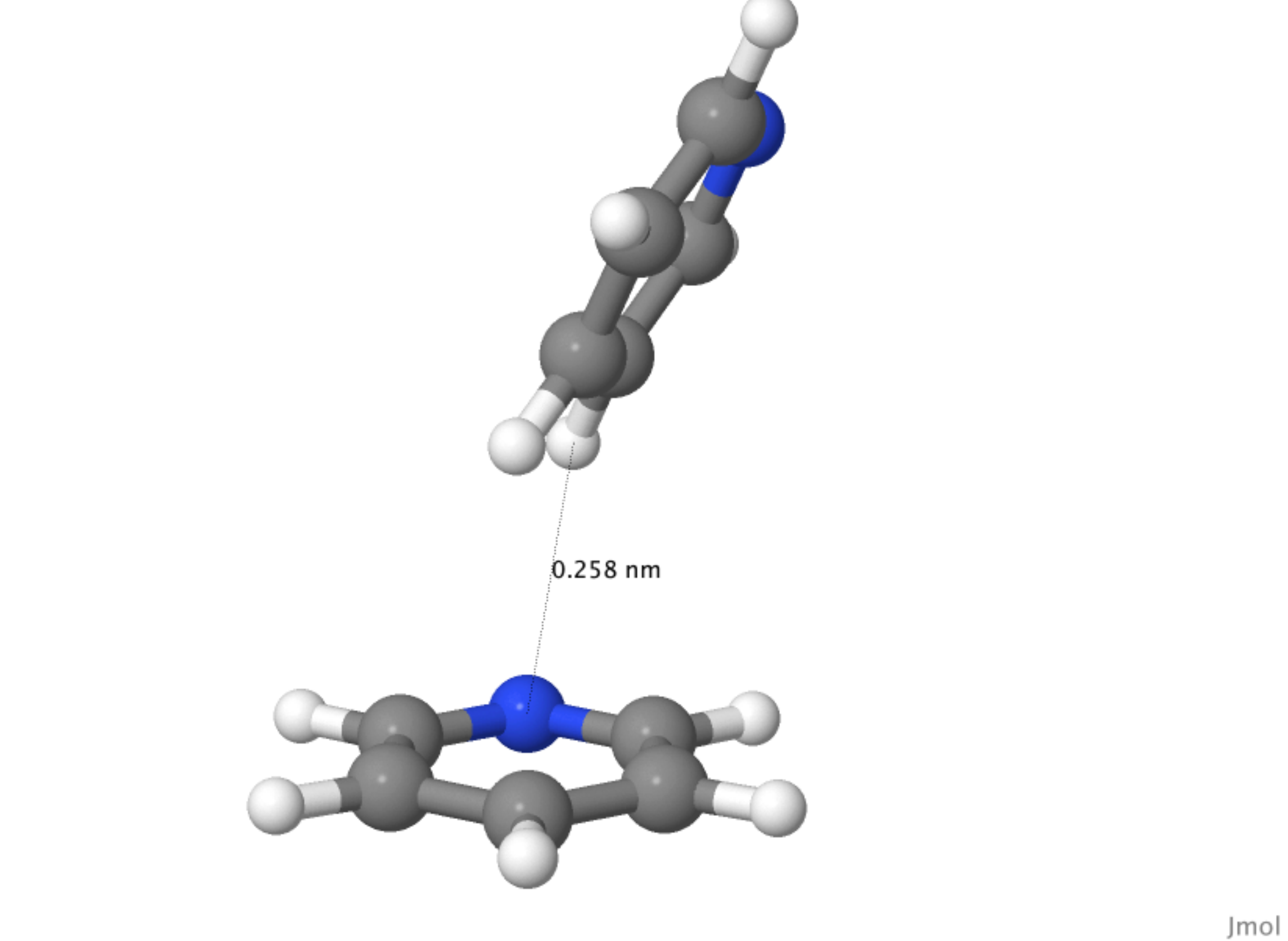}}{(7) bT}
        \stackunder[-1pt]{\includegraphics[scale=0.13]{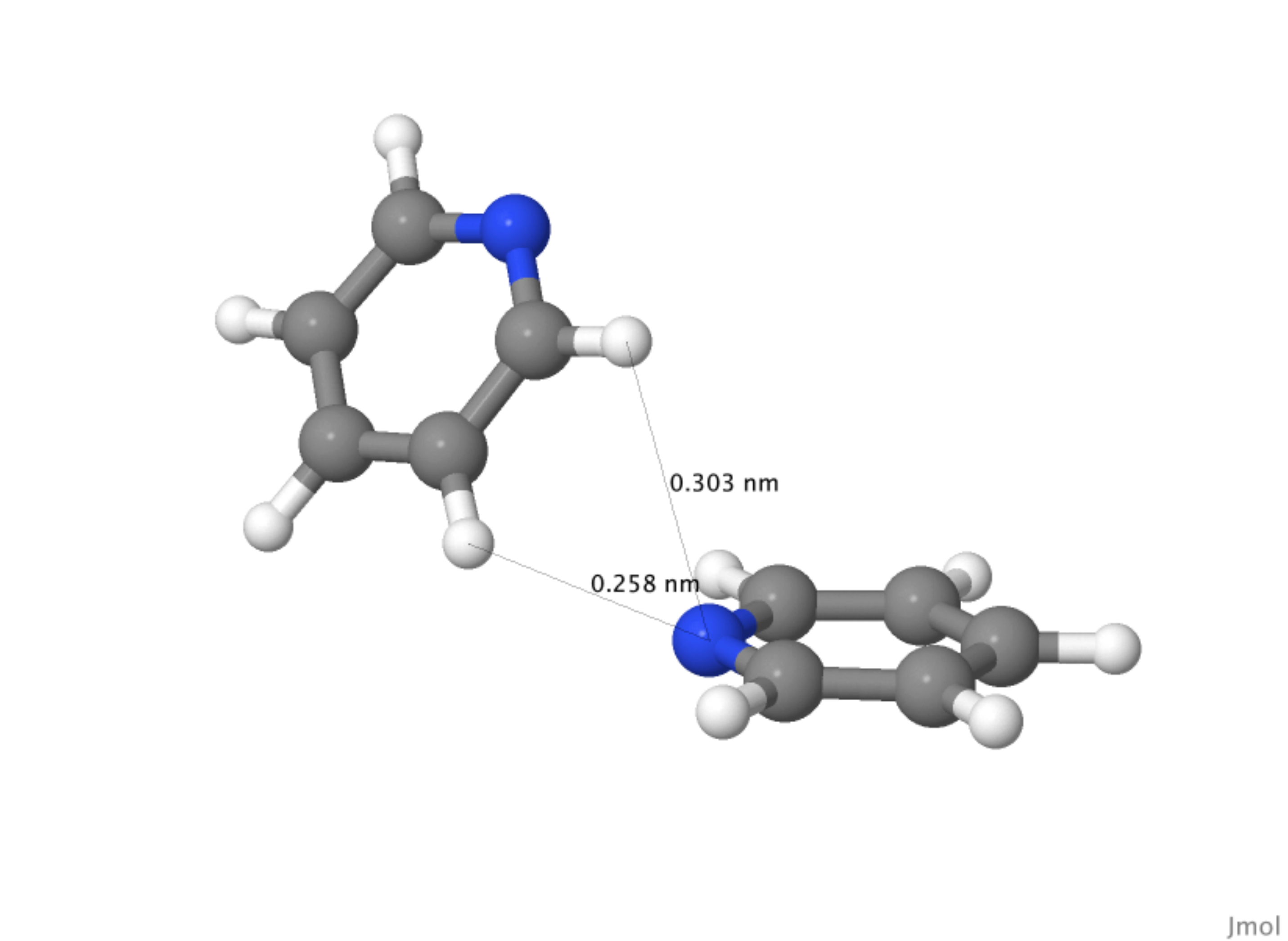}}{(8) Hb3}
    \end{center}
    \caption{
        Structures of pyridine dimers at stable minima on the three PESs.
        The structures are ordered according to their energies calculated 
        using SAPT(DFT). 
        These images have been produced using the {\sc Jmol} program \cite{Jmol}.
    }
    \label{fig:minima-structures}
\end{figure}

\begin{figure}
    % Fig 14
    \begin{center}
        \includegraphics[scale=0.45]{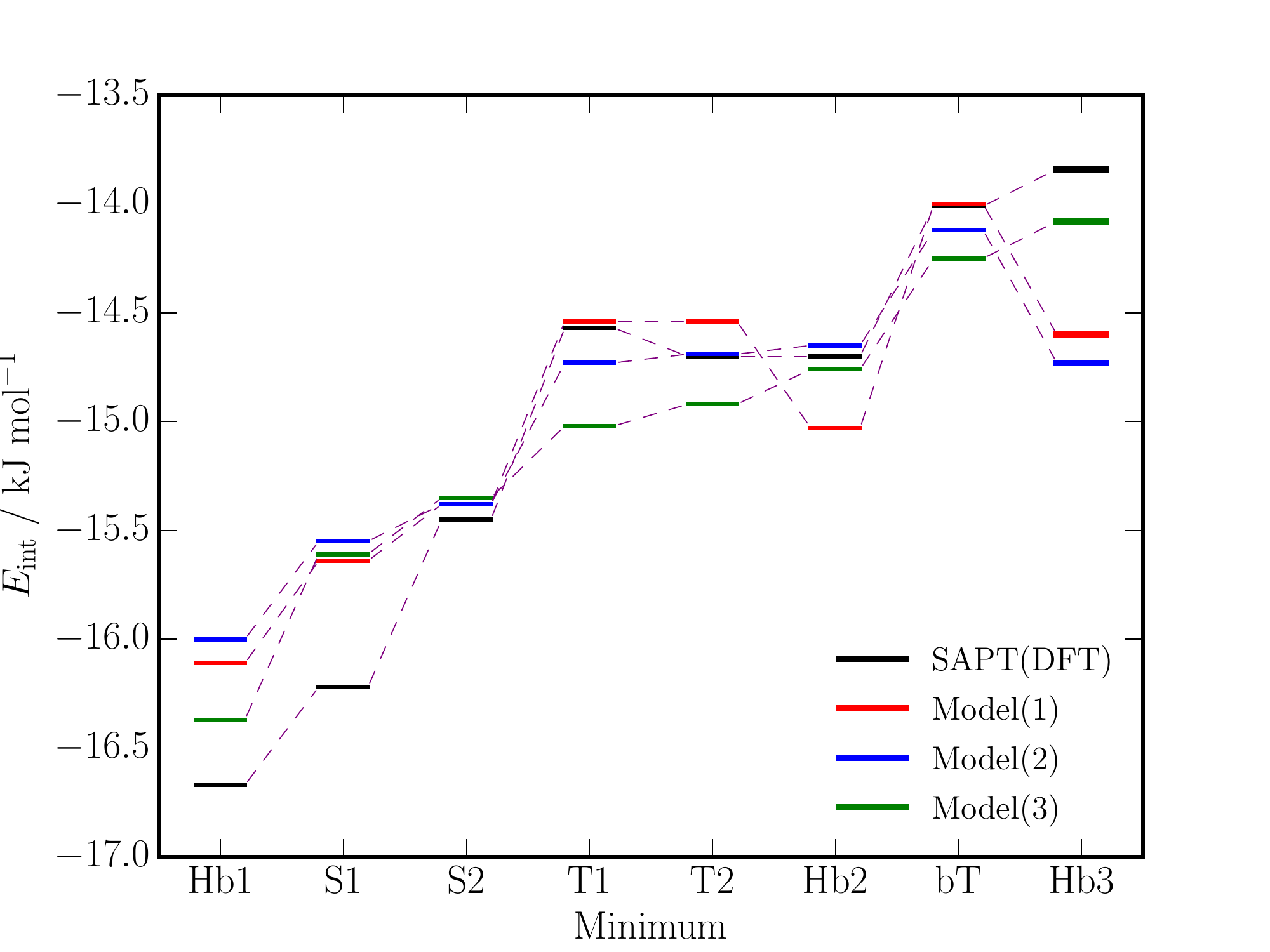}
    \end{center}
    \caption{
        Visualisation of the data in Table~\ref{tab:Eint-minima}. 
        The energies of stable dimers on the three PESs are displayed as 
        solid horizontal bars. The dashed lines link the energies levels
        associated with each of the three models. 
        SAPT(DFT) reference energies have been calculated at the dimer
        geometries from Model(3).
        Data for Model(3)-DF-DMA4 are not shown here.
    }
    \label{fig:Eint-minima-corr}
\end{figure}

The agreement between the ISA-based models (relaxed and unrelaxed)
is made even clearer in Figure~\ref{fig:Eint-sections} where
we display PES sections at representative minima. 
The agreement between SAPT(DFT) and all models --- including the 
unrelaxed Model(3) --- for the minima is generally very good, both 
in the overall shape of the PESs and the location and depth of
the radial minimum. 
Plots for the remainder of the minima can be found in the SI.

In Table~\ref{tab:minima-freq} we report the lowest harmonic vibrational 
frequencies at these minima. 
These frequencies give us an indication of how different the shapes of 
the three PESs are at the stable minima configurations. 
There is generally a good agreement between the minima on all ISA-based models,
but the frequencies seem to vary more with the models than the corresponding
energies. This may reflect the importance of the anisotropy in determining
the shape of the PES.
This agreement, though imperfect, is reassuring as it gives us some confidence
that the minima we observe are real and not artifacts of the fitting function used.
The largest differences are between the ISA-based models and the DF-based Model(3)-DF-DMA4.
The lowest vibrational frequencies of the Hb1 and S1 minima are only half as large as
the corresponding frequencies for Model(3), indicating that the shape of the PES
of Model(3)-DF-DMA4 differs from that of Model(3) in the regions of these minima.
This should not be a surprise given the rather significant differences
in the AIM shapes from the ISA- and DF-based density partitioning schemes 
as shown in Figures~\ref{fig:pyr-DF-atoms} and \ref{fig:pyr-ISA-atoms}.

\begin{table*}
    \setlength{\tabcolsep}{7pt}
    \caption[Pyr2-Eint-minima]{
        Interaction energies (\kJpermol) of the pyridine dimers at the energy minima reported
        in Figure~\ref{fig:minima-structures}. 
        The SAPT(DFT) reference energies have been calculated at the dimer 
        geometries obtained on the relaxed Model(3) PES. 
        The energies reported for all models are for the stationary points
        on the model PES, therefore the dimer geometries at which the energies
        are evaluated will depend on the model and will differ from the 
        geometries used to obtain the SAPT(DFT) reference energies.
        Where a structure is not supported as a minimum we report in parentheses 
        the structure it relaxes into.
        Thus Model(2) does not support the Hb3 structure which instead relaxes to the T1 
        minimum on this PES.
        Structures on the Model(3)-DF-DMA4 surfaces that are only approximately the
        same as those on the other surfaces are indicated by an asterisk.

%        All energies are in \kJpermol.
        % AJM 16 May
    }
    \label{tab:Eint-minima} 
    \begin{center}
        \begin{tabular}{c.......}
            \toprule
            Minimum & \multicolumn{1}{c}{SAPT(DFT)} 
                               & \multicolumn{1}{c}{Model(1)} 
                                          & \multicolumn{1}{c}{Model(2)} 
                                                     & \multicolumn{2}{c}{Model(3)} 
                                                                            &\multicolumn{2}{c}{Model(3)-DF-DMA4} \\
                    &          &\multicolumn{1}{c}{Relaxed} 
                                          &\multicolumn{1}{c}{Relaxed} 
                                                     & \multicolumn{1}{c}{No Relax}
                                                                & \multicolumn{1}{c}{Relaxed} 
                                                                            &\multicolumn{1}{c}{No Relax}
                                                                                       &\multicolumn{1}{c}{Relaxed} \\
            \midrule
            Hb1     & -16.67   & -16.11   & -16.00   & -17.28   & -16.37    &-14.38    & -15.04  \\
            S1      & -16.22   & -15.64   & -15.55   & -14.54   & -15.61    &-13.60^{*}& -15.46  \\
            S2      & -15.45   & -15.38   & -15.38   & -14.17   & -15.35    &-12.71^{*}& -14.42^{*}\\ 
            T1      & -14.57   & -14.54   & -14.73   & -14.65   & -15.02    &-14.63    & -14.84^{*}\\
            T2      & -14.70   & -14.54   & -14.69   & -14.68   & -14.92    &\multicolumn{1}{c}{(Hb2)}&\multicolumn{1}{c}{(Hb2)}\\
            Hb2     & -14.70   & -15.03   & -14.65   & -14.57   & -14.76    &-15.19    & -15.61  \\
            bT      & -14.01   & -14.00   & -14.12   & -13.97   & -14.25    &\multicolumn{1}{c}{(Hb2)}&\multicolumn{1}{c}{(Hb2)}\\
            Hb3     & -13.84   & -14.60   &\multicolumn{1}{c}{(T1)}
                                                         & -13.88   & -14.08    &-14.00    &\multicolumn{1}{c}{(T1$^{*}$)}\\
            \bottomrule
        \end{tabular}
    \end{center}
\end{table*}

\begin{table*}
    \setlength{\tabcolsep}{7pt}
    \caption[Pyr2-minima-freq]{
        Lowest harmonic vibrational frequencies for the minima on the relaxed model PESs. 
        For Model(3) we also include data for the unrelaxed version of this model.
        Model(2) does not support the Hb3 minimum.
        All frequencies are reported in cm$^{-1}$.
        \label{tab:minima-freq} 
        % AJM 16 May
    }
    \begin{center}
        \begin{tabular}{c......}
            \toprule
            Minimum & \multicolumn{1}{c}{Model(1)} 
                                       & \multicolumn{1}{c}{Model(2)} 
                                                         & \multicolumn{2}{c}{Model(3)} 
                                                                                  & \multicolumn{2}{c}{Model(3)-DF-DMA4} \\
                    & \multicolumn{1}{c}{Relaxed}  
                                       & \multicolumn{1}{c}{Relaxed}  
                                                         & \multicolumn{1}{c}{No Relax}  
                                                                      & \multicolumn{1}{c}{Relaxed}
                                                                                  & \multicolumn{1}{c}{No Relax}  
                                                                                             & \multicolumn{1}{c}{Relaxed}   \\ 
            \midrule
            Hb1     &  15.96           &  12.76          &  15.79     &  15.08    &  7.01    &  7.70  \\
            S1      &   6.04           &   3.83          &   4.79     &   6.69    &  1.94    &  3.42  \\
            S2      &   9.99           &  10.53          &   8.98     &  11.24    &  9.60    & 11.81  \\
            T1      &   3.74           &   4.45          &   9.57     &   6.62    &  5.37    &  8.22  \\
            T2      &   1.94           &   3.89          &   7.38     &   6.07    &\multicolumn{1}{c}{---} & \multicolumn{1}{c}{---} \\
            Hb2     &  12.05           &   9.94          &  12.15     &  12.19    & 11.92    & 10.91  \\
            bT      &   5.55           &   7.43          &   3.13     &   6.28    &\multicolumn{1}{c}{---} & \multicolumn{1}{c}{---} \\
            Hb3     &  12.07           &\multicolumn{1}{c}{---}& 11.36&  10.88    &  6.88    & \multicolumn{1}{c}{---} \\
            \bottomrule
        \end{tabular}
    \end{center}
\end{table*}

\subsection{Second virial coefficients}
\label{sec:virial}

\begin{figure}
    % Fig 15
    \begin{center}
        \includegraphics[scale=0.45]{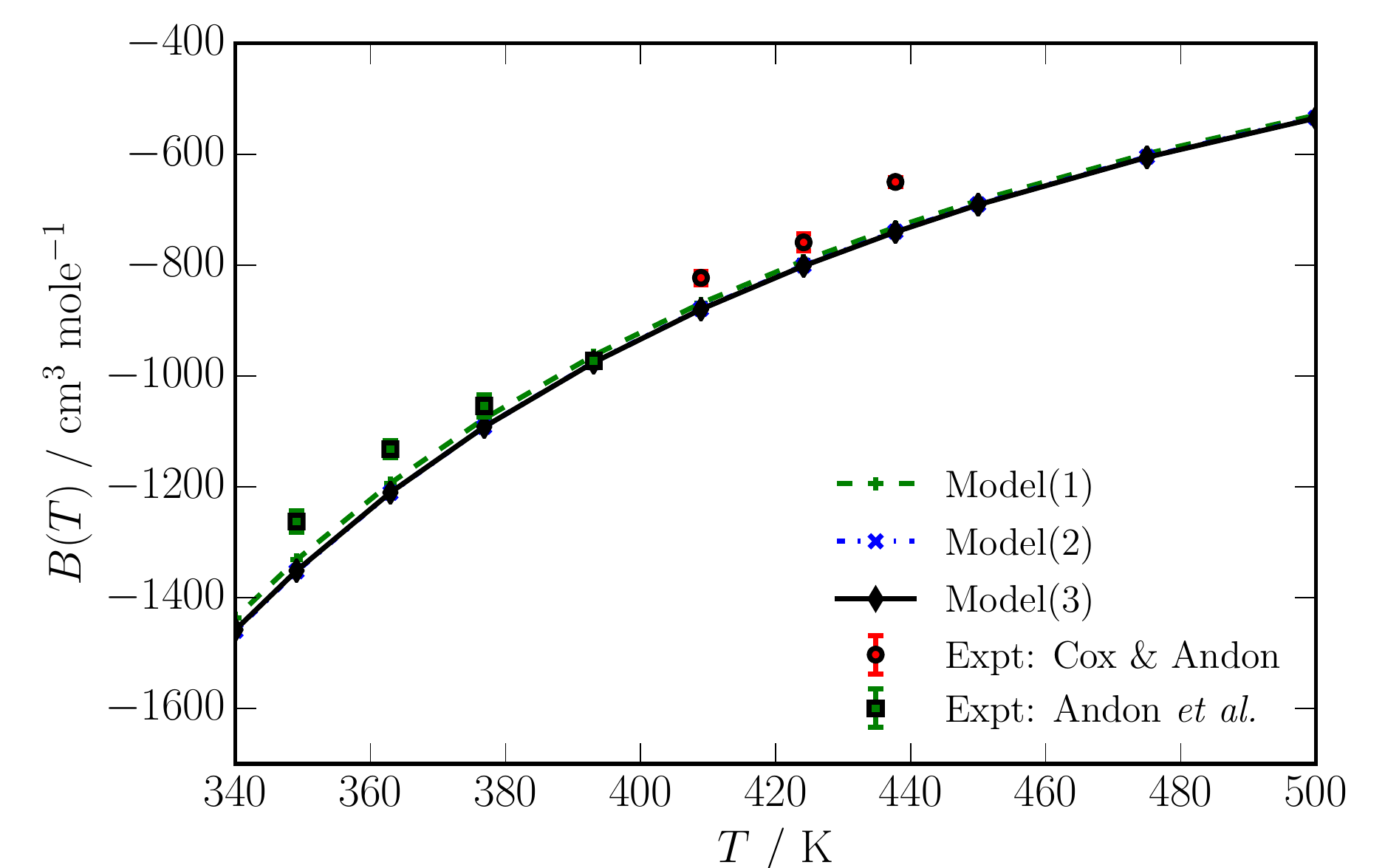}
    \end{center}
    \caption{
        Classical second virial coefficients for pyridine. The experimental data
        and error bars 
        are from Andon \etal \cite{AndonCHM57} and Cox \& Andon \cite{CoxA58}.
        Quantum corrections contribute very little and would not make a visible 
        difference on the scale of this graph.
    }
    \label{fig:virial}
\end{figure}

The second pressure virial coefficient $B(T)$ represents a necessary,
but not sufficient, test of the quality of the two-body PES.
As the virial coefficients average over the PES, it is possible to
construct an infinity of PESs that yield the correct values of $B(T)$
in a finite temperature range.  Nevertheless, it is important that 
any model PES reproduces the experimental values as a minimum requirement.
In Figure~\ref{fig:virial} we display second virial coefficients calculated
for the pyridine dimer. We have calculated $B(T)$ at a range of temperatures
using the \Orient program. Only the Classical results are presented as the
quantum corrections were found to be insignificant over the range of temperatures
reported here. 
We used a stochastic integration sampling algorithm with 102 radial steps
and 262,144 dimer orientations in order to integrate $B(T)$ sufficiently 
accurately. 
From Figure~\ref{fig:virial} we see that all three models show
good agreement with the experimental data of Andon \etal \cite{AndonCHM57} 
and Cox \& Andon \cite{CoxA58}. As the models all slightly overestimate $B(T)$ 
across the temperature range of the figure, they may, on the whole, be somewhat 
too attractive. We will return to this issue later in this paper.

\section{Analysis \& Discussion}
\label{sec:analysis}

\subsection{Polarization damping revisited}
\label{sec:pol-damping-revisited}

\begin{figure}
    % Fig 16
    \begin{center}
        \includegraphics[scale=0.45]{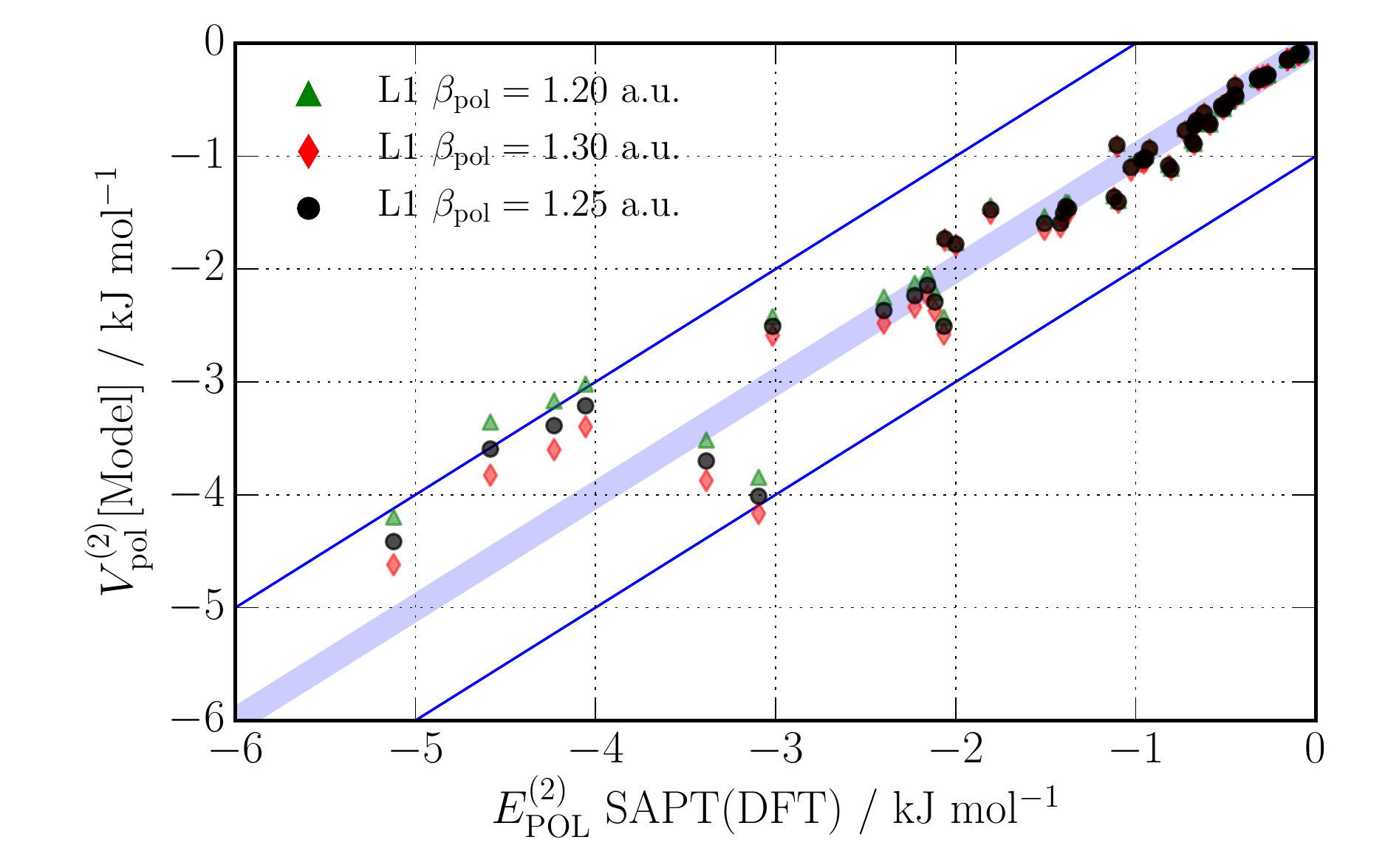}
    \end{center}
    \caption{
        Second-order polarization energies from regularised SAPT(DFT) 
        compared with the L1 polarization model. 
        The energies have been calculated using minimum energy 
        dimer configurations obtained on the Model(3) PES.
        Dimers with attractive total interaction energies are indicated
        with filled symbols, and those with repulsive energies with 
        open symbols.
        The thin blue lines indicate the $\pm 1$ \kJpermol error limits
        and the blue bar is present just as a visual aid.
    }
    \label{fig:pol2-all-minima}
\end{figure}

In developing the damping model for our polarizability models in 
\S\ref{A-sec:polarization} in \paperA we recognised an uncertainty in our choice
for damping model. This arose because the damping
parameter \betapol depends on the choice of dimer
configurations used to determine it. 
Here we re-examine this issue by assessing the damping models against 
data obtained at the eight minimum energy dimer orientations at 
various separations.
In Figure~\ref{fig:pol2-all-minima} we compare the second-order polarization
energies from the polarization models described in \S\ref{A-sec:polarization}
with second-order polarization energies from regularised SAPT(DFT), \EPOL{2}, 
It should be apparent that while our choices for the damping models
are reasonable, with errors typically less than 1 \kJpermol for the
attractive dimers, there is a systematic over-damping, with the polarization
energies of some (repulsive energy) dimers underestimated by as much as 
2.5 \kJpermol. 
This problem can be largely remedied by increasing the value of \betapol.
In the same figure we also display polarization energies calculated with 
the anisotropic L2 polarization model with $\betapol=1.0$ a.u.\ 
This small increase causes a significant improvement to the match between 
the model and \EPOL{2}.

In this manner, we are able to determine a new set of models with the 
appropriate polarization damping chosen self-consistently. 
As we emphasised in \S\ref{A-sec:polarization}, the choice of 
\betapol does not affect the quality of the two-body potential. 
Indeed, Model(3) with this change to the damping is nearly identical
in every respect to the original model. The effects will however be 
manifest in the many-body polarization energies.
We are currently investigating this issue.

\subsection{Multipole model rank reduction}
\label{sec:rank-reduction}

\begin{figure}
    % Fig 17
    \begin{center}
        \includegraphics[scale=0.45]{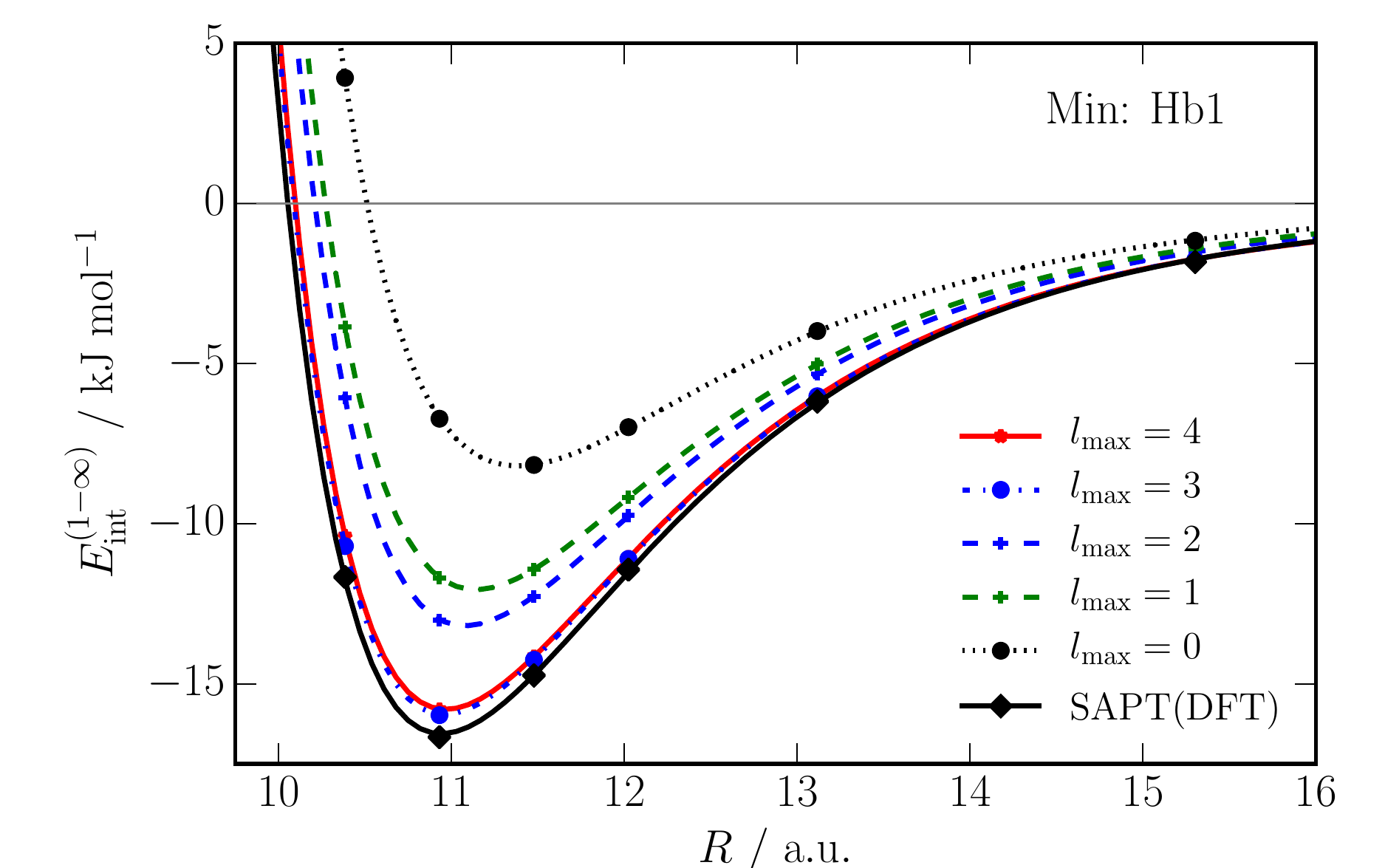}
        \includegraphics[scale=0.45]{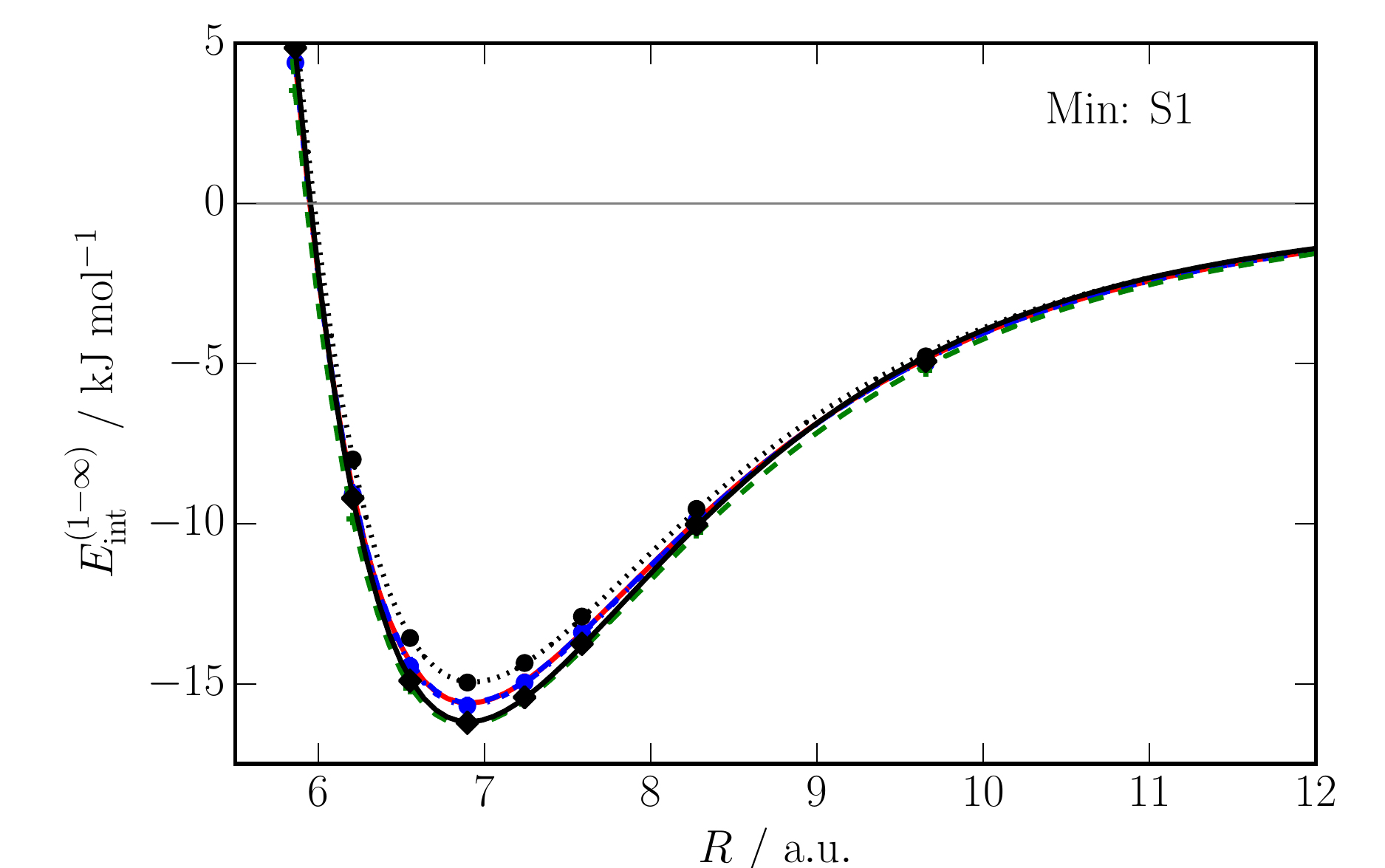}
        \includegraphics[scale=0.45]{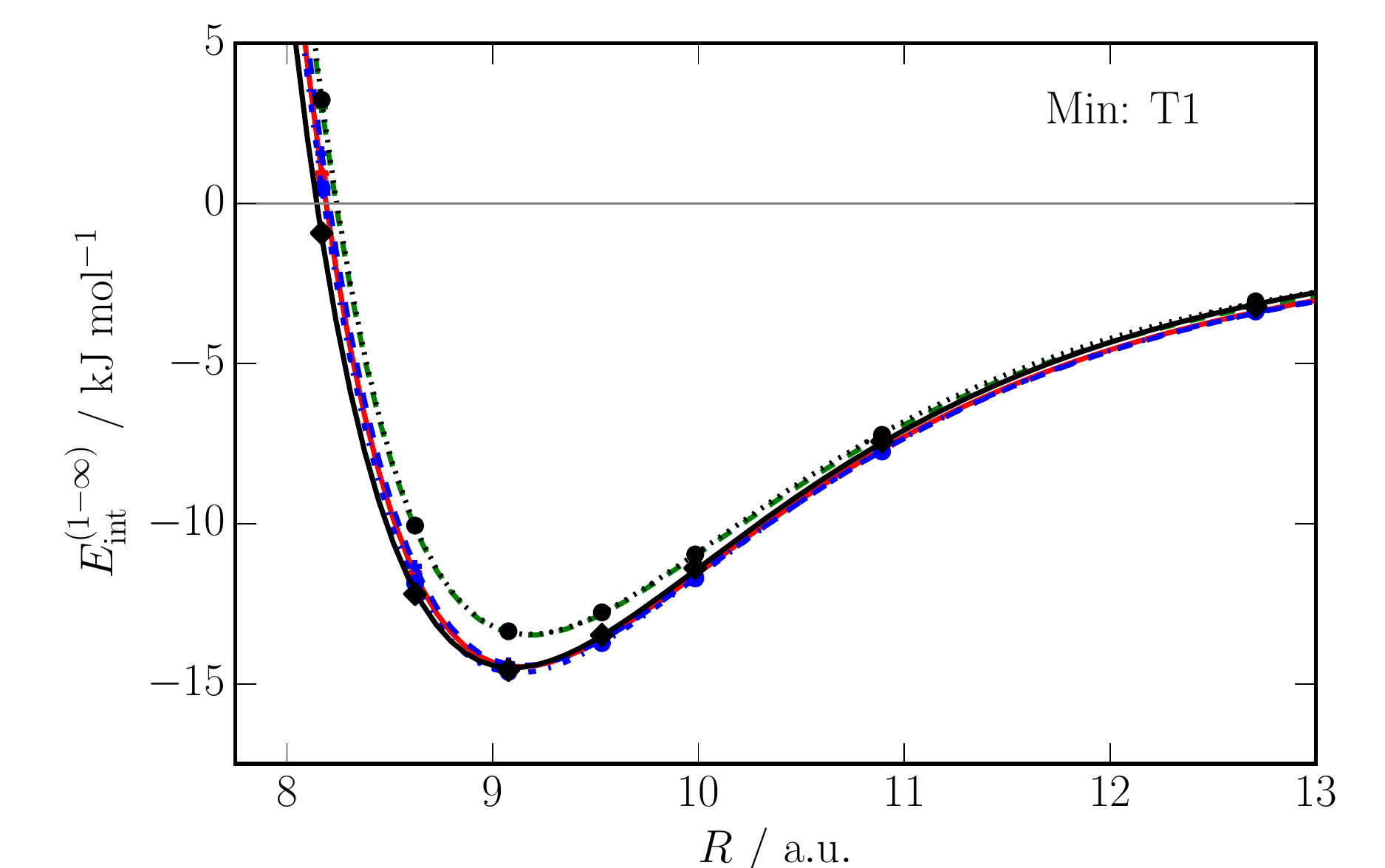}
    \end{center}
    \caption{
        The effect of rank reduction of the multipole model for Model(1).
    }
    \label{fig:rank-reduction}
\end{figure}

Our simplest model, Model(1), contains anisotropic terms only in the 
ISA-DMA multipole and the polarization models. 
In \S\ref{A-sec:electrostatics} in \paperA we have argued
that the ISA-DMA model shows better convergence properties than
the usual DMA procedure of Stone \cite{Stone05}. Based on that discussion
and the results presented in Figure~\ref{A-fig:DMA4-ISA-elst-convergence}  in \paperA , 
we may ask whether we can truncate the rank of the ISA-DMA model without
incurring a significant loss in accuracy. 
In Figure~\ref{fig:rank-reduction} we display interaction energy profiles
for Model(1) using the ISA-DMA model at various ranks. 
As before, these calculations have been performed at two representative dimer
orientations: Hb1 and S1. At the doubly hydrogen-bonded Hb1 orientation there is
no appreciable change on reducing rank to $l=3$, but any further reduction
results in a significant change in the PES with the interaction energy getting
systematically smaller (in magnitude).
At the dispersion-bound S1 orientation there is almost no change
to the model when the rank of the multipole expansion is reduced all the way
to $l=0$ (charges only).
This is perhaps to be expected as the electrostatic interaction is
relatively insignificant for the S1 (and S2) complexes.
What is surprising is that the T1 complex also shows a relative insensitivity
to the rank of the multipole expansion.

The behaviour of the models at the doubly-hydrogen-bonded Hb1 dimer configuration
needs some explanation. The rank of multipoles on the hydrogen atoms do not
appear to matter as the model interaction energies do not alter
significantly if only rank 0 (charge) terms are included on these atoms.
However, the nitrogen and carbon atoms appear to need the octopolar terms
to model the electrostatic interaction correctly in this
configuration. 
At least for the nitrogen atom this should not be surprising as the 
octopolar terms are needed to describe the effects from the lone pairs,
but it is surprising that the carbon atoms also require these terms.
In any case, it may be possible to improve the quality of the charge-only
model by including additional sites around the nitrogen and carbon atoms to 
account for these terms in much the same way as is done for the oxygen atom
in water models.
If successful, this would provide us with a route to construct a fully isotropic
interaction model for pyridine and other systems. 
This would be important as, with some exceptions such as the \Orient and \DMACRYS \cite{PriceLWHPKD10}
programs, simulation programs cannot normally use
potentials with anisotropic terms, a restriction that
significantly limits the usage of the accurate potentials we are 
able to develop.

\section{Conclusions \& Directions}
\label{sec:conclusions}

We have described a robust and relatively easy to implement algorithm for 
developing accurate intermolecular potentials in which most of the potential
parameters are derived from the charge density and density response functions,
and the remaining, short-range, parameters are robustly determined by associating
these with specific atom-pairs using a basis-space implementation of
the iterative stockholder atoms (ISA) algorithm. 
With this algorithm, accurate, many-body potentials can be derived using a 
relatively small number of dimer energies calculated using SAPT(DFT).
This significantly reduces the computational cost of the approach. 
Importantly, as all of the long-range and most of the short-range 
parameters are \emph{derived}, the predictive power of the resulting
potentials is significant. 

One of the major obstacles to intermolecular potential development has been
the derivation of the short-range parameters. We have demonstrated that
these can be relatively easily and robustly derived from the non-interacting
charge densities using the distributed density-overlap model based the ISA.
In this manner, even the atomic anisotropy terms, which are usually poorly
defined in a direct fit, are robustly determined with a relatively
small amount of computational effort. 
Using these techniques on the pyridine dimer, we have demonstrated that
features such as the density distortions due to the $\pi$-bonding on the
carbon atoms, and the lone pair on the nitrogen atom in pyridine are
well-defined using our approach. Indeed, only terms with a physical
origin are present in this approach. 

The main features of the methodology we describe in the paper are:
\begin{itemize}
    \item \emph{Efficient use of data}: The potentials are derived using a 
        hierarchy of data sets; the more extensive data sets include only
        first-order energies and can be very easily calculated, while the
        second-order energies are included through a significantly smaller
        data set.
    \item \emph{Priors}: We use the first and most extensive data set to determine
        prior values for most of the short-range parameters. 
        These priors may subsequently be modified using the second, smaller data set.
        These steps may be repeated thus leading to a multi-stage procedure which
        significantly reduces the amount of data needed to tune the potential.
    \item \emph{ISA}: The short-range parameters are determined using the ISA method
        for partitioning the molecular densities into atomic contributions.
        The BS-ISA algorithm allows this to be performed using extensive basis sets
        with a well-defined basis set limit. 
        The ISA atoms are as close to spherical as is possible and account for 
        charge movement within the molecule, consequently the resulting short-range
        repulsion parameters may be expected to be free from basis set artifacts, and
        be the most isotropic possible.
        This compares favourably with the density-fitting-based partitioning
        scheme we have proposed in earlier papers \cite{StoneM07,MisquittaWSP08}
        which does not fulfil either of these properties.
        Indeed the \rms errors made by the ISA-based models are half as much as those
        from the density-fitting-based models.
    \item \emph{Long-range models}: The long-range parameters of the potentials
        are determined using distributed multipoles, polarizabilities and 
        dispersion coefficients. The ISA-DMA multipoles are obtained from the BS-ISA
        approach and have been demonstrated to exhibit systematic convergence with
        rank. The WSM distribution scheme has been used to calculate the distributed 
        polarizabilities and dispersion coefficients, the latter of which we have tuned
        to SAPT(DFT) dispersion energies.
    \item \emph{Predictive power}: Most of the parameters are derived from or fitted
        to molecular properties, consequently they are physically meaningful and the 
        resulting potentials exhibit a considerable predictive power.
    \item \emph{Hierarchy of models}: The methods we have described allow us to
        determine potentials of various levels of complexity in a meaningful manner.
        These may be fully isotropic at the atom--atom level or
        contain as much anisotropy as is needed. 
\end{itemize}

We have used these techniques to develop a set of potentials of varying
levels of detail for the pyridine dimer. The simplest of these include
only isotropic short-range terms, and the most detailed includes all 
significant anisotropy terms up to rank two. The predictive power of these
potentials is quite significant and all are able to predict SAPT(DFT) interaction
energies for low energy dimers not included in the fit. 
As a consequence, the potentials are robust to the inclusion of additional
data: parameters alter very little on relaxation, and features on the 
potential energy landscape change only slightly.
This robustness is particularly important in the development of multidimensional
potentials, as we will generally be unable to sample dimer configuration space
adequately, especially for larger monomers.

We have compared our newly derived pyridine potentials to the rather limited
set of data available in the literature. 
Of the eight stable minima found on the Model(3) PES, the double hydrogen-bonded
Hb1 dimer has been found in previous DFT+D work by Piacenza and Grimme \cite{PiacenzaG05},
and the CCSD(T) energy for this structure \cite{HohensteinS09} differs from our
SAPT(DFT) interaction energy by only 7\%. 
The two other hydrogen-bonded structures, Hb2 and Hb3, have not been seen before.
Both the stacked structures, S1 and S2, have been found earlier \cite{PiacenzaG05}.
Of the three T-shaped structures, only the bT structure resembles a previously
found structure \cite{PiacenzaG05}, while the T1 and T2 structures appear to be
unique to the models developed in this paper.
As the DFT+D method cannot be relied on to correctly describe the subtle balance of 
dispersion, electrostatic, polarization and charge-transfer interactions seen in the 
eight dimers of pyridine, it is possible that the set of eight minima we have found
are a more accurate representation of this system. Further tests are needed at the
CCSD(T) level of theory if we are to be sure of this.

In this paper we have provided solutions to some of the most significant issues
related to potential development, and, as a consequence, have inevitably exposed
other minor issues that need resolving. Some of these are:
\begin{itemize}
    \item The WSM method for deriving distributed polarization and dispersion 
        models is a good one, but it is based on a less than ideal partitioning
        method \cite{MisquittaS06} that seems to result in some artifacts in the
        models and a small, but undesirable basis-set dependence.
    \item The current damping of the dispersion model based on molecular ionisation 
        potentials only is less than ideal and there is good reason to expect a 
        site--site damping model to perform better.
    \item More needs to be done to understand the origin of the polarization 
        damping. Like the dispersion damping, here too it is clear that the 
        damping model needs to depend on the pair of interacting sites,
        but there is evidence \cite{Misquitta13a} that the polarization damping
        differs strongly from that used for the dispersion.
        This is probably the least understood issue at present.
    \item The resulting potentials are for rigid monomers only.
        However, as the potential parameters are closely associated with the
        properties of the atoms in the interacting molecules through either
        the ISA or the DF-based partitioning methods, it is possible that
        these models may be applicable to flexible monomers.
        This conjecture needs to be tested.
    \item One of the most serious limitations of the approach we have described here
        is that there are very few simulation programs capable of using these potentials.
        Most simulation programs use the simpler Lennard-Jones models with point-charge
        electrostatic models. However, distributed multipoles are being increasingly
        available in simulations codes: both \OpenMM \cite{OpenMM-EastmanEtAl13}
        and \DLPOLY \cite{TodorovSTD06} allow the use of distribute multipoles
        and simple polarization models, but only the \Orient \cite{Orient} 
        and \DMACRYS \cite{PriceLWHPKD10} programs currently support the use of
        the anisotropy terms present in our more complex potentials.
        We do not doubt that this situation will change as potential
        development using the methods described in this paper becomes more streamlined
        and easy to use, and as we accumulate evidence that these more elaborate
        potentials do result in higher predictive accuracy.
\end{itemize}

It should be apparent that the ISA --- in particular, the BS-ISA algorithm --- 
plays a central role in the methodology we have described. 
Consequently it should come as no surprise that some of the issues listed above
may be resolved using data extracted from the ISA atomic densities.
In a forthcoming paper \cite{VanVleetMSS15} we will describe how the
dispersion damping issue may be resolved using the ISA, and also how even
more of the short-range parameters may be derived rather than fitted.

However, there are issues with the models we have presented here. 
Second virial coefficients are well reproduced using our isotropic and 
anisotropic potentials, though all three models give $B(T)$ somewhat too 
negative. This indicates that the models are somewhat too attractive
on the average. We have established that there are indeed regions of configuration
space where all potentials systematically overbind and these are associated
with stacked-like configurations. While we do not fully understand the
origin of the problem, it is possible that the additivity assumption
we have made in the definition of $\rho_{ab}$ in eq.~\eqref{eq:shape-func-additive}
is inappropriate, and also that the SAPT(DFT) interaction energies are themselves
too attractive for these configurations due to the known problems with the
\deltaHF term for dispersion-bound systems \cite{PodeszwaS05,PatkowskiSJ06}.
We are actively engaged in understanding these issues.

% \section{Programs}
% \label{sec:programs}
% 
% Many of the theoretical methods described in this paper are implemented
% in programs available for download. Some of these, together with their main
% uses in the present work, are:
% \begin{itemize}
%   \item \CamCASP 5.9 \cite{CamCASP}: 
%   Calculation of WSM polarizabilities, the dispersion models, the SAPT(DFT)
%   energies, and overlap models.
%   \item \ORIENT\ 4.8 \cite{Orient}: Localization of the
%   distributed polarizabilities, calculation of dimer energies using the
%   electrostatic, polarization and dispersion models, visualization of the
%   energy maps, fitting to obtain the analytic atom--atom potentials,
%   and basin-hopping simulations.
%   \item {\sc Dalton 2.0} \cite{DALTON2}: DFT calculations. 
%   A patch \cite{SAPT2008} is needed to enable {\sc Dalton 2.0} 
%   to work with \CamCASP.
% \end{itemize}

\section{Acknowledgements}
AJM would like to thank Prof Sally Price for initiating and motivating
this project and supporting it, particularly in its early stages,
Dr Richard Wheatley for useful discussions related to the ISA,
and Lauretta Schwartz for 
preliminary work on rank reduction of the multipole model.
We would like to thank Mary J. Van Vleet for useful comments on the manuscript.
AJM would also like to thank Queen Mary University of London for
computing resources, the Thomas Young Centre for a stimulating
environment, and the Cambridge University Library for generous resources.

This work was partially funded by EPSRC grant EP/C539109/1.

\section{Supplementary Information}
The SI included with this paper contains details of the three potentials
derived in this paper. Additionally, plots referenced but not included 
in this paper are provided in the SI.

% Now include the BIBLIOGRAPHY.bib file that contains all references and uses
% the abbreviations from the above file.
\setlength\bibsep{2pt}
%\bibliography{BIBLIOGRAPHY}
\providecommand{\latin}[1]{#1}
\providecommand*\mcitethebibliography{\thebibliography}
\csname @ifundefined\endcsname{endmcitethebibliography}
  {\let\endmcitethebibliography\endthebibliography}{}

\end{document}